\documentclass{aa}  
\usepackage{graphicx}
\usepackage{xcolor}

\usepackage{txfonts}
\usepackage[colorlinks=true,linkcolor=blue,citecolor=blue,urlcolor=blue]{hyperref}
\usepackage{threeparttable, tablefootnote}
\usepackage{xspace}
\newcommand{\citenobracket}[1]{%
  \citeauthor{#1} \citeyear{#1}%
}

\newcommand{\eg}{{e.g.}\xspace}

\newcommand{\lcdm}{\ensuremath{\small \textnormal{$\Lambda$CDM}}\xspace}

\newcommand{\sm}{\ensuremath{\,{\rm M_{\odot}}}\xspace}
\newcommand{\smyr}{\ensuremath{\,{\rm M_{\odot} \,yr^{-1}}}\xspace}
\newcommand{\slu}{\ensuremath{\,{\rm L_{\odot}}}\xspace}

\newcommand{\si}{\ensuremath{\,{\!\sim\!}\,}\xspace}

\newcommand{\kpc}{\ensuremath{\,{\rm kpc}}\xspace}

\newcommand{\ikpccube}{\ensuremath{\,{\rm kpc^{-3}}}\xspace}
\newcommand{\pc}{\ensuremath{\,{\rm pc}}\xspace}

\newcommand{\iyr}{\ensuremath{\,{\rm yr^{-1}}}\xspace}
\newcommand{\Myr}{\ensuremath{\,{\rm Myr}}\xspace}
\newcommand{\Gyr}{\ensuremath{\,{\rm Gyr}}\xspace}
\newcommand{\kms}{\ensuremath{\,{\rm km}\,{\rm s}^{\rm -1}}\xspace}
\newcommand{\s}{\ensuremath{\,{\rm s}}\xspace}
\newcommand{\km}{\ensuremath{\,{\rm km}}\xspace}
\newcommand{\kelvin}{\ensuremath{\,{\rm K}}\xspace}

\newcommand{\dex}{\ensuremath{\,{\rm dex}}\xspace}
\newcommand{\as}{\ensuremath{\,{\rm arcsec}}\xspace}

\newcommand{\icmsq}{\ensuremath{\,{\rm cm^{-2}}}\xspace}
\newcommand{\icmcube}{\ensuremath{\,{\rm cm^{-3}}}\xspace}

\newcommand{\dg}{\ensuremath{^{\circ}}\xspace}
\newcommand{\HI}{{\textnormal{H}}{\small \textnormal{I}}\xspace}

\newcommand{\Rvir}{\ensuremath{R_{\rm vir}}\xspace}

\newcommand{\chA}{}
\newcommand{\chII}{}

\renewcommand{\arraystretch}{1.3}

\defcitealias{Blana2020}{B20}
\defcitealias{Bland-Hawthorn2016}{BG16}
\defcitealias{Adams2018}{AO18}

\newcommand{\coB}[1]{}

\newcommand{\chIII}{}

\let\ch\chII
\newcommand{\LEt}[1]{}
\begin{document}     

\title{The Milky Way satellite galaxy Leo~T: A perturbed cored dwarf}

\author{\href{http://orcid.org/0000-0003-2139-0944}{Mat\'ias Bla\~na}\inst{\ref{inst1}}
\and \href{http://orcid.org/0000-0001-6879-9822}{Andreas Burkert}\inst{\ref{inst2},\ref{inst3},\ref{inst4}}
\and \href{http://orcid.org/0000-0002-3989-4115}{Michael Fellhauer}\inst{\ref{inst5}}
\and \href{http://orcid.org/0000-0002-9019-9951}{Diego Calder\'on}\inst{\ref{inst6}}
\and \href{http://orcid.org/0000-0002-8759-941X}{Manuel Behrendt}\inst{\ref{inst2},\ref{inst3},\ref{inst4}}
\and \href{http://orcid.org/0000-0003-1318-8631}{Marc Schartmann}\inst{\ref{inst2},\ref{inst3},\ref{inst4}}
}

\institute{Instituto de Alta Investigaci\'on, Universidad de Tarapac\'a, Casilla 7D, Arica, Chile \email{matias.blana.astronomy@gmail.com}\label{inst1} 
\and Max-Planck-Institut f\"ur extraterrestrische Physik, Gießenbachstraße 1, D-85748 Garching bei M\"unchen, Germany \label{inst2},
\and Universit\"ats-Sternwarte, Fakult\"at f\"ur Physik, Ludwig-Maximilians-Universit\"at M\"unchen, Scheinerstraße 1, D-81679 M\"unchen, Germany \label{inst3},
\and Excellence Cluster ORIGINS, Boltzmannstr 2, D-85748 Garching bei M\"unchen, Germany\label{inst4}
\and Departamento de Astronom\'ia, Universidad de Concepci\'on, Avenida Esteban Iturra s/n, Casilla 160-C, 4030000 Concepci\'on, Chile\label{inst5}
\and Hamburger Sternwarte, Universit\"at Hamburg, Gojenbergsweg 112, 21029 Hamburg, Germany,\label{inst6}
}

\date{Received 24 July 2024; Accepted 15 October 2024}
% \abstract{}{}{}{}{} 
% 5 {} token are mandatory
 
\abstract{
% \LEt{***The English in this work is already very good and it did not require full
% language editing. I corrected the Abstract and Introduction and then the Conclusion of the
% paper in detail. Please review the corrections I made throughout (see diff file) and the
% notes I added (flagged with ***), and apply the corrections to the whole paper. For more
% details, see https://www.aanda.org/images/stories/author/EnglishGuide-2021.pdf }\\\\\LEt{***General notes: \\a) I edited to UK English convention. \\b) A\&A uses the past
% tense to describe the specific steps used in a paper and the present tense to
% describe general methods and recent findings. Please make sure this is followed
% throughout. For details, see Sect. 6 of the Language Guide: https://www.aanda.org/
% for-authors/language-editing/6-verb-tenses \\c) Instrument and programme names are
% introduced (when appropriate) at first use in the main text. All other abbreviations
% and acronyms are introduced at first use, once in the Abstract and again in the
% main text. All abbreviations should be used consistently. Please check and amend
% as necessary throughout.\\d) question mark in the title: In a formal paper, direct
% questions should be avoided, and question marks removed from titles and headings. I added the word "possible" (also in running title), or it could be simply "A perturbed cored dwarf".}\\\\
The impact of the dynamical state of gas-rich satellite galaxies at the early moments of their infall into their host systems and the relation to their quenching process are not completely understood at the  low-mass regime.
Two such nearby systems are the infalling Milky Way (MW) dwarfs Leo~T and Phoenix located near the MW virial radius at $414\kpc\,(1.4 \Rvir)$, both of which present intriguing offsets between their gaseous and stellar distributions.
 Here we present hydrodynamic simulations with {\sc ramses} to \chII{reproduce the observed dynamics of Leo~T:} its $80\pc$ stellar-\HI offset and the 35\pc offset between its older ($\gtrsim5\Gyr $) and younger  ($\sim\!200\!-\! 1000\Myr$) stellar population.
We considered internal and environmental properties such as stellar winds, two \HI components, cored and cuspy dark matter profiles, and different satellite orbits considering the MW circumgalactic medium.
We find that the models that best match the observed morphology of the gas and stars include mild stellar winds that interact with the \HI generating the observed offset, and dark matter profiles with extended cores. 
\chII{The latter} allow  \ch{long oscillations of the off-centred} younger stellar component, due to long mixing timescales ($\gtrsim200\Myr$), and the slow precession of near-closed orbits \chIII{in} the cored potentials; instead, cuspy and compact cored dark matter models result in the rapid mixing of the material ($\lesssim 200\Myr$). 
These models predict that non-equilibrium substructures, such as spatial and kinematic offsets, are likely to persist in cored low-mass dwarfs and to remain detectable on long timescales in systems with recent star formation.}
%{}{}{}{}
\titlerunning{The Milky Way satellite galaxy Leo~T: A perturbed cored dwarf}
\authorrunning{Bla\~na et al.}
\maketitle

%%%%%%%%%%%%%%%%%%%%%%%%%%%%%%%%%%%%%%%%%%%%%%%%%%%%%%%%%%%%%%%%%%%%%%%%%%%%%%%%%%%%
%%%%%%%%%%%%%%%%%%%%%%%%%%%%%%%%%%%%%%%%%%%%%%%%%%%%%%%%%%%%%%%%%%%%%%%%%%%%%%%%%%%%
%%%%%%%%%%%%%%%%%%%%%%%%%%%%%%%%%%%%%%%%%%%%%%%%%%%%%%%%%%%%%%%%%%%%%%%%%%%%%%%%%%%%
\section{Introduction}
\label{sec:intro}
\ch{Dynamical equilibrium is a fundamental and common assumption in the dynamical and mass modelling  of observed galaxies, and  of the material used to trace the potential \citep{Binney2008}.
However, this material could be under different degrees of perturbation, depending on the specific system.}
The \textit{Gaia} observatory, for example, has revolutionised the understanding of our home galaxy, revealing a number of transient stellar substructures in the Milky Way (MW)  or `digging up' satellite galaxy `fossils'\LEt{***perhaps finding artefacts? or do you mean actual excavation of Galactic material here on Earth?:}\coB{REP: the MW community usually talks about galactic "archaeology" for its similarities to real archaeology, while "digging up" substructures of accreted galaxies from huge databases (Gaia, etc). would it be correct/allowed its use here then?} from previous accretion events \citep{Belokurov2018, Deason2018, Haywood2018, Helmi2018}.
The most massive dwarf galaxy satellites, the Magellanic Clouds, show substructures in a much more obvious state of perturbation.
Furthermore, \textit{Gaia} has also allowed us to determine
\ch{the internal kinematic structure of the closest dwarf spheroidal galaxies (dSphs) ($\lesssim 150\kpc$) \citep{Martinez-Garcia2021,Qi2022,Tolstoy2023}, 
as well as their orbital properties around the MW halo \citep{Fritz2018,McConnachie2020,McConnachie2021}}. 
However, the more distant population of satellites, located in the outskirts of the MW, near or beyond its virial radius ($\Rvir$), has significantly larger proper motion errors given their large distances, which only allow us to determine some constraints on their
orbital properties, determining, for example, if some of these are falling into the Galaxy for the first time
\chII{(\citenobracket{Blana2020}, hereafter \citetalias{Blana2020}; \citenobracket{McConnachie2020,McConnachie2021}).}

Furthermore, this outskirts population, with Leo~T, Phoenix, Eridanus II, and others as members,
is particularly interesting for studying the accretion history of the MW at later times \citep[][and \citetalias{Blana2020}]{Diemer2017a, Diemer2020,Deason2020,Fritz2020,Diemer2021,Bakels2021,ONeil2021}, and also for understanding the quenching process of satellites that fall into their host \citep{Rodriguez2019, Simpson2018,Buck2019}, allowing us to understand the transformation of gas-rich dwarf galaxies into dSph galaxies.

In addition, dwarf galaxies are the most abundant type of galaxy and play an important role as laboratories to study the nature of dark matter, which remains one of the most important challenges in astrophysics today. 
Dwarf galaxies can help us probe dark-matter models at the low-mass regime and address studies such as the cuspy versus cored dark-matter profile subject.
We have known for some time that dwarfs can contain cored dark-matter densities \citep{Burkert1995} instead of the predicted cuspy Navarro-Frenk-White (NFW) profile produced in $\lcdm$ dark matter-only  simulations \citep{Navarro1996}.
Many theories have addressed this phenomenon, suggesting that it is a consequence of gravitational interactions with the baryonic mass components \citep{Navarro1996b, Read2006, Ogiya2014a, Ogiya2014b}
or an intrinsic property of dark matter such as warm dark matter \citep{Feng2010, Bullock2017}, fuzzy dark matter \citep{Hu2000,Burkert2020b}, self-interacting dark matter \citep{Burkert2000,Markevitch2004,Harvey2018,Nishikawa2020}, or more common external mechanisms such as tidal stripping and shocking due to interactions with the host or a nearby galaxy.

Therefore, it is of interest to identify pristine isolated dwarfs falling for the first time on the MW and to distinguish them from backsplash satellites \citepalias{Blana2020}, which could have been affected by the MW in the past, or even by the Andromeda galaxy in the case of more exotic `renegade' satellites \citep{Knebe2011,Fouquet2012,TeyssierM2012} \citep[also called Hermeian satellites;][]{Newton2021}, or to distinguish them from satellites that arrive within smaller groups such as the Magellanic Cloud subgroup \citep{Sales2017,Kallivayalil2018,Pardy2020}.\\

In this paper we complement the satellite orbital study of Leo~T performed in \citetalias{Blana2020}, 
\ch{in which the authors explored first infall and backsplash orbital solutions, where both solutions lay within the proper motions measurements given the large observational errors \citep{McConnachie2020}.}
Here we investigate the internal dynamics of Leo~T and possible perturbations generated by the interaction with the environment as it enters the MW, or due to internal processes. 
In particular, we focus on qualitatively reproducing \chII{the main properties of 
the exquisite \HI observations of \citet[][hereafter \citetalias{Adams2018}]{Adams2018}}, and the
observed offset between the distributions of the \HI and the younger and older stellar populations.
The article continues with Sect. \ref{sec:mod} where we describe the  set-up of our models based on the main observational properties of Leo~T, its environment, and the scenarios that we explored. In Sect. \ref{sec:res} we present the results, with separate subsections for the analysis of environmental effects and the analysis of internal processes. We conclude with a summary and discussion in Sect. \ref{sec:sum}.

\begin{table*}[ht!]
\centering
    \scriptsize
    % \tiny    
    % \centering
    % \raggedleft
    \caption{Main parameters of the simulation set-up.}
    \hskip-2.cm
    \setlength{\tabcolsep}{4pt} % Default value: 6pt
    \renewcommand{\arraystretch}{1} % Default value: 1
    \begin{tabular}{ccclcclcccccc}\hline
    1 & 2 & 3 & 4 & 5 & 6 & 7 & 8 & 9 & 10 & 11 & 12 & 13 \\\hline
    \shortstack{Model\\Name} & \shortstack{Simulation\\Name} & \shortstack{Scenario\\ \,} & \shortstack{Dwarf\\Model} & \shortstack{DM\\ profile} & \shortstack{$M_{300}$\\$10^6\sm$}  & \shortstack{Initial Offset \\ Type, PA[\dg]}  
    & \shortstack{$\dot{M}$\\$10^{-5}\smyr$}  & \shortstack{$w$\\$\kms$} 
    & \shortstack{Dwarf gas\\component} & \shortstack{Orbital F.V.\\$u_{\rm t}$,$|V|$ km/s} & \shortstack{Motion\\$\theta_{\rm sky}[\dg]$}
    & \shortstack{Orbit\\type}  \\\hline
    % %%%%%%%%%%%%%%%%%%%%%%%%%%%%%%%%%%%%%%%%%%%%%%%%%%%%%%%%%%%%%%%%%%%%%%%%%%%%%%%%%%
    %%%%%%%%%%%%%%%% NEW TABLE  %%%%%%%%%%%%%%%%%%%%%%%%%%%%%%%%%%%%%%%%%%%%%%%%%%%%%%
    %%%%%%%%%%%%%%%%%%%%%%%%%%%%%%%%%%%%%%%%%%%%%%%%%%%%%%%%%%%%%%%%%%%%%%%%%%%%%%%%%%
    % \hline\hline\hline\hline\hline\hline
     M1 & S4M4 & I & D1 &  Burkert   & 1.7 &  -    & - & - &   WNM     &   50, 83 & 52 & BA \\ 
     M2 & S4M5 & I & D1 & Burkert    & 1.7 &  -    & - & - &   WNM     &  100, 120 & 33 & BA--FI \\
     M3 & S4M6 & I & D1 & Burkert    & 1.7 &  -    & - & - &   WNM     &  200, 210 & 18 & FI \\     
     M4 & S4M7 & I & D1 & Burkert    & 1.7 &  -    & - & - &   WNM     &  300, 307 & 12 & FI \\
     M5 & S4M9 & I & D1 & Burkert    & 1.7 &  -    & - & - &   WNM     &  100, 120 & 33 & BA--FI+P \\
     M6 & S11M12 & I & D2 & Burkert    & 8.0 &  -    & - & - &   WNM     &  200, 210 & 18 & FI \\     
     M7 & S11M11 & I & D2 & Burkert    & 8.0 &  -    & - & - &   WNM     &  300, 307 & 12 & FI \\\hline
     %%%%%%%%%%%%%%%%%%%%%%%%%%%%%%%%%%%%%%%%%%%%%%%%%%%%%%%%%%%%%%%%%%%%%%%%%%%%%%%%%%%%%%%%%%%%%%%%%
     M8 & A1M1 & II & D1  & Burkert   & 1.7 &  stellar(r),  46  & 10 & 20  &   WNM     &  200, 210 & 18 & FI  \\
     M9 & A1M6 & II & D2  & Burkert   & 8.0 &  stellar(r),  46  & 20 & 30  &   WNM     &  200, 210 & 18 & FI \\
     M10 & S1M12& II & D1t & Burkert   & 1.7 &  stellar(r), ~~0  & 10 & 20  &   WNM+CNM &  200, 210 & 18 & FI \\ 
     M11 & SZM4 & II & D3  & Burkert   & 15.5 &  stellar(r),  46 & 40 & 50  &   WNM     &  200, 210 & 18 & FI \\
     M12 & A1M7 & II & D4  & NFW       & 8.0 &  stellar(r),  46  & 20 & 30  &   WNM     &  200, 210 & 18 & FI \\
     M13 & S7M3 & II & D4  & NFW       & 8.0 &  stellar(r),  46  & 10 & 20  &   WNM     &  200, 210 & 18 & FI \\ 
     M14 & S7M4 & II & D4  & NFW       & 8.0 &  stellar(c),  46  & 10 & 20  &   WNM     &  200, 210 & 18 & FI  \\\hline
     %%%%%%%%%%%%%%%%%%%%%%%%%%%%%%%%%%%%%%%%%%%%%%%%%%%%%%%%%%%%%%%%%%%%%%%%%%%%%%%%%%%%%%%%%%%%%%%%%
     M15 & S10M1 & III & D1  & Burkert  & 1.7 &  gas(r),  46   & - & -  &   WNM &  200, 210 & 18 & FI   \\
     M16 & S10M2 & III & D2  & Burkert  & 8.0 &  gas(r),  46   & - & - &  WNM &  200, 210 & 18 & FI     \\
     M17 & S10M3 & III & D4  & NFW      & 8.0 &  gas(r),  46   & - & - &  WNM &  200, 210 & 18 & FI     \\ 
     M18 & S8M2 & III  & D1t & Burkert  & 1.7 &  gas(r),  46   & -  & -  &   WNM+CNM &  200, 210 & 18 & FI \\
     M19 & S8M7 & III  & D2t & Burkert  & 8.0 &  gas(r),  46   & -  & -  &   WNM+CNM &  200, 210 & 18 & FI \\
     M20 & S8M8 & III  & D4t & NFW      & 8.0 &  gas(r),  46   & -  & -  &   WNM+CNM &  200, 210 & 18 & FI \\
     M21 & S8M6 & III  & D4t & NFW      & 8.0 &  gas(c),  46   & -  & -  &   WNM+CNM &  200, 210 & 18 & FI \\\hline   
    \end{tabular}
    \label{tab:setup}
    \begin{tablenotes}
    \small
    \item {Columns:} 1) Model name; 2) Run name; 3) Scenario; 4) Dwarf models with different mass distributions, where the letter `t'  indicates a set-up with two gas components; 5) Dark matter profile; 6) Dynamical mass within 300\pc;
    7) Offset of stellar or gaseous component, with  offset type  radial (r),  circular (c), or none (-), \chA{and the position angle of the initial offset}; 
    8) Stellar winds outflow mass rate; 
    9) Velocity of stellar winds outflows; 
    10) Included gaseous components (warm neutral medium WNM, cold neutral medium CNM);
    11) FV: magnitude of the final GSR tangential velocity ($u_{\rm t}$) and total velocity ($V$) of the orbits; 
    12) Angle between the LOS and tangential (sky) velocities of the satellite orbit (e.g. $\theta_{\rm sky}=0\dg$   corresponds to a tangential velocity only, with the satellite moving on the plane of the sky);
    13) Orbit type as backsplash orbit (BA), first infall (FI), or in the limit between the two solutions (BA--FI), and a comment if a perturbation (+P) is included in the IGM gas density. Videos of the simulations can be found at this link \href{https://www.youtube.com/playlist?list=PLoegIUZ3yJ9FuDldEF-USJWd-o_hyLavl}{{(here)}}.     \end{tablenotes}
\end{table*}
\LEt{***I will now do a deep edit of the Conclusion and the figure and table captions, and then will
look quickly through the rest of the paper for any big problems. I will not necessarily mark
everything (or at first instance), so please check carefully throughout and apply any
corrections I have made to the entire paper }
\section{Modelling Leo~T and its environment} 
\label{sec:mod}
Here we describe the set-up of the simulations, starting with the environment Leo~T and its orbit, followed by the dynamical and mass models for the dwarf galaxy models, and finally the set-up of the three main scenarios (I,II,III) that we designed to explore possible solutions for the current dynamical state of the gas and stars in Leo~T. In Table \ref{tab:setup} are summarised the main parameters for each model and simulation.\footnote{Videos of the simulations can be found at this link \href{https://www.youtube.com/playlist?list=PLoegIUZ3yJ9FuDldEF-USJWd-o_hyLavl}{{(here)}}}

\subsection{The satellite's environment, \ch{orbit, and orientation}} 
\label{sec:mod:env}

\LEt{***A\&A uses the serial comma (Oxford comma) between three or more items in
a list: a, b, and c; x, y, or z. Please check and amend throughout as appropriate. }Given that Leo~T is located at a large distance from the Galactic centre with $r^{\rm GC}\!=\!414\kpc\sim 1.4\Rvir$ \citepalias{Blana2020} \citep[or a heliocentric distance of $D^{\sun}=409^{+29}_{-27}\kpc$][]{Clementini2012}, the weak effects of the MW gravitational tidal forces can be neglected. For example, a satellite with a virial mass of $10^7\sm$ or $10^8\sm$ would result in a Jacobi radius that is larger than its virial radius.
Furthermore, even the backsplash orbital solutions found by \citetalias{Blana2020} for Leo~T place the pericentre passages between 6 and 8\Gyr ago, while here we focus only in the last 1 to 2\Gyr of evolution, being still beyond the MW virial radius (see their Figs. 5 and 9). 
This allows us to implement a wind tunnel box simulation approximation that significantly reduces the computational time and cost of each run, 
which is performed with grid-based adaptive mesh refinement code {\sc ramses} \citep{Teyssier2002}. 
The domain is a box with sides of 20\kpc that reaches a maximum resolution of 2 - 4\pc (see Sect. \ref{sec:mod:num} for more details).
As the satellite moves in the outskirts of the MW, it feels the ram pressure of the environment. 
We compute the resulting wind speed of this environment based on the precomputed orbital positions
and velocities for different orbits of Leo~T using the software {\sc delorean} \citepalias{Blana2020}.
\ch{The wind is injected on one side of the box, while the remaining sides have open boundary conditions.}\\
\\
% The MW coronal wind velocity is derived from orbits calculated 2\Gyr into the past integrated back into the present.
% VEl WNM GSR −64.419372kms
% Vel Stars GSR −65.919372kms
% Vel CNM GSR −66.619372kms
The line-of-sight (LOS) velocity of the satellite is well determined, being $v^{\rm GSR}_{\rm los,\star}=-65.9\pm2.0\kms$ for the stars
in the Galactic Standard of Rest (GSR), or $v^{\sun}_{\rm los,\star}=38.1\pm2.0\kms$ in the heliocentric system \citep{Simon2007} and for the \HI gas is $v^{\rm GSR}_{\rm los,gas}=-64.4\pm0.1\kms$ ($v^{\sun}_{\rm los,gas}=39.6\pm0.1\kms$) \citepalias{Adams2018}.
\ch{However, the stellar proper motions have large errors given the large distance to the satellite ($\mu_{\alpha} \cos\delta=0.10^{+0.67}_{-0.69}\,{\rm mas/yr}$ and $\mu_{\delta}=0.10^{+0.67}_{-0.69}\,{\rm mas/yr}$), which result in tangential velocity measurement with errors $\Delta \sim\!5$ times larger than the estimated value \citep[see their Table 4]{McConnachie2020}}. 
The orbital analysis of \citetalias{Blana2020} finds that the observational constrains encompass the backsplash orbital solutions as well as the first infall solutions.
Therefore, we explore the orbits for both solutions \ch{where \citetalias{Blana2020} determined that the range of magnitude of the (GSR) tangential velocity values ($u_{\rm t}$) that allow backsplash solutions is $u_{\rm t}\leq69^{+47}_{-36}\kms$, with the uncertainty range given by the different MW and M31 mass models. The selected orbits are:}
\ch{
\begin{itemize}
    \item Backsplash orbit (BA): we use an orbit with a final GSR tangential velocity of $u_{\rm t}=50\kms$, that combined with the 
    $v^{\rm GSR}_{\rm los,\star}=-65.9\kms$ results in a final velocity of $V=83\kms$. As we are interested in the more recent history of Leo~T we model the last 2\Gyr of evolution for Scenario I, and the last 1\Gyr for Scenario~II and III, which are defined in Sect. \ref{sec:mod:scen}. As this is a wide BA orbit, the apocentre is at 500\kpc from the MW, which was reached -3\Gyr ago with a velocity of 34\kms, while -1\Gyr ago it reached a velocity of 58\kms, revealing the slow increase of velocity at those distances.
    \item First infall orbit (FI): here we explore two orbits that result in Leo~T being on its first infall onto the MW. 
    One has a final velocity of $u_{\rm t}=200\kms$ and $V=210\kms$, and velocities of 195 and 182\kms at -1 and -2\Gyr ago, respectively. 
    The extreme fast FI orbit $u_{\rm t}=300\kms$ and $V=307\kms$, allowed by the large observational proper motion errors, has velocities of 292 and 280\kms at -1 and -2\Gyr ago, respectively. 
    \item BA--FI orbit: here we take an orbit at the limit of BA and FI orbits, with a final velocity of $u_{\rm t}=100\kms$ and $V=120\kms$, and velocities of 100 and 86\kms at -1 and -2\Gyr ago, respectively. 
\end{itemize}
}
\ch{These different tangential velocities result in trajectories that cross the plane of the sky with different entrance vectors and possible orientations for Leo~T on the sky.
Therefore, we define here this `angle of attack' derived from the GSR tangential and LOS velocities,  
\begin{eqnarray}
\theta_{\rm sky}&=&90\dg-\arctan({u_{\rm t}/v_{\rm los})},
\label{eq:angle}
\end{eqnarray}
which defines the instantaneous motion of the satellite with respect to the plane of the sky where, for example, $\theta_{\rm sky}=0\dg$ would only correspond to a motion in the plane of the sky  and $\theta_{\rm sky}=90\dg$ only to an inward radial motion  and perpendicular to the plane of the sky.
For simplicity and that the direction of motion of Leo~T is not well constrained, we decided to project $\theta_{\rm sky}$ along the negative vertical axis of the images, meaning that in this frame the relative motion of the Galactic wind would move into the plane of the sky and in the direction of the positive vertical axis when $u_{\rm t}>0\kms$.
Following \citetalias{Blana2020} we adopt $v^{\rm GSR}_{\rm los}=-65.9\kms$ from the stellar velocity, 
where for simplicity here we assume that this value corresponds to the galactocentric radial velocity, 
as the large distance to Leo~T (420\kpc) results in an angular difference of the helio-galactocentric vector of only 0.9$\dg$.}
It is worth mentioning that the explored velocities encompass a range of values of \citet{Emerick2016}, who showed how the wind properties affect the overall properties of the gas stripping in Leo~T-type systems.  
Furthermore, \citet{Tonnesen2019} showed that a variable wind driven by the orbital history of a satellite around its host can affect its gas morphology, when compared to constant winds.\\
\\
\chII{We need to model the gas of the environment of Leo~T, which is currently at a galactocentric distance of
$r=415\kpc \sim 1.4\Rvir$ from the MW centre. 
At these distances, the environment changes from the circumgalactic medium (CGM) to the intergalactic medium (IGM) at the convention of $\sim 1\Rvir$ \citep{Tumlinson2017}.
Here we  name\LEt{***The present tense (rather than the future) is advised for generalities and when you discuss what can be found within your own work }\coB{Ok.} the medium density model as IGM, given that the satellite orbits explored for Leo~T in \citetalias[][(see their Fig. 5)]{Blana2020} place it in the IGM region in the last 2\Gyr for both orbital solutions (BA and FI).}
We model the density of this medium $\rho^{\rm MW}_{\rm IGM}\left(r\right)$ with an extrapolation of the hot halo beta model from \citet{Salem2015}, where the density varies as a function of the orbital distance $r$, while keeping the temperature constant to $T=10^{6.3}{\rm K}$.
This temperature would correspond to a sound speed of $207\kms$ for a fully ionised plasma of hydrogen with a cosmic helium mass fraction of 0.24. 
Extrapolating two hot halo models to Leo~T's current position the mass and number $(n)$ densities would correspond to $\rho^{\rm MW}_{\text{IGM}}=84\sm\ikpccube\,(2.6\times 10^{-6}\icmcube)$ \citep{Salem2015}, or $120\sm\ikpccube\,(3.6\times 10^{-6}\icmcube)$ \citep{Miller2015} \citepalias[see Fig. 3 in][]{Blana2020}.\\

In addition, hot gas halos can present substructures and density perturbations in their spatial distribution. We notice that Leo~T is located at $1.4\Rvir$ distance from the MW, a location where cosmological simulations predict the presence of shocks and a splashback (caustic) radius, which is a transition zone in the shape of the distribution of dark matter and gas \citep{Hurier2019,Diemer2020a,Deason2020,Aung2021,ONeil2021}.
Cosmological simulations of a MW-type galaxy also show strong gas density perturbations over the background density of more than one order magnitude, resulting from a substructure in the hot halo \citep{Dolag2015}. 
\chII{In addition, cosmological simulations suggest the presence of a colder phase component in the IGM ($10^3\!<\!T/\kelvin\!<\!10^5$, $-8\!<\!\log(n[{\rm cm}^{-3}])\!<\!-4$) associated with cosmic web filaments \citep{Galarraga-Espinosa2021}; however, additional physical processes, such as thermal conduction, may impact these results \citep[][]{Nipoti2004,Armillotta2017,Kooij2021}.}
Motivated by these findings we also explore models where we include strong gas density perturbations on top of the MW hot halo beta profile to test how our dwarf models react to these strong density variations while aiming at reproducing the gas-stellar densities offset observed in Leo~T. For simplicity, we adopt a sinusoidal perturbation of the form 
\begin{eqnarray}
\rho_{\rm env} &=& \rho^{\rm MW}_{\rm IGM} + \delta\rho\,\, {\rm, where}\\
\delta\rho &=& \epsilon\, \rho^{\rm MW}_{\rm IGM} \left(|\sin(r)| + \sin(r)\right)/2,
\label{eq:env}
\end{eqnarray}
which avoids negative values when going below the local IGM density value, with a maximum amplitude a factor $\epsilon$ times over the local IGM density. We note that we also include perturbations in the temperature to impose a local isobaric state, in order to avoid pressure gradients that could generate additional undesired secondary perturbations in the wind.

\subsection{Dwarf galaxy models}
\label{sec:mod:dwarf}
\LEt{***The specific steps you took for this particular research should be in the past simple
(We used, We extrapolated, The observation was performed). Descriptions of
methodology, universal truths, constants, and findings should be in the present
simple (The first step in the process is, Water boils at 100 degrees C, We find). Please check and amend as appropriate throughout. }
We built a sample of dwarf galaxy models for Leo~T with stellar, gaseous, and dark matter properties based on observational and theoretical estimates from the literature, which were then used to run wind tunnel simulations under various conditions to explore a range of different parameters, summarised in Table \ref{tab:setup}. The dark matter distribution is the most uncertain mass component, making this the main quantity that differs between dwarf models.
We used full N-body modelling for the stellar and dark-matter distributions, which allowed us to explore different types of perturbation to reproduce the gaseous and stellar morphology observed in Leo~T, particularly its \HI-stellar offsets. 
We generated the initial conditions for the dwarf models with the software {\sc dice} \citep{Perret2014,Perret2016}, which were then inserted into the wind tunnel models that were performed with {\sc ramses}. These models are defined as follows:

\begin{figure}
\begin{center}
\includegraphics[width=9.5cm]{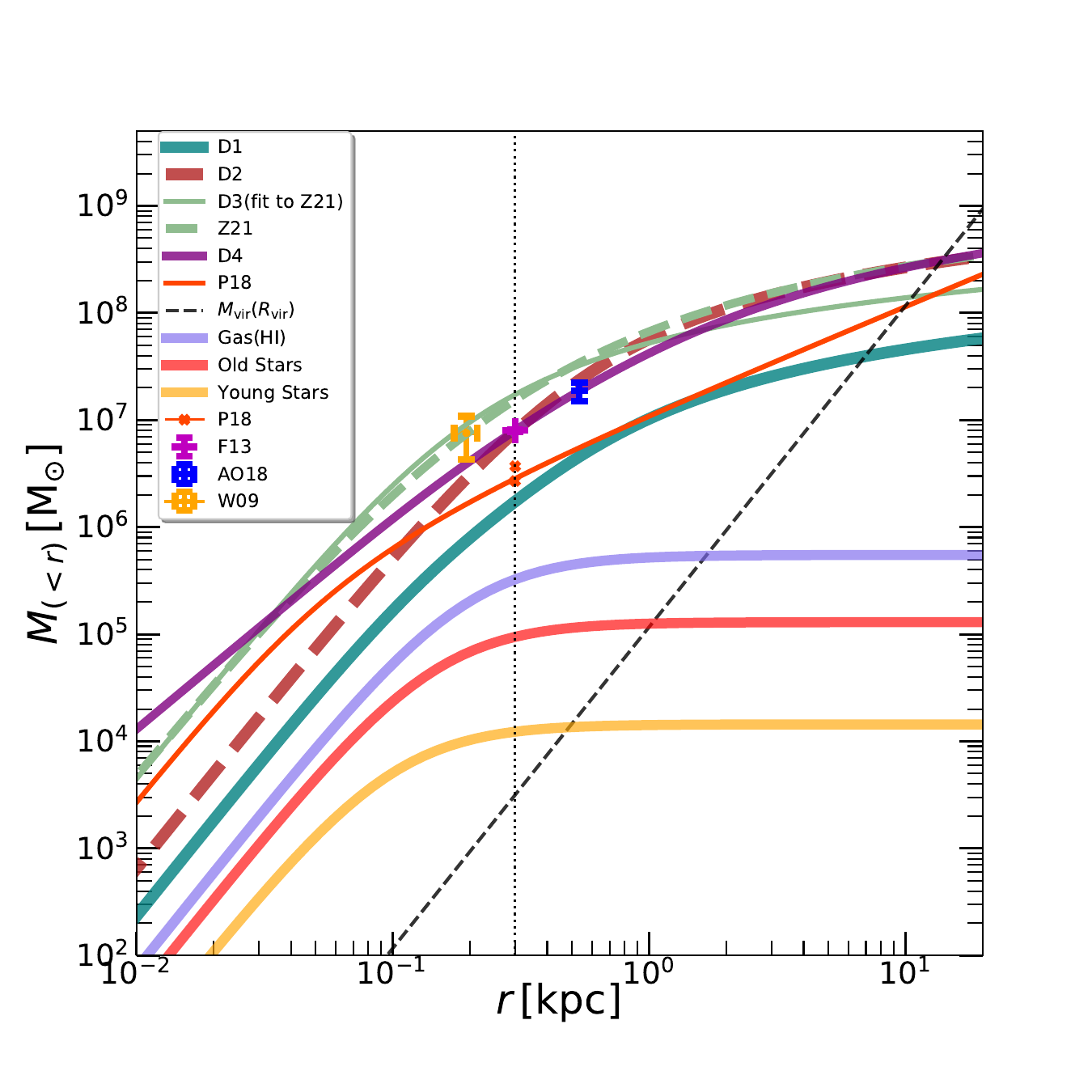}\\
\vspace{-0.5cm}
\caption{Cumulative mass radial profiles of the dwarf galaxy models used for Leo~T shown in Table \ref{tab:setup}: Our fiducial Burkert model D1, the massive Burkert model D2, and the NFW model D3.
We also show the pseudo-isothermal dark halo model of \citet{Patra2018} (P18) fitted to \HI kinematic data of \citetalias{Adams2018}, and the model of \citet{Zoutendijk2021} (Z21) with our adapted model D4 fitted to match a Burkert model within 400\pc. 
The models include gaseous and stellar components;  we also plotted them separately modelled by Plummer mass profiles with parameters determined from observations \citep{DeJong2008, Weisz2012, Adams2018, Blana2020}. 
We include dynamical mass estimates derived from gaseous kinematics \citep[F13, A18, P18;][]{Faerman2013,Adams2018,Patra2018} and stellar kinematic observations \citep[W09;][]{Walker2009a}.
The dashed black line indicates the relation between the virial mass ($M_{\rm vir}$) and radius ($R_{\rm vir}$) where we truncate the halos.
The vertical dotted line marks the standard mass radius $M_{300}$ at $300\pc$ \citep{Strigari2008}, used here to compare the masses between dwarf models in Table \ref{tab:setup}.}
\label{fig:fig_massprof}
\end{center}
\end{figure}

\subsubsection{Dark matter component}
\label{sec:mod:dwarf:dm}
 Due to the large distance to Leo~T and its faint stellar distribution, the determination of its dark-matter distribution and concentration is difficult to constrain.
Many dark matter mass estimates in the literature explore cuspy density profiles with NFW models. However, cored dark-matter profiles are probed with different functional forms such as the Burkert model, the cored NFW model, the soliton model, or the pseudo-isothermal model. 
This makes discriminating between different cored models difficult.
\citet{Simon2007} measured stellar kinematics of 19 stars, determining a velocity dispersion of $7.5\pm1.6\kms$. Using analytical mass estimators they obtained a dynamical mass of $8.2\pm3.6\times10^6\sm$ within the half-light radius, $R_{\rm h}^{\rm V} = 73\as\,(145\pc)$, which here we re-scaled to $7.6\times10^6\sm$, considering the new closer distance estimates of \citet{Clementini2012}.
% while re-calculating it to the latest heliocentric distance measurement of Leo~T \citetalias{Blana2020} find $7.6\pm3.3\times10^6\sm$.
\citet{Faerman2013} fit hydrostatic models to the \HI observations of \citet{RyanWeber2008} and find $M_{300}=8.0\pm0.2\times10^6\sm$ within 300\pc, while \citetalias{Adams2018} use their global \HI velocity dispersion and a dynamical equilibrium (virial) relation to find a value of
$M_{400}=19\times10^6\sm$ within 400\pc.
Furthermore, \citet{Patra2018} finds the lowest dark-matter mass estimates to date, which are derived from fitting hydrostatic models to the \HI density and velocity dispersion data of \citetalias{Adams2018}. The author also used pseudo-isothermal dark-halo models, finding models with dark-matter masses of $M_{300}=2.7-3.7\times10^6\sm$ measured at 300\pc for a wide range of estimates for the core radius and central density of the dark-matter halo.
Dwarf galaxy evolution simulations find dark matter cores with sizes comparable to their stellar half-mass radius, which are produced by the gravitational perturbations of the baryonic components that are driven by star formation processes, which then gravitationally perturb the dark matter centre \citep{Read2006,Ogiya2014a}.
Moreover, recent mass estimates of \citet{Zoutendijk2021} used MUSE observations to increase the number of stellar radial velocity measurements by up to 75 members to fit cuspy (NFW) and different cored models, finding more massive and compact cores with $r_{\rm c}\!=\!66\pc$ for the model profiles of \citet{Lin2016}.\\
In summary, the dynamical mass estimates of Leo~T in the literature reveal a degree of degeneracy between the scale length and the central halo density of the dark-matter models, which would require more observations to solve.
Therefore, here we proceeded to explore dwarf galaxy models with a range of masses and densities with cuspy and cored dark-matter models with different core sizes, as shown in Table \ref{tab:setup}.
The mass profiles of these models are shown in Fig.~\ref{fig:fig_massprof}, which were derived from fits to the dynamical mass estimates from the literature.
In the following, we explain the dark-matter models used for the different dwarf galaxy models, which is then followed by the set-ups of the stellar and gaseous components:\newline

\begin{enumerate}
    \item Dwarf Model D1 (Fiducial): We set our fiducial dwarf model D1
with a Burkert dark-matter density profile, which has a density core in the centre while beyond its scale length the density closely follows an NFW profile \citep{Burkert1995,Burkert2015}.
We \chIII{took} a dark-matter core radius comparable in size to the gas distribution with $r_0\!=\!400\pc$, and an enclosed mass at 1\kpc that matches the estimate of \citet{Patra2018}, who used a pseudo-isothermal halo. 
Therefore, the mass profile decreases faster in the inner region, having an enclosed mass at 300\pc of $M_{300}=1.7\times10^6\sm$. 
Although these values are arbitrary, they set a low-mass model, which allows us to explore the dynamics of models that are more sensitive to perturbations.\newline

\item Dwarf Models D2 and D3: We set model D2 with a more massive cored dwarf galaxy model that follows the mass estimates derived from \HI kinematics of \citet{Faerman2013} at 300\pc and \citetalias{Adams2018} at 400\pc, also choosing a core radius of $r_0\!=\!400\pc$ and $M_{300}=8\times10^6\sm$. 
Model D3 also has a Burkert halo, but fitted to match the halo model of \citet{Zoutendijk2021} within 300\pc obtaining $M_{300}=15.5\times10^6\sm$ and $r_0\!=\!142\pc$, as shown by the mass profiles in Fig.~\ref{fig:fig_massprof}. These more massive models better match the upper-mass estimates in the literature and help us explore the effect of perturbations on Leo~T for a more massive scenario. However, while model D3 agrees with the mass estimates from the stellar kinematics, it has $\sim2$ times more mass within 300\pc than the mass estimates from gas kinematics.\newline

\item {Dwarf Model D4:} Finally, we set dwarf model D4 with a cuspy dark matter density model using an NFW profile with the same mass at 300\pc as model D2 ($M_{300}=8\times10^6\sm$) with a scale length of $r_{\rm s}\!=\!770\pc$. We note that due to its high central density, within 30\pc it has a mass larger than that of the dwarf model D3, as seen in Fig.~\ref{fig:fig_massprof}.
\end{enumerate}
\chIII{These settings generate a diverse range of dark-matter profiles which allow us to probe the effects of external and internal perturbations with different magnitudes and configurations.}
The total virial mass in each dwarf model is determined in {\sc dice} by truncating the galaxy's cumulative mass profile at the virial radius $R_{200}$, where the mean density is 200 times higher than the density of cosmic matter, as shown in Fig.~\ref{fig:fig_massprof}. We use $10^6$ particles for the dark-matter component.

\subsubsection{Stellar component(s)}
\label{sec:mod:dwarf:st}
 \citet{DeJong2008} find an old stellar component in Leo~T with a luminosity of $\sim 10^5\slu$ and half light radius $R_{\rm h}=145\pc\,(73\as)$, as well as a younger component with a luminosity of $\sim 10^4\slu$, an age ranging $\sim200\Myr - 1000\Myr$,  and a slightly more concentrated spatial distribution with half light radius $R_{\rm h}^{\rm Y}=102\pc$.  
The existence of this younger component is consistent with their estimates of the star formation history in Leo~T and also with the study of \citet{Weisz2012} \citep[and the more recent][]{Vaz2023}, showing a bursty star formation history with a recent drop, attributed to a quenching or to an IMF sampling effect.
Moreover, spectroscopy of 19 stars revealed a metal-poor population with ${\rm [Fe/H]}=-2.19\pm0.10\pm(0.35)_{\sigma_{\rm [Fe/H}]}$ \citep{Simon2007}, 
while recent estimates find ${\rm [Fe/H]}\!=\!-1.53\pm0.05 \pm (0.26\pm0.06)_{\sigma_{\rm [Fe/H]}}$ \citep{Vaz2023}.
We set up all dwarf models with both stellar distributions using Plummer profiles based on the observed photometric structural parameters determined in \citet{Irwin2007, DeJong2008}, considering 90\% of the total stellar mass in the old stellar component with $1.3\times10^5\sm$ \citep{Weisz2012}, and the younger stellar component with $1.4\times10^4\sm$ \citet{DeJong2008}, each with a particle mass resolution of $1\sm$. 
\citet{DeJong2008} also reported an intriguing offset between the old and the younger stellar components of 
\chII{about $\Delta R_{\rm o|y}\!=\! 35.7\pc\,(18\as)$ in projection, significant to within 2 sigma and that could be even larger in de-projection.
Furthermore, this spatial offset has a phase-space counterpart, where recent observations revealed that the older stellar component has a velocity and dispersion of
$v_{\rm los}^{\odot}\!=\!39.39^{+1.32}_{-1.29}\kms$ and $\sigma_{\rm los}\!=\!7.07^{+1.29}_{-1.12}\kms$, 
while the younger stellar population has
$v_{\rm los}^{\odot}\!=\!39.33^{+2.09}_{-2.14}\kms$ and $\sigma_{\rm los}\!=\!2.31^{+2.68}_{-1.65}\kms$} \citep{Vaz2023}.
\chII{Such offsets are plausible if younger stars were formed in star clusters displaced from the dwarf's centre, as seen, for example, in the MW dwarf galaxy Eridanus II with a star cluster shifted $23-45\pc$ in projection from the dwarf centre \citep{Crnojevic2016,Amorisco2017, Simon2021}. 
Furthermore, clusters in disruption within cored dwarf galaxies can present stellar substructures for long timescales \citep{AlarconJara2018}.
Depending on the scenario considered in Sect. \ref{sec:mod:scen}, we  provide an initial offset to the younger stellar component and study its decaying time, or explore mechanisms that could generate such offsets.}\\

\begin{figure*}
\begin{center}
\includegraphics[width=8.5cm]{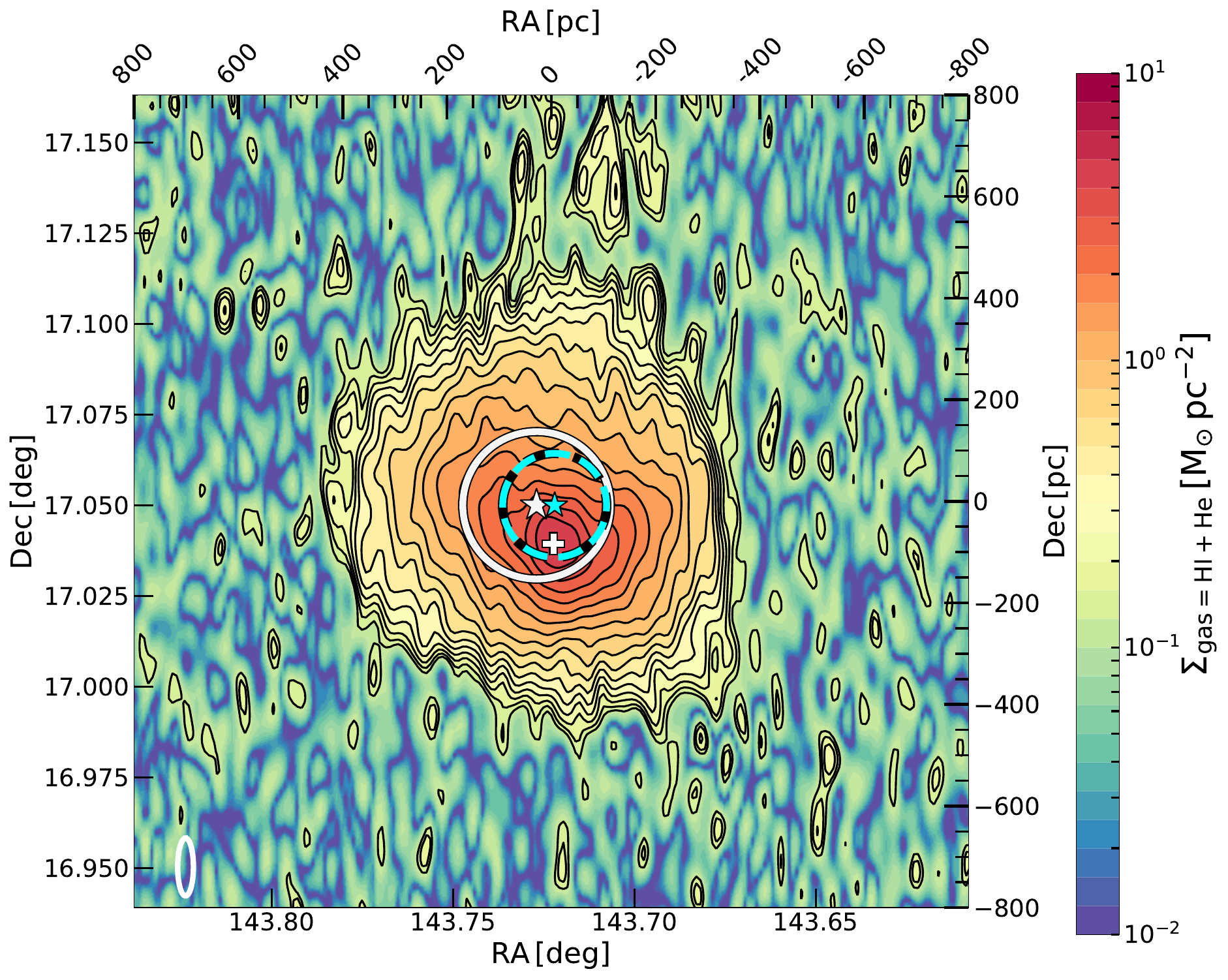}
\includegraphics[width=8.5cm]{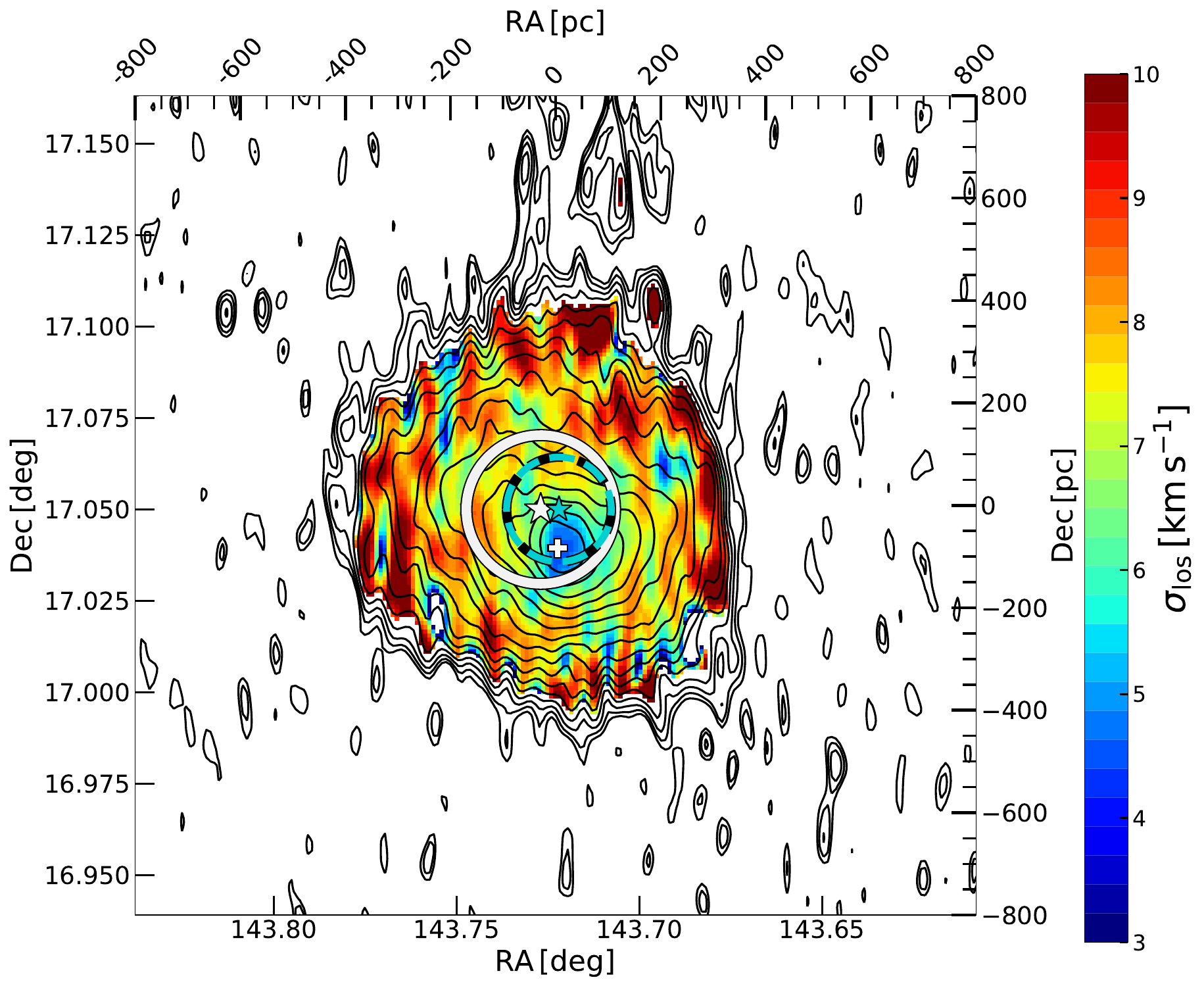}

\includegraphics[width=8.5cm]{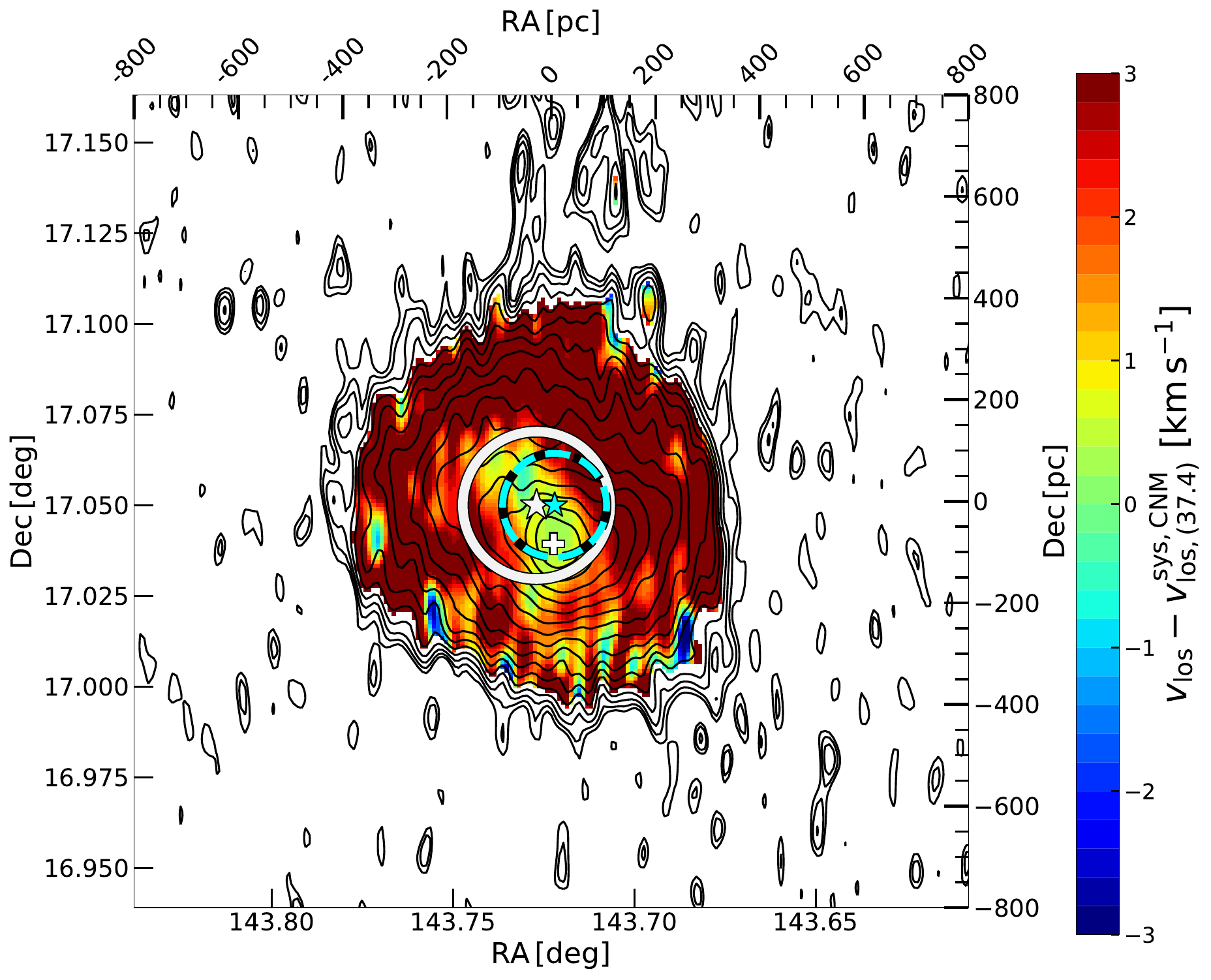}
\includegraphics[width=8.5cm]{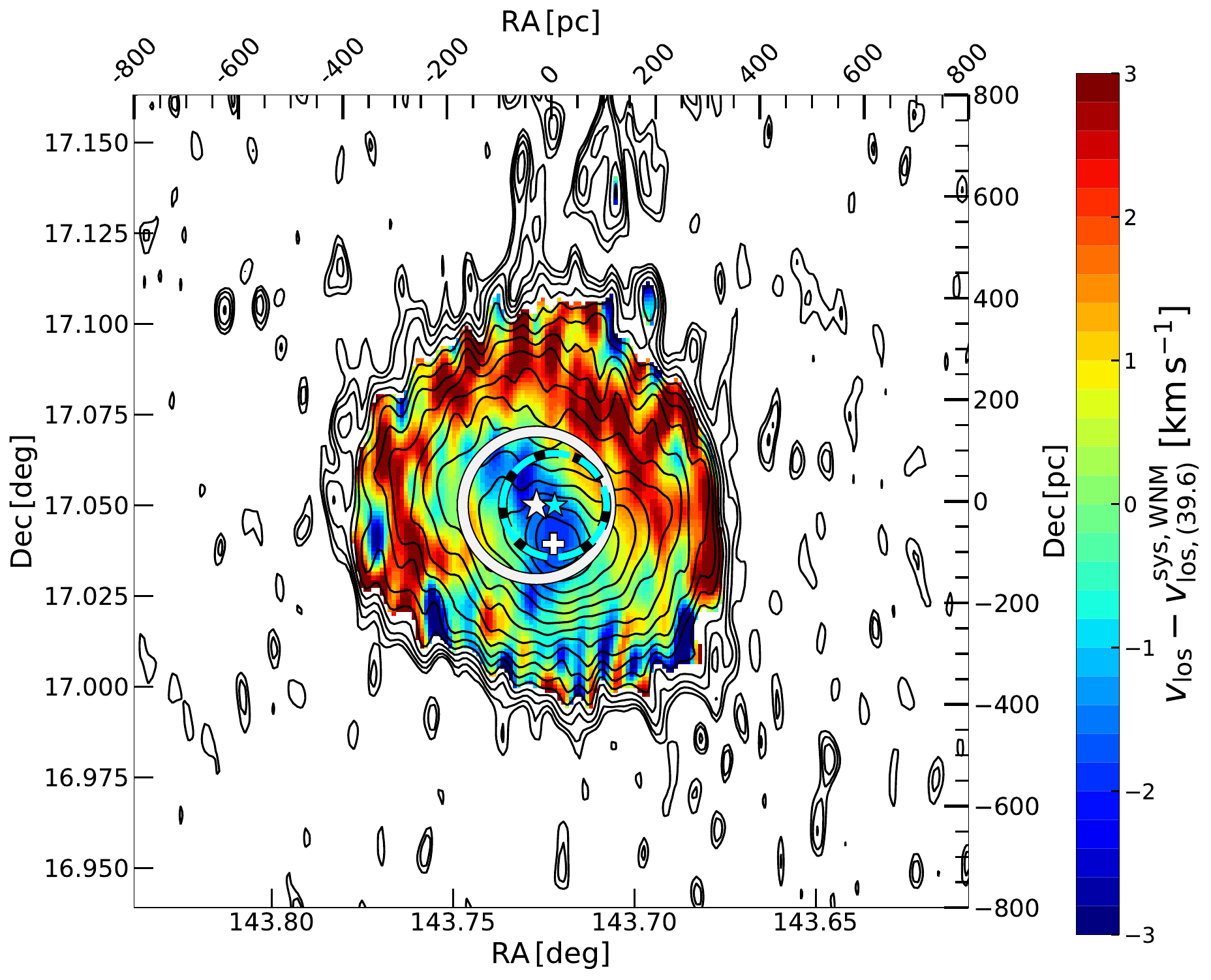}
%\vspace{-0.3cm}
\caption{\HI observations from \citetalias{Adams2018} taken with the Westerbork Synthesis Radio Telescope (WSRT), 
showing the \HI gas surface mass density map (top left),
the \HI line-of-sight velocity dispersion map (top right),
the \HI line-of-sight velocity map where the heliocentric systemic CNM velocity is subtracted ($37.4\kms$ (bottom left), 
and the \HI line-of-sight heliocentric velocity map where the heliocentric systemic WNM velocity is subtracted ($39.6\kms$) (bottom right). 
The stellar heliocentric systemic velocity lies between the two gas velocity estimates, with $38\pm2\kms$.
All panels show the \HI iso-contours separated by 0.1\dex, the \HI peak marked with a white cross.
The old and younger stellar population positions and their respective (projected) half-light radii of $1.22'\,(145\pc),\, 0.86'\,(102\pc)$ are marked with stars and circles coloured in white (old stars) and cyan (younger stars), according to \citet{DeJong2008}. 
\chA{\citetalias{Adams2018} fitted an ellipse to the external HI contours ($R\sim400\pc$), finding its centre near the old stellar component position.\LEt{***missing info?:}\coB{yes, thanks }}
Here   $\Sigma_{\rm gas}\!=\!1\sm\pc^{-2}$ corresponds   to $N_{\HI}\!=\!1.2\!\times\!10^{20}\icmsq$. 
\chA{The WSRT beam size of $(15.7\times57.3)\as=(31,113)\pc$ is displayed with the white ellipse in the lower left corner of the top left panel.}}
\label{fig:LeoT_HI}
\end{center}
\end{figure*}

\subsubsection{Gaseous component(s)}
\label{sec:mod:dwarf:gas}
While a first glance of Leo~T shows a \HI gas morphology with a global spherical distribution centred near the older stellar distribution, the deeper interferometric radio observations revealed a more complex distribution and kinematics \citep{RyanWeber2008}. 
In particular, the rich \HI observations of \citetalias{Adams2018}, shown in Fig.~\ref{fig:LeoT_HI}, reveal a gas distribution with $M_{\rm gas}\!\!=\!\!5.2\pm0.5\!\times\!10^5\sm$[\footnote{where we have rescaled their value to an updated heliocentric distance, as reported in \citetalias{Blana2020}}] that extends to $\sim400\pc$ in radius with a velocity dispersion of $8.1\kms$.\LEt{***perhaps move footnote marker to the end of the sentence so that it does not look like a squared symbol? (M$^2$) } 
\chA{They found that the outer \HI isocontours ($R\sim400\pc$) are round, with the centre of a fitted ellipse roughly overlaying with the stellar (optical) centre \citepalias[see Fig. 1 in][]{Adams2018},
 while the inner region ($R<200\pc$) the isocontours are flatter and shifted south and have a major axis almost perpendicular to the North, with its \HI surface density peak shifted to the south from the stellar centre by $\si80\pc$ in projection (see Fig.~\ref{fig:LeoT_HI}). They also determined that the total \HI flux centre is between the optical and the \HI peak centres, as expected.}
Furthermore, the \HI velocity map in Fig.\ref{fig:LeoT_HI} shows a velocity difference of $\sim4\kms$ between the central region ($R\sim150\pc$) and the outer region ($\sim 400\pc$).
The dispersion map also shows strong variations and a kinetically colder state of the density peak.
Consequently, the set-up for the gas component(s) and properties are the following:

\begin{enumerate}
\item {\HI Warm neutral medium}: 
We set up a gas component for our fiducial dwarf model D1 along with models D2, D3 and D4 mentioned in Table \ref{tab:setup} using the structural Plummer parameters from the profiles fitted to the azimuthally averaged radial profile (\citetalias{Blana2020}) of the \HI observational data \citepalias{Adams2018}, namely a Plummer radius and \HI mass of $193.9\pc$ and $4.1\times10^5\sm$ (or $5.5\times10^5\sm$ with Helium), respectively.
\chII{This profile has a Plummer \HI number and mass densities of $n^{\HI}_{\rm Pl}\!=\!0.54\icmcube$ and $\rho^{\HI}_{\rm Pl}\!=\!1.35\!\times\!10^7\!\sm\ikpccube$, and central column number and mass surface densities of $N_{\rm Pl}^{\HI}\!=\!4.43\!\times\!^{20}\icmsq$ and $\Sigma_{\rm Pl}^{\HI}\!=\!3.48\sm\pc^{-2}$, respectively.}\\

\item {Equation of state (EoS)}: 
\chA{The gas temperature profiles of the initial conditions of the models} are set iteratively in the code {\sc dice} to reach the hydrostatic equilibrium solution, where we checked that these have values below $10^4\kelvin$, as suggested by the thermal equilibrium models for the \HI warm neutral medium \citep[WNM; see Fig. 30.2 in][]{Draine2011}.
\chA{For our fiducial simulation models with \textsc{ramses} shown in Table \ref{tab:setup} we adopted an adiabatic EoS for two main reasons.}
First, as we want to explore hydrodynamical solutions where mechanical de/compression could explain the perturbed shape of the gas, we must remove instabilities that could be driven by radiative processes.
And second, observations suggest that the gas in Leo~T has been stable to thermal collapse   
in the last \chII{$<1\Gyr$} without signatures of (significant) star formation in that period.
For example, \citet{RyanWeber2008} estimates that Leo~T's \HI distribution is stable to collapse, finding a Jeans mass an order of magnitude larger than the measured enclosed dynamical mass.

In addition, the scarcity of information on the ISM chemical composition in Leo~T makes a precise estimation of the thermal stability of the gas difficult. 
The stellar metallicity estimates could be used as a proxy for the gas metallicity; however, dwarf galaxies can expel large fractions of their metals due to their low escape velocities.
Estimates in the literature consider low-density and metal-poor environments for Leo~T, finding long cooling timescale values:
$\tau_{\rm cool}\sim 1\Gyr$ where \citep{Wadekar2019} considered a low cooling rate, 
or $\tau_{\rm cool}\sim400\Myr$ considering a gas metallicity of $[Z/H]=-1$ and the cooling functions of \citet{Kim2021}, which agrees with the more general estimates for a metal-poor WNM gas \citep{Glover2013}. 
Furthermore, currently there are no molecular hydrogen measurements available for Leo T (see paragraph point 4), with this being an important source of cooling lines in cool gas.
Here we estimate some values for $\tau_{\rm cool}$ defined as in \citet[][see Eq.8.94]{MoBoWh2010}, and written as
\begin{eqnarray}
\tau_{\rm cool} &= \frac{3/2\, k_{\rm B} T}{n\Lambda}
\label{eq:tcool}
,\end{eqnarray}
with the Boltzmann constant $k_{\rm B}=1.38\times10^{-16}[\rm {erg/\kelvin }]$.
\chII{An accurate estimate of $\tau_{\rm cool}$ can be affected by uncertainties in Leo~T's gas properties, such as its 3D density and temperature distributions, and lack of chemical composition (no gas metallicity estimates, nor ${\rm H_{2}}$ nor dust information available). 
An MCMC $\tau_{\rm cool}$ estimate with $10^6$ draws from uniform random distributions for the parameters in Eq.~\ref{eq:tcool} using ranges from observed and theoretical values ($n=0.1 - 0.5\icmcube$, $T\!<\!10^4\kelvin$) and cooling function $\Lambda$ values from the literature (for $Z/H<-1$, and no heating) are in the range $\Lambda\!=\! 10^{-27}-10^{-26.8}[\rm {erg\,cm^3\,s^{-1}}]$ \citep[][see their Figs.~3 and 18]{Kim2023} or \citet[][their Fig.~4]{Maio2007} results in a distribution of $\tau_{\rm cool}\!=\!100^{+67}_{-31}\Myr$ with a maximum of $\tau_{\rm cool}^{\rm max}=310\Myr$.
However, these cooling times could be longer when heating terms are added in the cooling function,
which would be comparable to the dynamical time scale of the evolution we considered here.
For example, a cooling time of 1\Gyr would require a cooling function value of $\Lambda\!=\!10^{-28}{\rm erg\,cm^{3}\,s^{-1}}$ for $n\sim 0.5\icmcube$ and $T\!=\!8000\kelvin$.}
Nonetheless, we \chIII{ran} test simulations that use an isothermal EoS \ch{which assumes a cooling-heating equilibrium by construction}, and models with heating-cooling equations where we tested \ch{gas metallicities ($[Z/H]$) between -3 and -2.} The tests \chIII{showed} in general a behaviour of the gas that is similar to the adiabatic cases, except for models with gas metallicities higher than $[Z/H]>-1$, where the cooling of the implementation is more effective and collapses the gas distribution in less than 100\Myr \ch{~(see more details in Sect. \ref{sec:mod:num})}. 
Therefore, given that Leo~T shows a mean (stellar) metallicity lower than -1.5 and that there is no current star formation, nor had important star formation events in the last $\sim200\!-\! 1000\Myr$ \citep{DeJong2008,Weisz2012,Vaz2023}, \chII{it suggests that the gas should be currently radiatively stable at a global scale, supporting the adiabatic approximation.} Further exploration of these processes on a longer timescale is \chIII{out of the scope of this work} and it will be addressed in a future publication, \chIII{while here} we focus instead on the present dynamical state of this dwarf. \newline

\item {Cold neutral medium}: In addition, the \HI in the interstellar medium (ISM) of a galaxy could be constituted of two coexisting regimes in heating-cooling equilibrium, a WNM and a cold neutral medium (CNM) \citep{Draine2011}.
Therefore, \citetalias{Adams2018} decomposed the \HI spectra into two Gaussian components, resulting in a less massive CNM component with a velocity dispersion of $\sim 2.5\kms$ with 10\% ($5\times10^4\sm$) of the more massive WNM component, the latter having a dispersion of $7.5\kms$.
Furthermore, they also found slightly different systemic (heliocentric) velocities, with $39.6\kms$ for the WNM and $37.4\kms$ for the CNM.
Therefore, we also set up models with two gas components, indicated in Table \ref{tab:setup}, including a second central denser component with constant densities \chII{between $11-20 \icmcube$ $(3.7-6.6\times10^8\sm\ikpccube)$} and an initial temperature of $300\kelvin$ \chII{which are set in equilibrium with the WNM in {\sc dice} by producing a temperature gradient, where we checked that its temperature has values lower than $T\!<\!1000\kelvin$. Later, the gas is evolved with {\sc ramses} with an adiabatic EoS.}
% den0_unif= 6.593056E+08 Msun/kpc^3 = 2E+01 n/cm^3
% rcut_unif= 30.711771079694596 pc
% den0_unif= 3.730194E+08 Msun/kpc^3 = 1.131552E+01 n/cm^3
We set the initial CNM mass at a factor of 2 higher than the estimates of \citetalias{Adams2018} (exchanging this by the WNM mass to keep the total gas mass constant), as perturbations can heat the CNM by mechanical compression heating or gas mixing with the WNM, \ch{reducing its amount by the end of the simulation,} and also to enhance the effects of its interaction with the lower density component.\\

\item {Current Star Formation \& Molecular gas:}
 \chII{the authors of \citetalias{Adams2018} suggest that, despite the potential presence of a CNM component, there would be little or no molecular gas, as the observed surface density \citepalias[\chII{$\Sigma_{\rm gas}\si 4.6\sm\pc^{-2}\,(10^{0.66}\sm\pc^{-2})\si5.7\!\times\!10^{20}\icmsq)$}][]{Adams2018,Blana2020} is at the lower limit of the 
 Kennicutt–Schmidt relation \citep{Kennicutt2012}. 
 Furthermore, recent ${\rm H\alpha}$ observations place an upper limit star formation rate estimate of ${\rm SFR}\!<\!8\!\times\!10^{-6}\sm\iyr$ and SFR surface density $<\!\!10^{-5}\sm\iyr\kpc^{-2}$ \citep{Vaz2023},\footnote{private communication} which agrees with other studies showing a lack of star formation in the last 200 - 1000\Myr \citep{DeJong2008,Weisz2012}, which adds to the lack of strong star formation signatures \LEt{***ratio or tracers and signatures? see note 20 } such as dust or OB stars. 
 This also puts Leo~T below the KS relation for metal-poor dwarf galaxies \citep{Filho2016}, and below empirical and theoretical threshold estimates for the transition from atomic to molecular gas \citep{Skillman1987a,Schaye2004, Krumholz2009, Elmegreen2018}.
  Of course, another possibility is that the molecular gas is in a diffuse state, or that Leo~T is currently forming dense cores at the bring of a star formation episode, which could be better revealed by future observations. Therefore, we leave the star formation modelling for a follow-up publication (Bla\~na\,et\,al.\,in\,prep.), while focussing here in reproducing the current dynamical state of Leo~T. }

\end{enumerate}

\subsection{Set-up of scenarios: Reproducing the \HI - stellar offsets} 
\label{sec:mod:scen}
We set up three\LEt{***Write out numerals lower than 11 when not directly used as a measurement
with the unit abbreviation following. For an example, see Sect. 2.7 of the language guide } different scenarios where we explored variations of the environmental and internal conditions to reproduce the observed properties of Leo~T, focussing on the $80\pc$ offset between the peak of the \HI distribution and the old stellar distribution, and the $35\pc$ offset between the older and younger stellar distributions.
\ch{Briefly these are: 
\chIII{In} Scenario~I we explored if it is possible to generate these offsets with environmental perturbations produced by the motion of the satellite in the IGM. 
\chIII{In} Scenario~II we generated the offsets with stellar winds from an offset (perturbed) AGB stellar population (no initial gas offset is given). 
Finally, in Scenario~III we imposed an initial gas offset and estimate if its decaying timescale is comparable to the last star formation episode in Leo~T ($\gtrsim200\Myr$). 
All scenarios include a Galactic wind produced by the motion of the satellite in the IGM.}
The simulation parameters that are associated with the different scenarios are described in Table \ref{tab:setup}. 
\chIII{The detailed descriptions of each scenario are in the following sections.}

\subsubsection{Scenario~I:  Interaction with the environment}
\label{sec:mod:scen:sI}
Here we explored the effects of the ram pressure of the IGM on the morphology and equilibrium of the gas distribution of the satellite alone. We considered slow wind velocities due to the slower trajectories of the backsplash orbital solutions and fast winds due to the first infall orbital solutions, both calculated 2\Gyr in the past until the present as explored in \citetalias{Blana2020}. 
No stellar winds are included here. 
More details of orbital properties are given in Sect. \ref{sec:mod:env}.
We also considered strong IGM density perturbations that mimic possible substructures in the outer regions of the MW.

\subsubsection{Scenario~II:  Stellar winds and initial offset of the younger stellar population}
\label{sec:mod:scen:sII}
\ch{Here we explored the effects of including internal perturbations to the dwarf models while having environmental perturbations due to the satellite orbit described in Sect. \ref{sec:mod:env}. 
\chA{To have a significant but no overwhelming ram pressure effects from the Galactic wind, we chose the FI orbit with a final velocity of 210\kms, which implies an attack angle on the sky of 
$\theta_{\rm sky}\!=\!18\dg$ for all these models (see Table \ref{tab:setup}).}}
In addition to the younger stellar population and its offset reported by \citet{DeJong2008}, \citet{Weisz2012} identified a handful of bright stars detected using Hubble Space Telescope observations, whose location on colour diagrams suggests that they are part of the red giant branch (RGB) population, or a population of intermediate-age asymptotic giant branch (AGB) stars. 
In particular, we focused on the second alternative \chIII{for presenting} the interesting possibility of a hydrodynamical interaction with the surrounding \HI gas in Leo~T, explaining the observed \HI offset. 
AGB stars have masses of $0.8 - 8\sm$, effective temperatures $>3000\kelvin$, and mass outflows of $\dot{M}\sim 10^{-8} - 10^{-4}\smyr$ of slow and adiabatically cooled winds with terminal velocities of $w\sim 3-30\kms$ \citep{Hofner2018}, which are comparable to the stellar and gaseous velocity and velocity dispersion measured in Leo~T ($\sim 4-9\kms$, see Fig.~\ref{fig:LeoT_HI}). Moreover, the wind velocity could be effectively increased by the orbital velocity of the AGB stars relative to the centre of Leo~T.

% \citep{VanDeSande2017,Hofner2018}.
Therefore, we explored the possible effects that winds of AGB star candidates could have in a system like Leo~T, while testing on different dwarf models shown in Table \ref{tab:setup}. 
\ch{For this we introduced a single wind source term that mimics the combined contribution of an AGB population winds with outflow and momentum parameters defined in the table mentioned above.}
We note that while each AGB star lifetime spans only 1-3\Myr, we aim to mimic the outflow produced by a population of $\sim10$ AGB stars, assumed to be constantly populated by ageing stars entering this stage.
\ch{Consequently, we expect the massive members of the AGB stellar population to be associated with the younger component.}
The stellar wind source treatment follows \citet{Calderon2020b,calderon2020a}, but with parameters adapted to slower and colder winds. The wind source outflows from a sphere with a 15\pc radius which is locked to follow the centre of mass of the younger stellar component.

In order to explore the offset between the gas and the stars we shift the younger stellar distribution from the centre of the dark matter distribution and measure its relaxation time decay into the dwarf's centre while measuring the magnitude of the oscillations. This is motivated by the 80\pc offset between the gas and the stars, \chII{and also by the $\Delta R_{\rm o|y}=35\pc$ offset between the younger and older stellar populations mentioned in Sect. \ref{sec:mod:dwarf:st}} and reported by \citep{DeJong2008,Vaz2023}.
It is likely that the older and more massive stellar population is in equilibrium, tracing the gravitational potential, which is dominated by dark matter, while the younger population could have formed in a gas cloud displaced from the centre during the last star formation episode $\sim 200-1000 \Myr$ ago \citep{DeJong2008, Weisz2012}, which would be currently mixing with the main stellar component.
\chII{The fragmentation of the main gas distribution into an inhomogenious distribution of clouds can imprint their kinematics into the newly formed star clusters, as shown in smaller scale high-resolution star formation simulations \citep{Bate2009b,Krumholz2011}.}
\newline
\indent \chA{We chose an initial offset for the younger stellar component of $r\!=\!141\pc$, 
which is slightly larger than the observed projected offsets (35 and 80\pc) and accounts for the expected orbital decay due to phase-mixing of the younger stellar component. 
The angle of the direction of the initial offset (IO) is set with the angle $\theta_{\rm IO}$ measured anticlockwise from the dwarf orbital trajectory vector (wind tunnel direction), which is perpendicular to the horizontal axis. 
This angle is related to the position angle as
\begin{eqnarray}
{\rm PA}&=\arctan{(\tan{\theta_{\rm IO}}\,\sec{\theta_{\rm sky}})},
\label{eq:pa}
\end{eqnarray}
which results from the rotation around the horizontal-axis by $\theta_{\rm sky}$.
This rotation means that the orbital trajectory axis is coplanar with the vertical axis in the figures.
For most models in Table \ref{tab:setup} we take $\theta_{\rm IO}\!=\!45\dg$ to explore an internal perturbation that is at an intermediate axis with respect to the vertical motion of the satellite in the IGM.
Although we tested other angels, we show one model with $\theta_{\rm IO}\!=\!0\dg$ (see Table \ref{tab:setup}).
The resulting position angle values are quite similar (${\rm PA}=46\dg$ and 0\dg, respectively), due to $\theta_{\rm sky}=18\dg$ for all models in this scenario.
The initial offsets of the shifted young stellar component are set with zero velocities to explore radial infalls.}
However, we also consider initial circular orbits for the shifted stellar component in dwarf models with NFW halos to explore how the infall-time decay changes, as indicated in Table \ref{tab:setup}.

\subsubsection{Scenario~III: Initial gaseous offset}
\label{sec:mod:scen:sIII}
We also considered a third scenario to explore whether the whole gas distribution could be oscillating and sloshing in the overall potential.\LEt{***ok? see note 2d re direct questions} This has been suggested, for example for the MW dwarf galaxy Phoenix~I, which exhibits an even larger offset of $\sim500\pc$ in projection between its stellar and gaseous components \citep{Young2007}, being this large considering its stellar half-light radius of 274\pc \citep{Battaglia2012}. \citet{Young2007} hypothesised that that offset could be the result of a previous supernova explosion that could have pushed the gas outside the dwarf's centre. 
Therefore, we explored a gas sloshing scenario on Leo~T by shifting the gas distribution $141\pc$ from the centre of the dark matter potential, \chA{${\rm PA}=45\dg$} from the north axis, 
and measured its properties while it decays back to the centre, 
\ch{while the satellite moves in the galactic wind with the FI orbit with final velocity.}
We explored models with cored and cuspy dark matter profiles, as indicated in Table \ref{tab:setup}.

%%% ====================================================================

\subsection{Additional physical and numerical parameters}
\label{sec:mod:num}
As mentioned in Section \ref{sec:mod:dwarf}, we set the initial conditions (IC) for the dwarf galaxy models with the software {\sc dice} \citep{Perret2014,Perret2016}, which are then used in the wind tunnel simulations using {\sc ramses} \citep{Teyssier2002} inside a 20\kpc side box. \ch{The software is compatible with {\sc ramses}, but designed for SPH codes, so it may experience some mild radial gas expansions that eventually settle, and that can be improved by increasing the number of iterations in the IC generation.}
As we are not interested here in modelling star formation scenarios with {\sc ramses}, we simply choose a geometrically nested variable spatial resolution centred on the dwarf location, which allows cell sizes down to 2 - 4\pc.
\ch{The fiducial set-up for the simulations includes self-gravity forces for all mass components, which is well resolved given the large number of particles for the stellar ($\sim 10^5$) and dark matter ($10^6$) components and the collisionless regime of the dwarf galaxy.}
The hydrodynamic equations are solved with the exact Riemann solver together with a MonCen slope limiter \citep{Toro2009}.
\chA{The simulation models in Table \ref{tab:setup} use the adiabatic EoS of an ideal gas with an adiabatic index of 5/3 for the monoatomic gas.  
However, we also tested simulations \chIII{with the setup of} dwarf model D1 in Scenario I (M3), but with an isothermal EoS, 
and three additional models with $[Z/H]=-3$, -2 and -1, respectively, that included the heating and cooling models available in \sc{ramses} (model courty) based on \citet{Few2012}.}

%%%%%%%%%%%%%%%%%%%%%%%%%%%%%%%%%%%%%%%%%%%%%%%%%%%%%%%%%%%%%%%%%%%
%%%%%%%%%%%%%%%%%%%%%%%%%%%%%%%%%%%%%%%%%%%%%%%%%%%%%%%%%%%%%%%%%%%
\section{Results}
\label{sec:res}

\subsection{Scenario I: Interaction with the environment}
\label{sec:res:env}

In this section we present how the dwarf galaxy models react to different environmental conditions of the wind produced by the relative motion of the satellite in the IGM surrounding the MW.
We focus on the distribution of the cold gas to reproduce three main features observed in Leo~T, visible in Fig.~\ref{fig:LeoT_HI}:
i) the general bullet-shape of the outer \HI contours of Leo~T,
ii) the faint \HI material that could be trailing the dwarf, 
iii) the offset between the peak of the gas surface density and the stellar centre.

\begin{figure*}[ht!]
\begin{center}
\includegraphics[width=5.94cm]{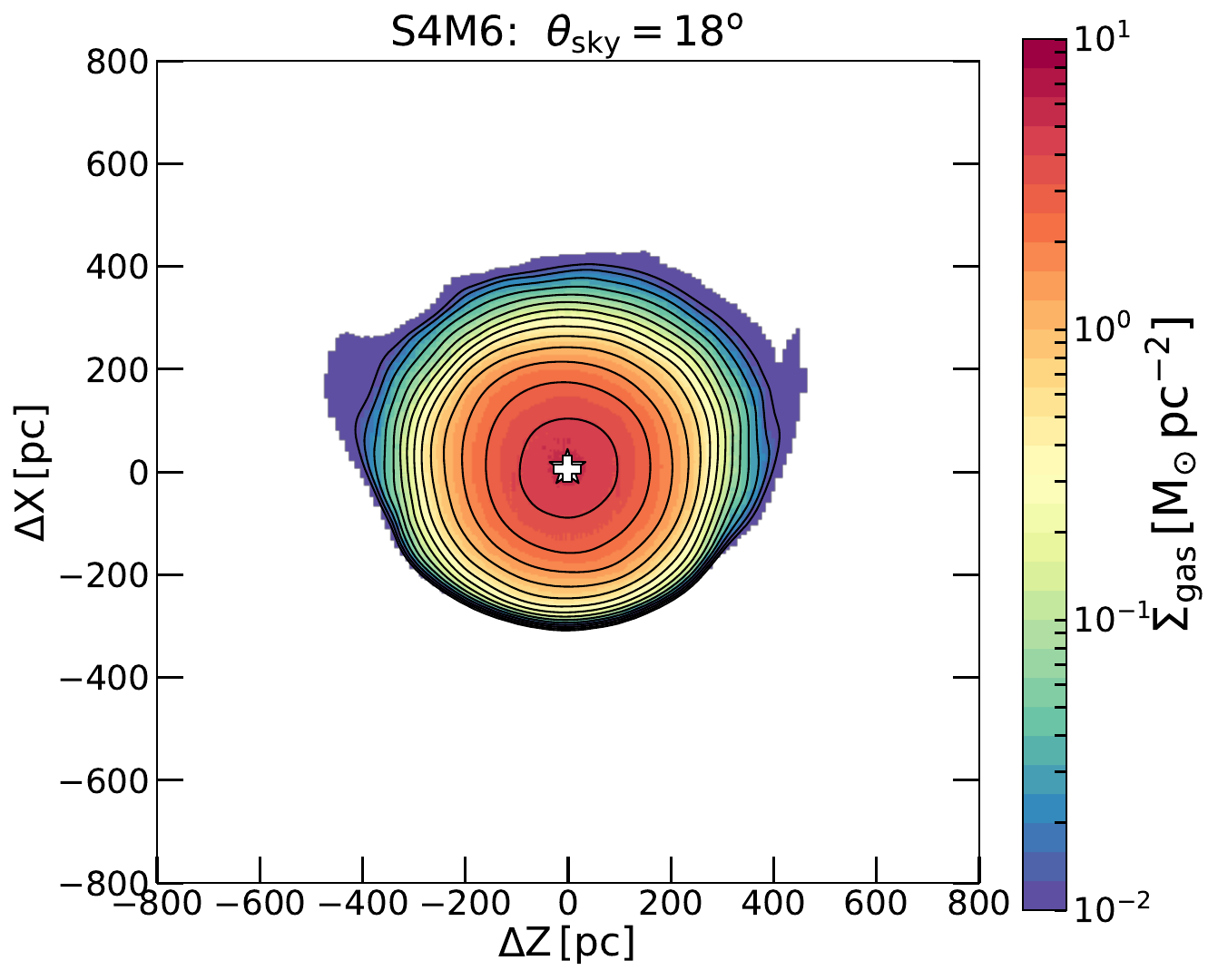}
\includegraphics[width=5.94cm]{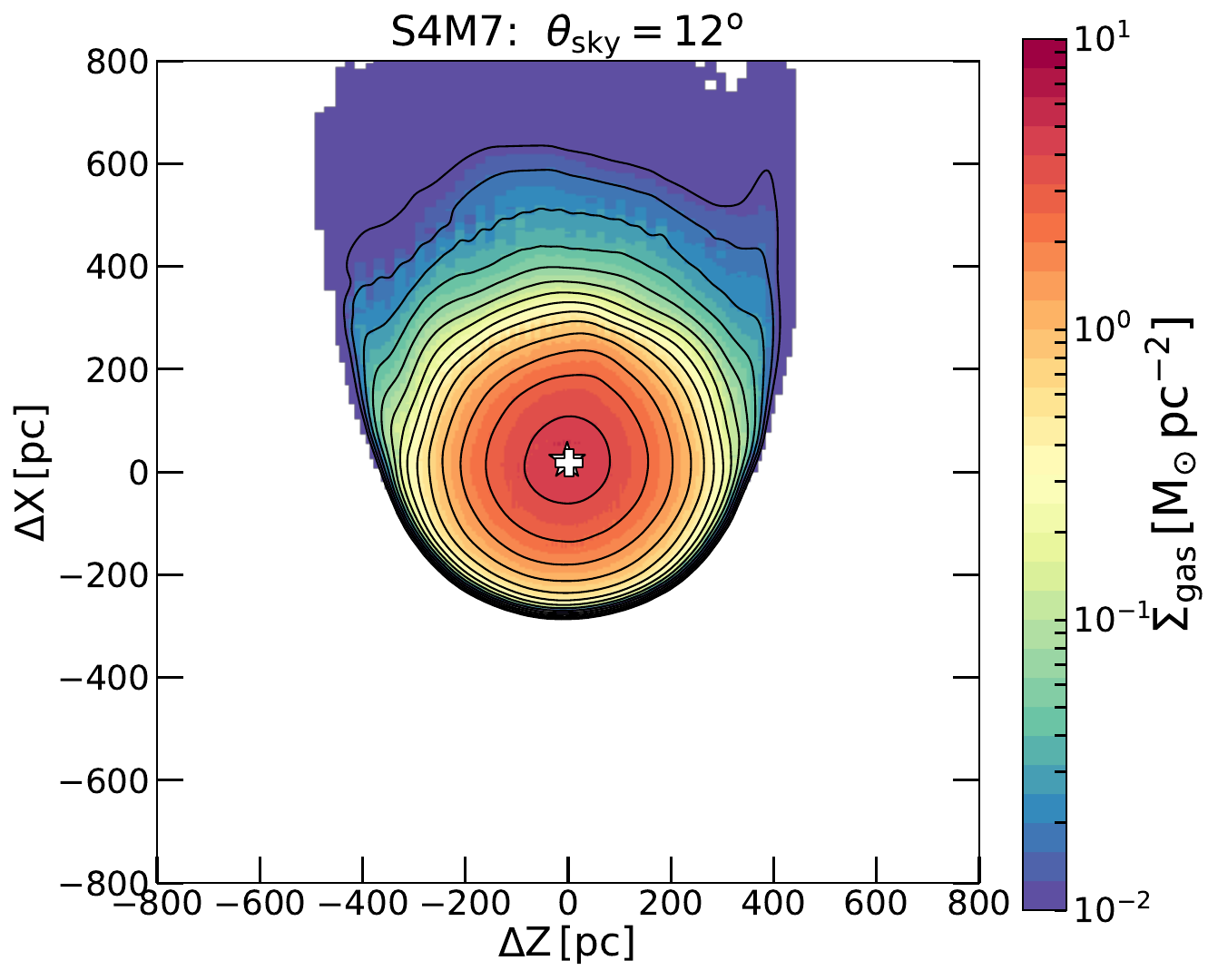}
\includegraphics[width=5.94cm]{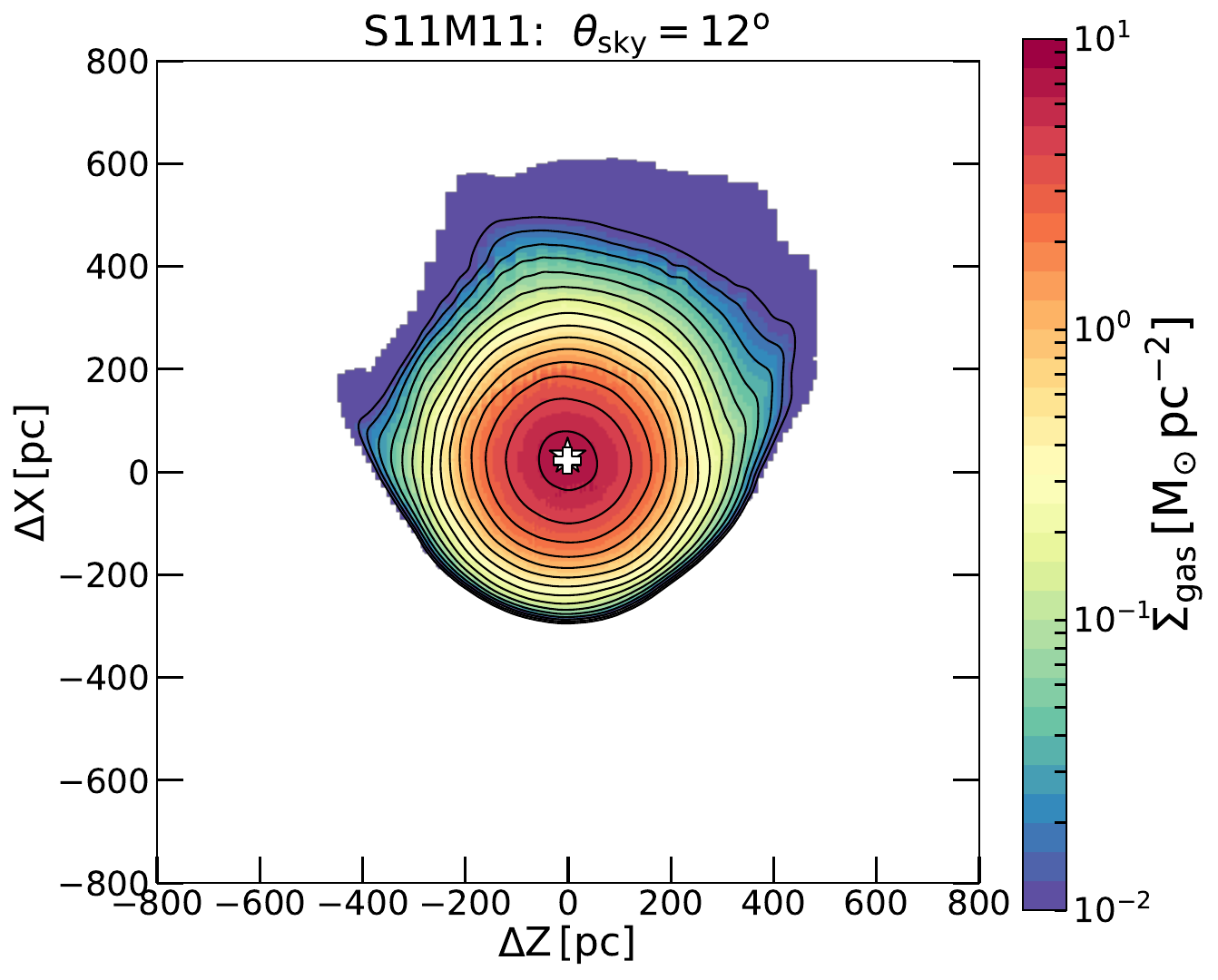}
\includegraphics[width=5.94cm]{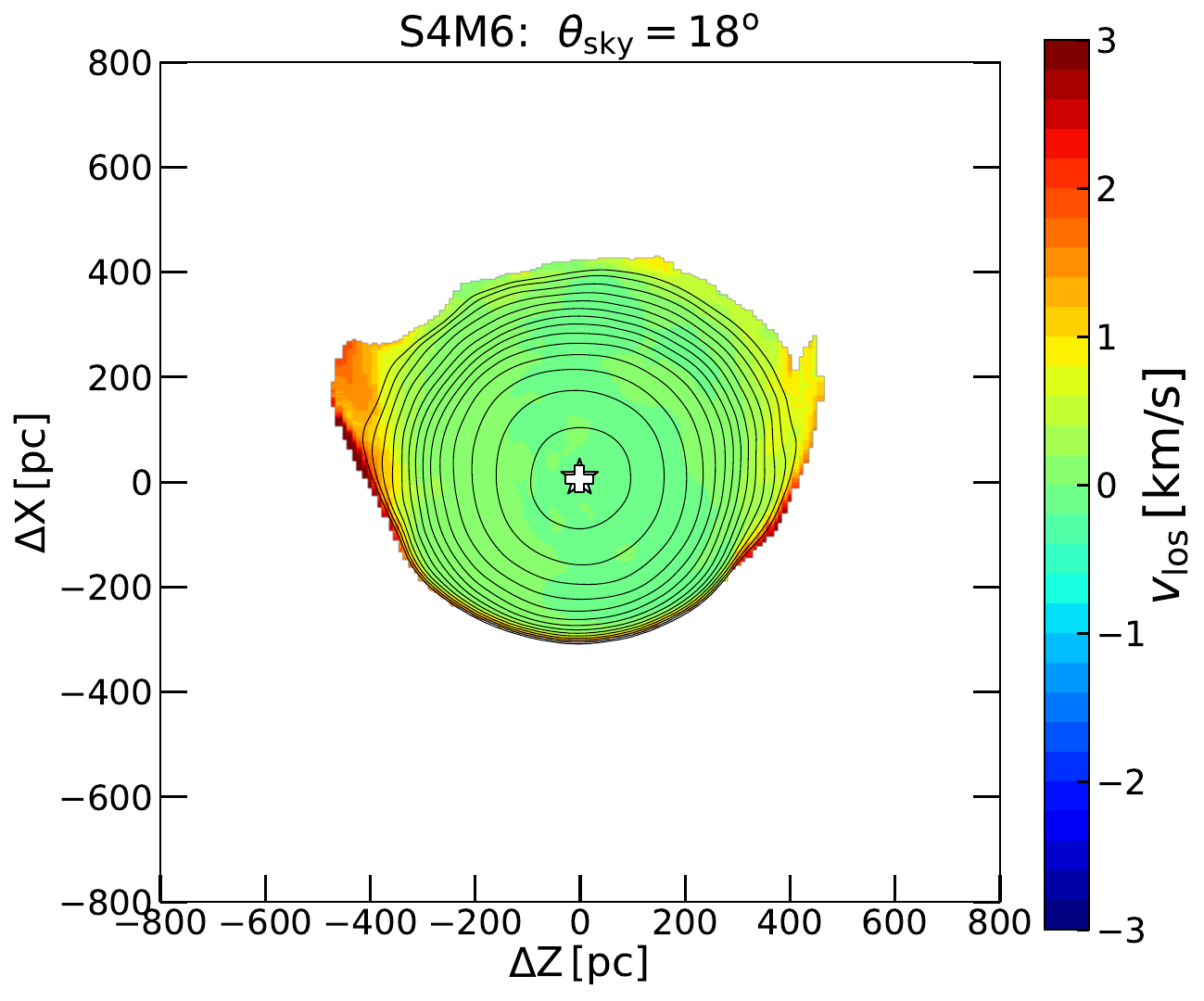}
\includegraphics[width=5.94cm]{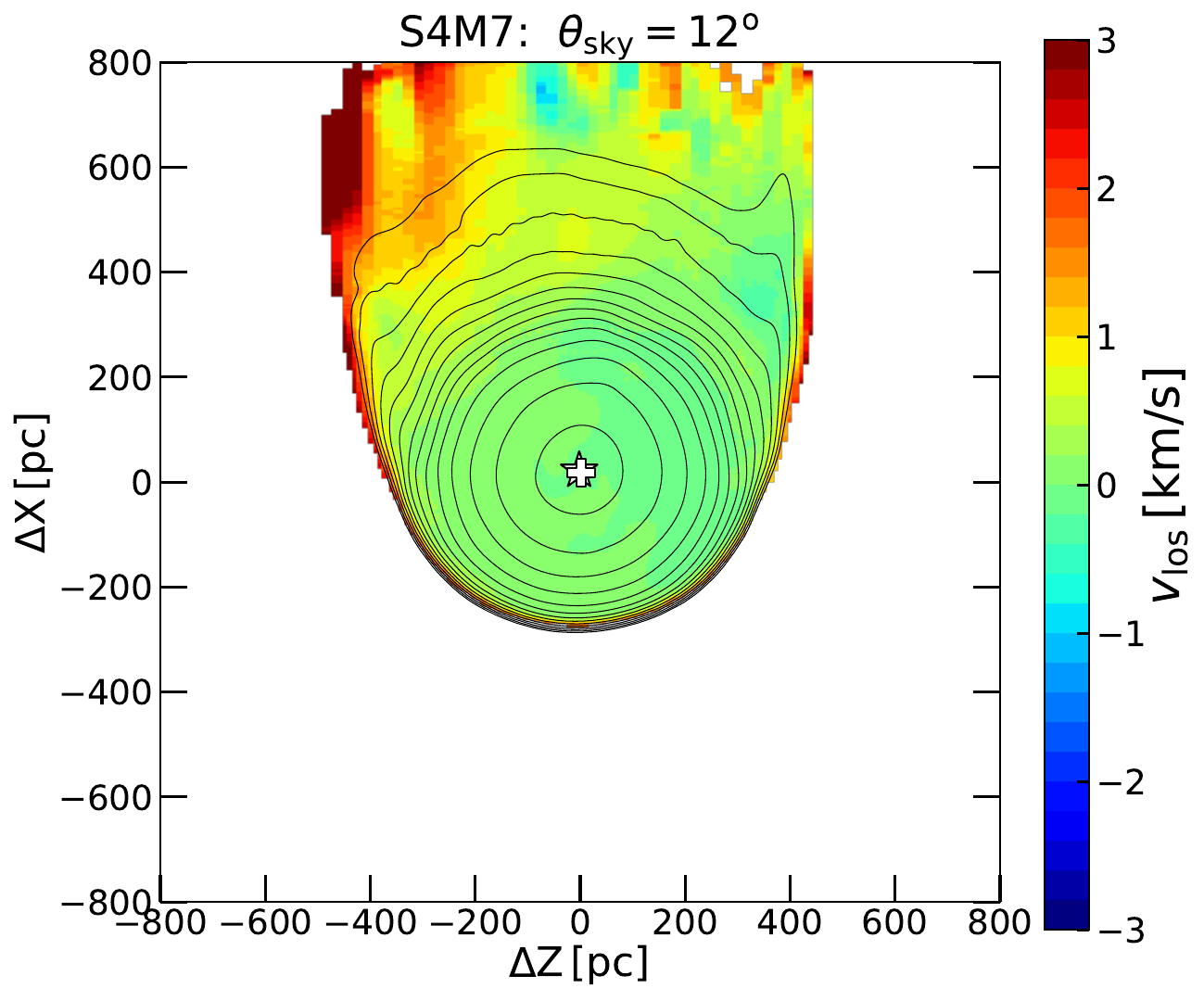}
\includegraphics[width=5.94cm]{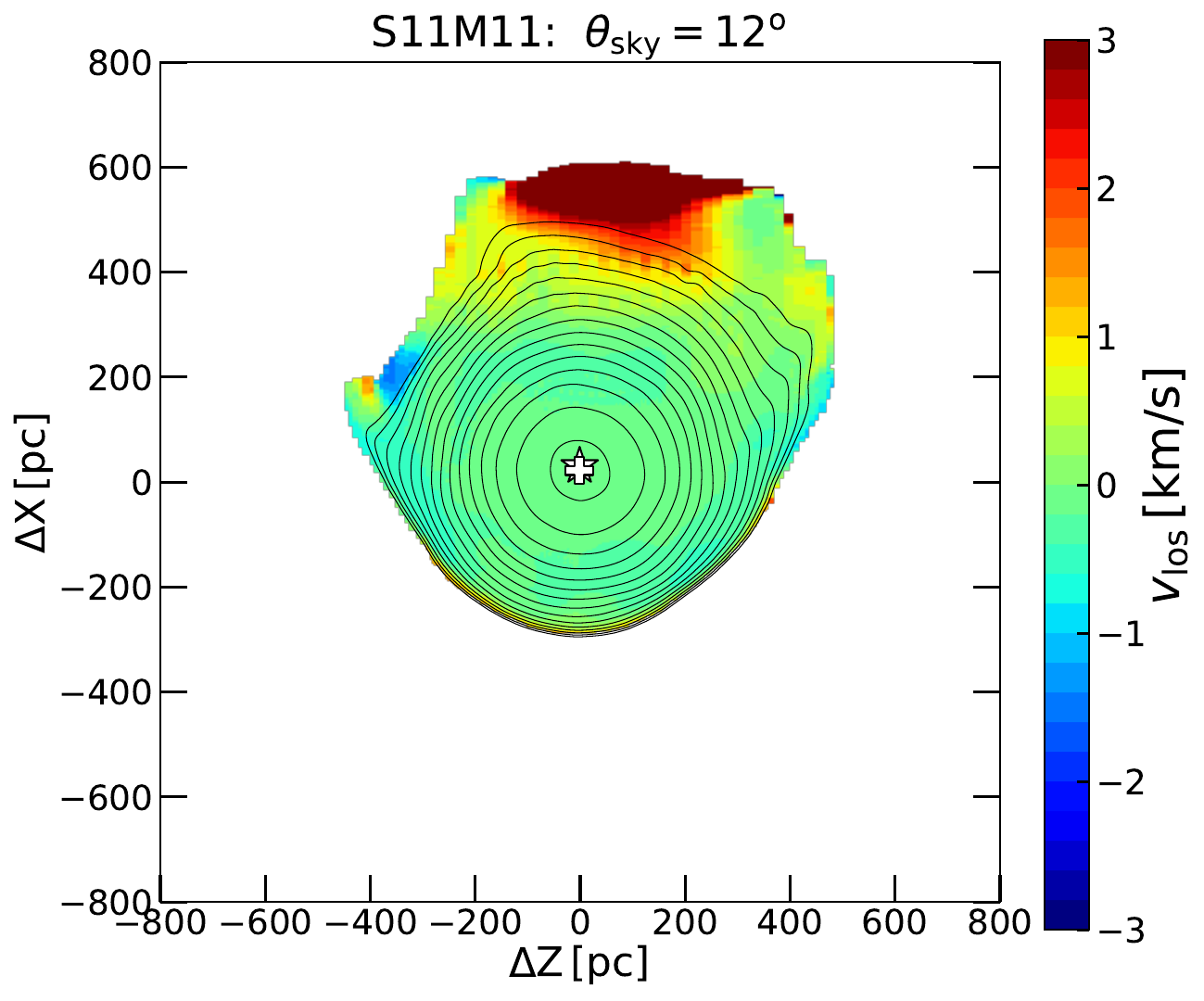}

\includegraphics[width=5.94cm]{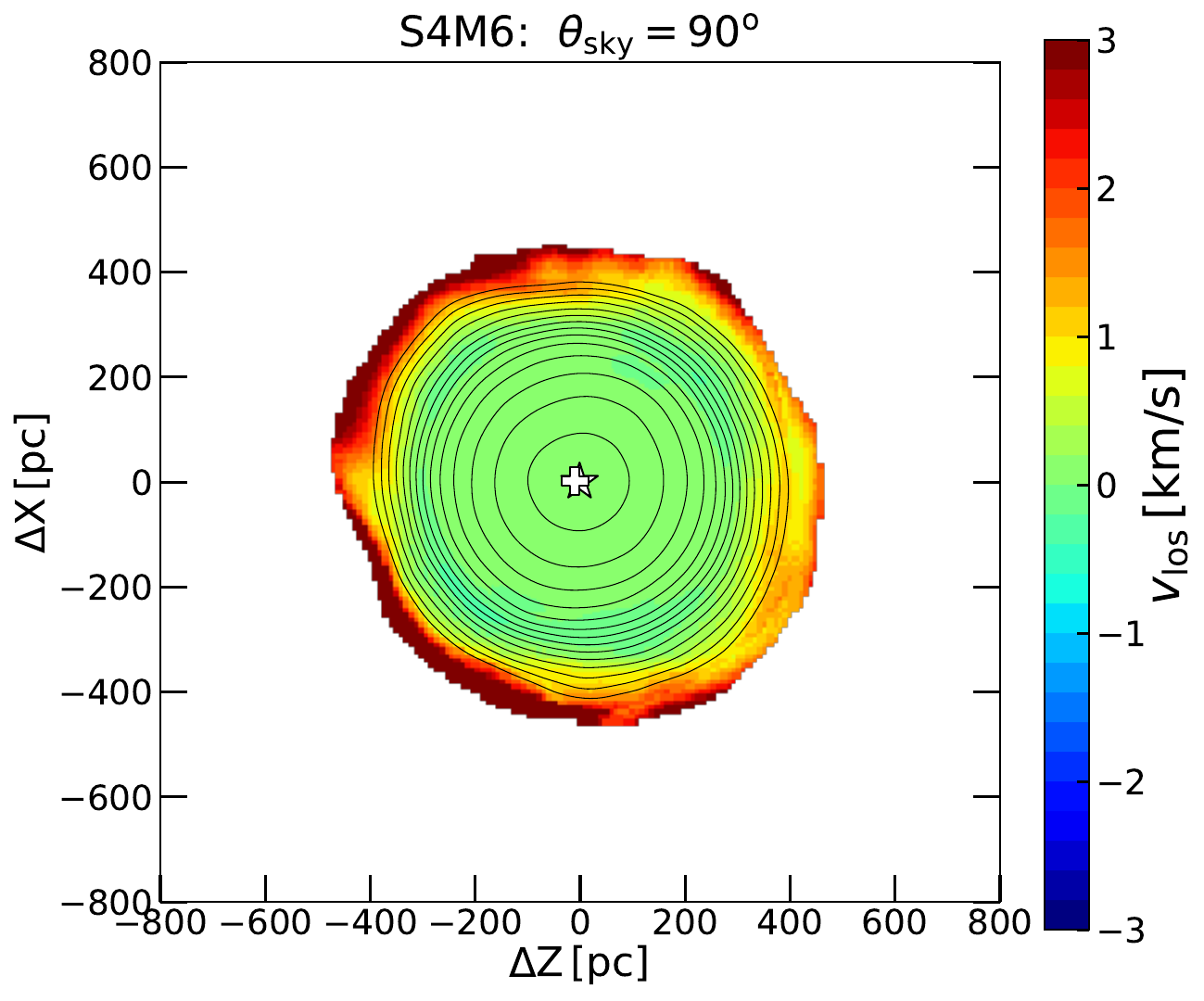}
\includegraphics[width=5.94cm]{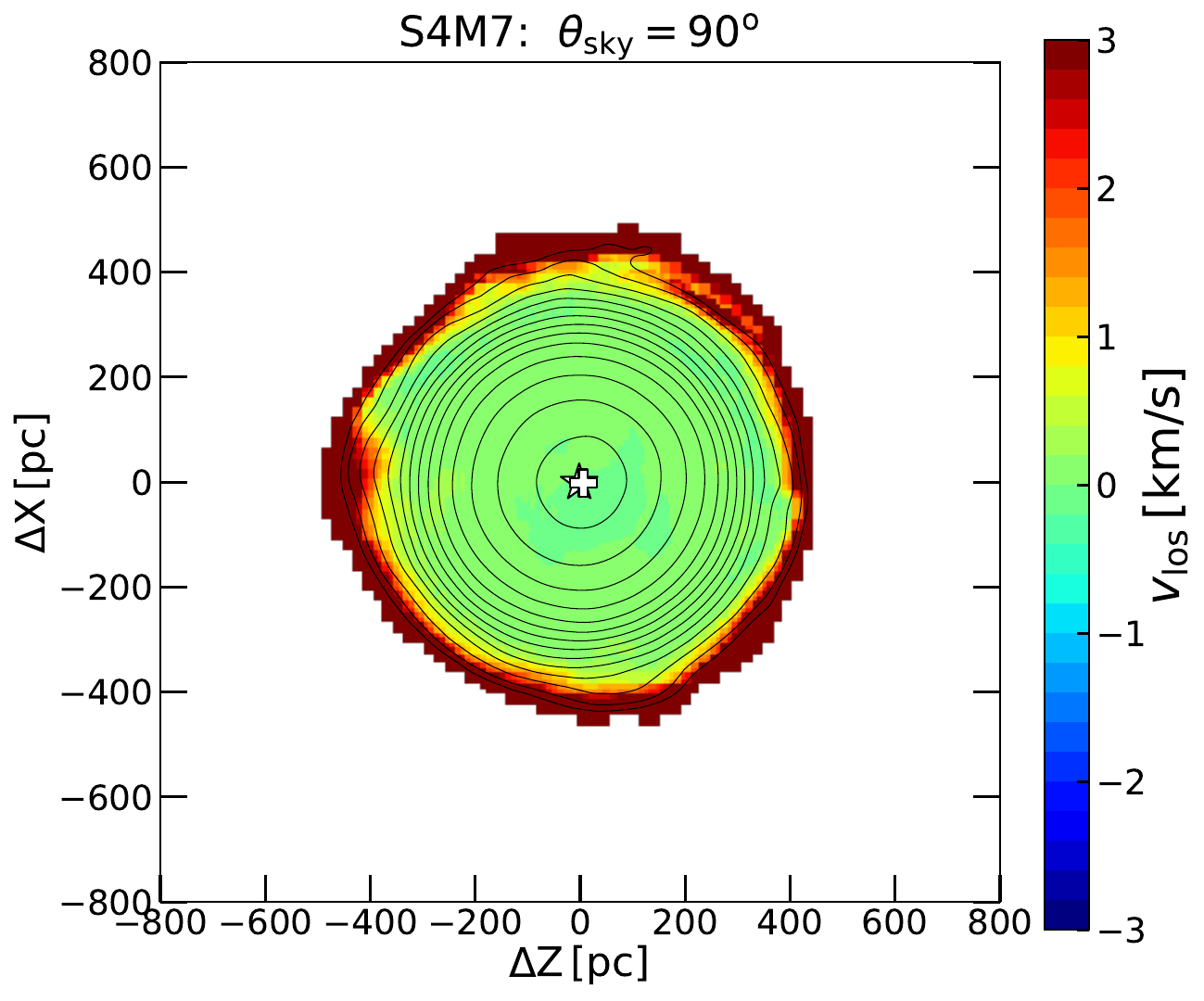}
\includegraphics[width=5.94cm]{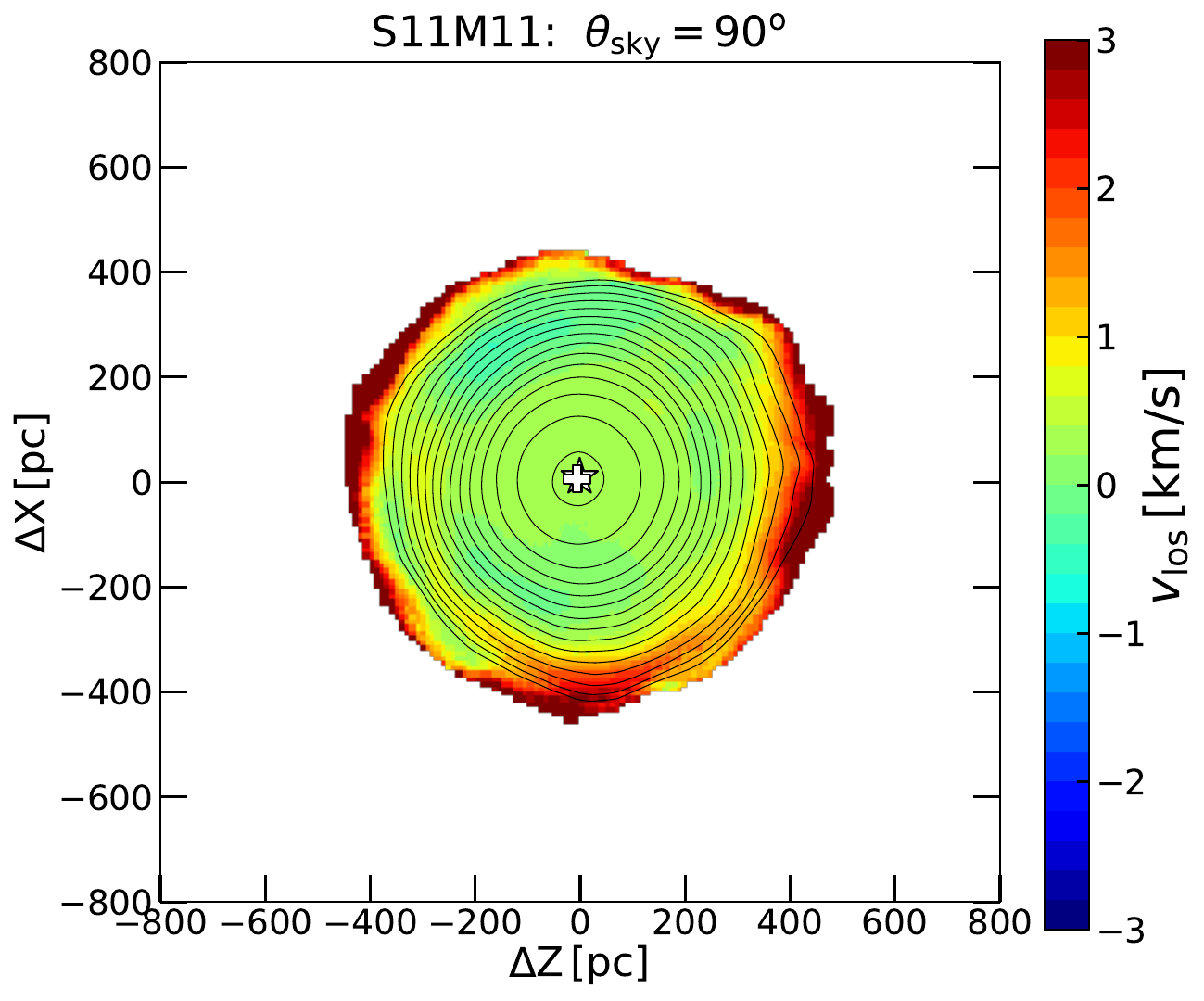}
% %\vspace{-0.75cm}
% \par\noindent\rule{\textwidth}{0.5pt}
%\vspace{-0.2cm}
\caption{\chII{Scenario I: Final snapshots of three models evolved from $-2\Gyr$ until  the current position of Leo~T.
We show two models that used the fiducial dwarf model D1, model M3 (S4M6) (left column) with a final tangential velocity of $u_{\rm t}=200\kms$ and model M4 (S4M7) (middle column) with $u_{\rm t}=300\kms$.
We also show model M7 (S11M11) (right column), which uses the more massive dwarf model D2 and $u_{\rm t}=300\kms$.
The first and second rows  show the surface mass and line-of-sight (LOS) velocity maps projected according to attack angles $\theta_{\rm sky}$ expected from the orbital orientation (Eq.\ref{eq:angle}).
We note that $v_{\rm los}$ are in the dark matter centre of mass rest frame.
In the bottom row we include images with a projection of $\theta_{\rm sky}=90\dg$ to reveal the morphology and the maximum velocity gradients along the orbits.
The centre of mass of the combined stellar and dark matter distributions is marked with white stars,
and the gas peaks are marked with white crosses.
To better represent the \HI observations we made the calculations with gas colder than $T<10^4\kelvin$. 
The $v_{\rm los}$ maps are in the rest frame of the dwarf's dark matter centre of mass,
and   the centre of mass of the stellar and dark matter components overlap in all panels.
The contours on the leading side may cover the velocity map due to the strong compression; this  can be avoided by zooming in the pdf version.
 We note   that $\Sigma_{\rm gas}\!=\!1\sm\pc^{-2}$ corresponds here to $N_{\HI}\!=\!1.2\!\times\!10^{20}\icmsq$, and the spacing between contours is $\log\Sigma/\sm\pc^{-2} = 0.2\dex$ for all simulation figures.
}}
\label{fig:fig_environment}
\end{center}
\end{figure*}

\subsubsection{Scenario I: Dwarf mass models and orbital velocity}
\label{sec:res:env:dwarf}
\chII{We start by showing the effects of the IGM wind considering different satellite orbits and dwarf mass models. 
To explore this, we choose the fiducial dwarf model D1 which has the lowest dark-matter content making
it the most sensitive dwarf model to external perturbations. 
\chIII{We also show comparisons with the more massive dwarf model D2.}
We do not include the more massive or concentrated dwarf galaxy models D3 or D4, as we find that their deeper gravitational potentials make the formation of instabilities, velocity gradients, and the gaseous tail more difficult.
For example, considering the current distance of Leo~T (412\kpc), an IGM density of $\rho_{\rm IGM}(412\kpc)\!=\!120\sm\kpc^{-3}\,(3.6\times10^{-6}\icmcube)$, and a satellite orbital velocity at that position (IGM wind) of $V_{sat}=300\kms$, we can calculate the instantaneous ram pressure \citep{Gunn1972}:}
\begin{eqnarray}
P_{\rm R} &=\rho_{\rm IGM} V_{\rm sat}^2
\label{eq:rp}
;\end{eqnarray}
\chII{which results in a value of $P_{\rm R}\!=\!10^7\sm\kpc^{-3}\km^2\s^{-2}$. 
On the other hand, taking a mean gas density of $1.79\times10^7\sm\ikpccube$, a core size of 300\pc, and central masses of Table \ref{tab:setup}
to estimate the restoring thermal pressure as in \citet{Mori2000}, 
\begin{eqnarray}
P_{\rm T}=G\,M_{\rm core} \rho_{\rm gas, core}\,(3r_{\rm core})^{-1}
\label{eq:tp}
,\end{eqnarray}
results in $P_{\rm T}\!\!=\!\!1.5\!\times\!10^{8}\sm\kpc^{-3} \km^2\s^{-2}$ for the dwarf model D1, 
and $P_{\rm T\!}\!=\!\!6.8\times\!10^{8}\sm\kpc^{-3} \km^2\s^{-2}$ for model D2, 
making them resilient to instantaneous ram pressure stripping effects, with model D2 having a $P_{\rm T}$ 4 times larger than D1.
However, the ram pressure of the IGM environment can still produce environmental perturbations driven by other mechanisms such as the
Kelvin-Helmholtz instability, which can strip the outer layers of gas of the dwarf through laminar or turbulent flows \citep{Mori2000}.  
% ('1.08E+07', '1.454186E+08', 'Msun/kpc3 (km/s)^2') ('1.08E+07', '6.84323E+08', 'Msun/kpc3 (km/s)^2').
Therefore, we proceed to explore a range of orbits for model D1 that have final tangential GSR velocities $u_{\rm t}$ of $50$, $100$, $200$, and $300\kms$ corresponding to models M1, M2, M3, and M4.
For model D2 we explore the fast orbits with $u_{\rm t}$ of $200$ and $300\kms$ in models M6 and M7.
This range includes backsplash and first-fall orbital solutions, as indicated in Table \ref{tab:setup}.}\\

\chII{As expected, slow orbits affect only mildly the outer morphology of the gas of the dwarf, 
given that the IGM density is low at that distance, which results in a weak ram pressure.
Therefore, in Fig.~\ref{fig:fig_environment} we show models M3, M4, and M7 selected
for having the most noticeable perturbations of their gas morphology and kinematics. 
However, we also show the models with slower orbits in Fig.~\ref{fig:fig_environment_A} that reveal
the much milder environmental effects of models with slow orbits. 
We note that all maps of the simulations are calculated with
gas with temperatures below $T\leq10^4\kelvin$ that better represent the \HI 
observations. 
Furthermore, if we include the IGM and hotter gas surrounding the satellite, 
we find only negligible changes of the morphology and kinematics of the edges of the dwarf models, as can be deduced by the density cuts in Fig.~\ref{fig:fig_profcut}.}\\

\chII{In general we see that the models match the overall extension the gas distribution in Leo~T.
As shown in Fig.~\ref{fig:fig_environment}, the gas of the dwarf is compressed on the leading face, whereas the trailing part of the dwarf is slightly more extended, generating a global mild bullet or trapezoidal-shaped morphology. 
This is already noticeable in dwarf model D1 with a final $u_{\rm t}\!=\!200\kms$ in model M3 and stronger in model M4 with 300\kms. 
This is more noticeable when looking at profile cuts of the density, pressure, and temperature, shown in Fig.~\ref{fig:fig_profcut}, where we can see that the asymmetry is generated not only in the shocked region in the leading face at 400\pc from the dwarf centre, but also in the inner region around the centre of the dwarf within 200\pc.
The more massive dwarf model D2 needs velocities as high as $\sim300\kms$ to generate this feature (model M7).
Slow orbits barely deform the outer layers of gas, where for the low-mass model D1 velocities below $u_{\rm t}\!<\!100\kms$ show a rounder shape with a mild leading compression, similar to dwarf model D2 with velocities lower than $200\kms$, as shown in Fig.~\ref{fig:fig_environment_A} with models M1, M2 and M6.}

\chII{We find that the resulting pressure gradient within the dwarf due to the IGM ram pressure is not strong enough to reproduce the offset between the gas and the centre of mass of the stars, as observed in Leo~T.
We also performed an inspection of the temporal evolution of all these simulations, searching whether the instabilities or the leading-trailing density asymmetry could generate an offset between the stellar centre and gas density peak, 
finding only weak transient perturbations, but not detecting any significant offset.}

\begin{figure*}[ht!]
\begin{center}
\includegraphics[width=8.5cm]{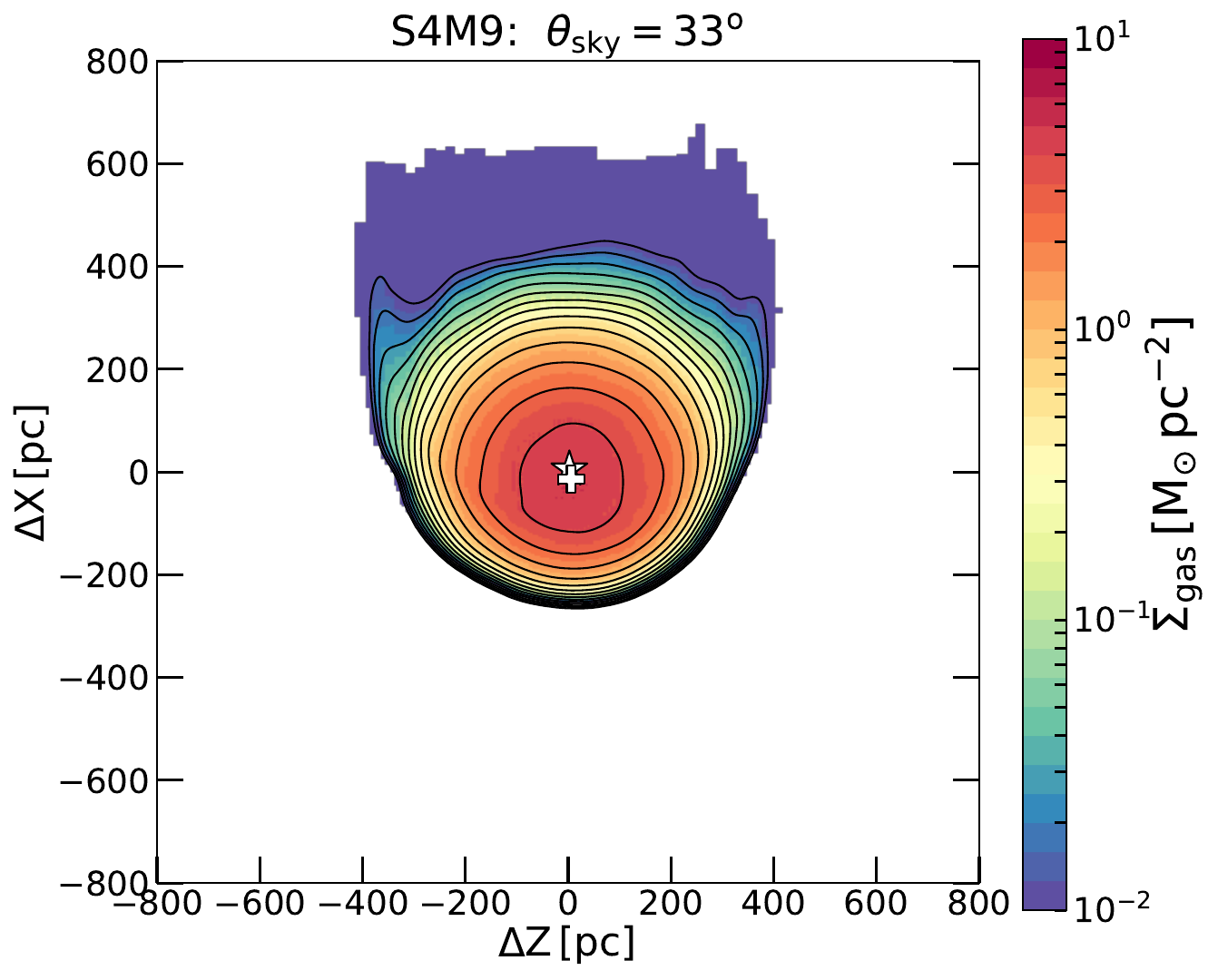}
\includegraphics[width=8.5cm]{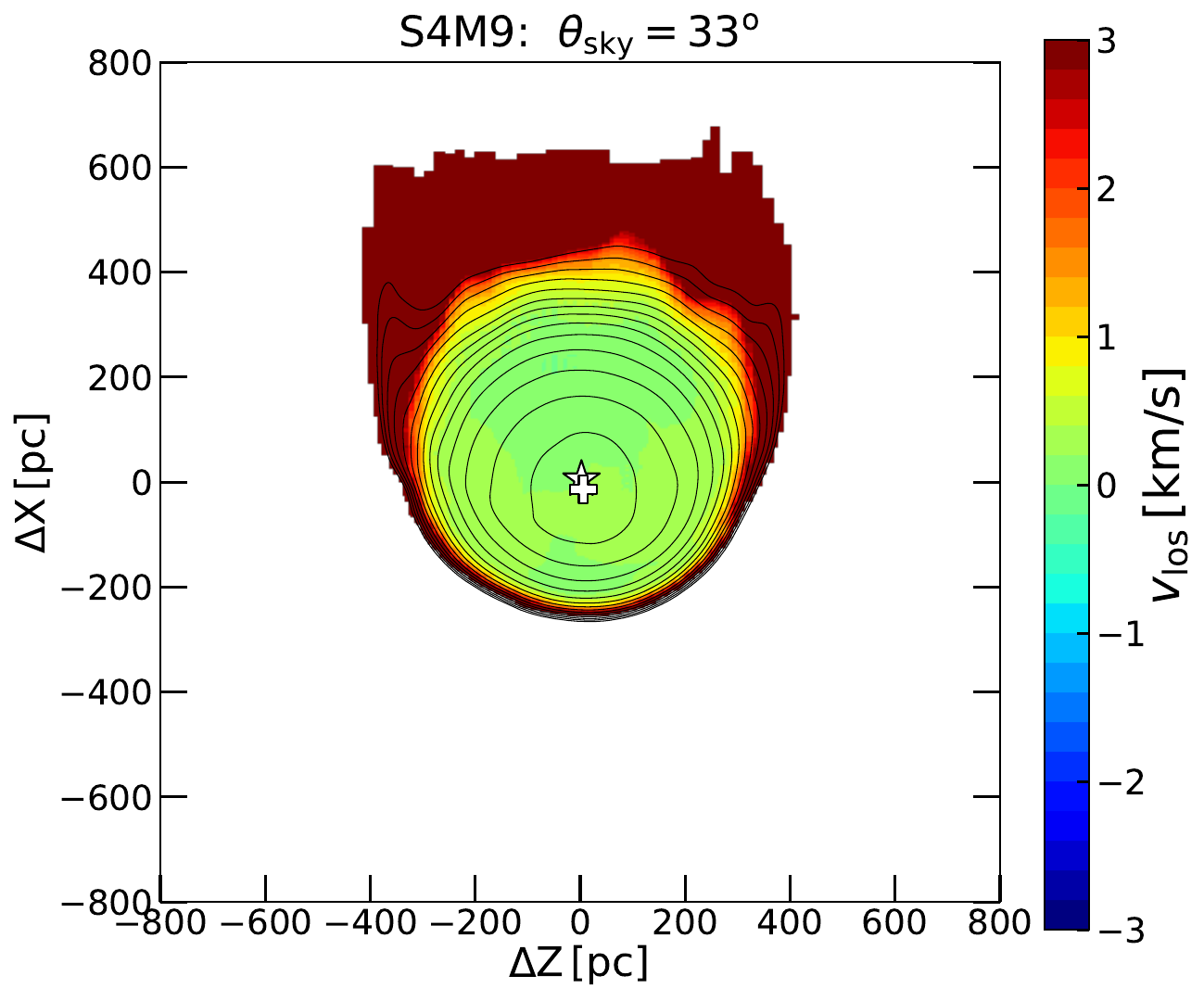}
\includegraphics[width=8.5cm]{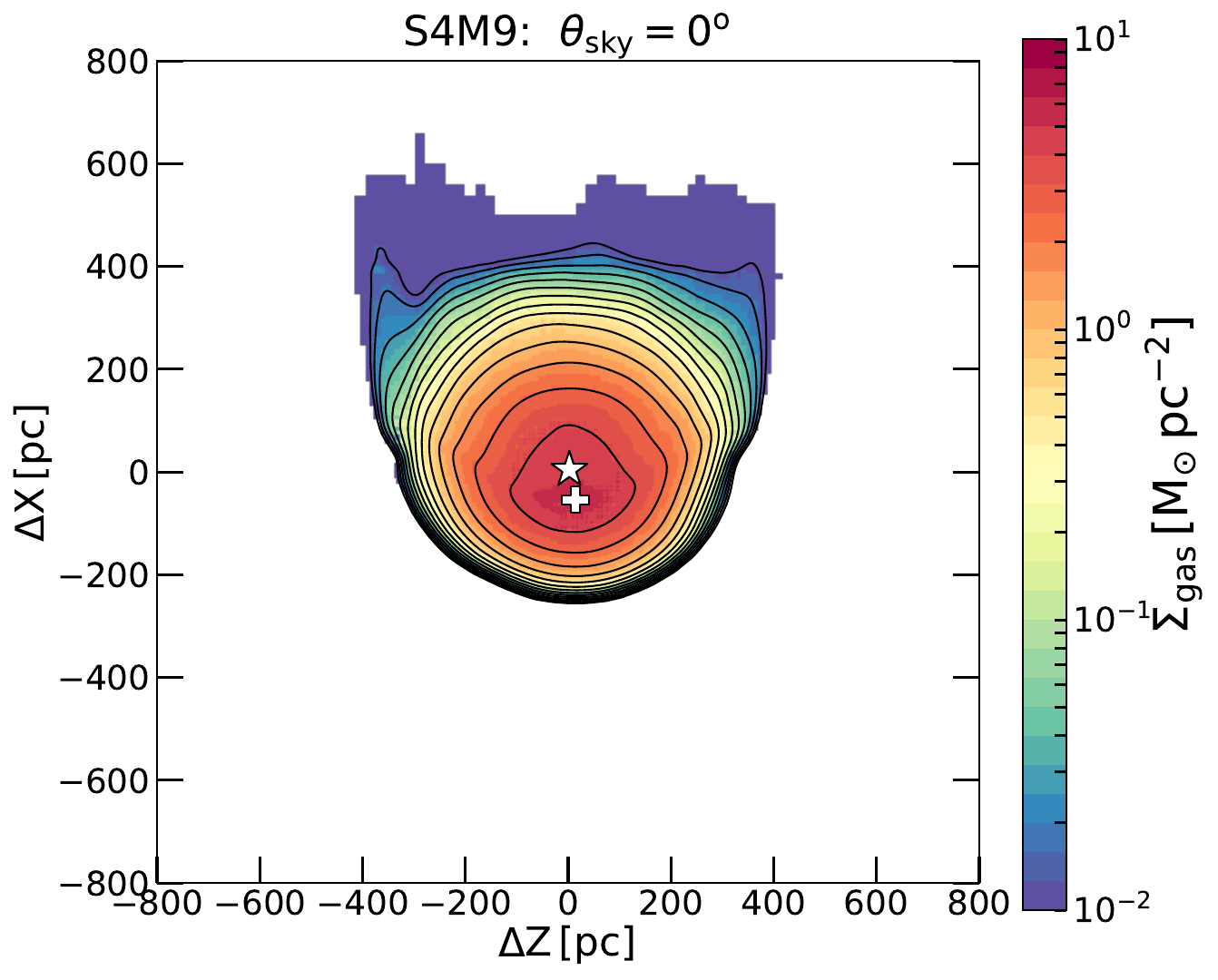}
\includegraphics[width=8.5cm]{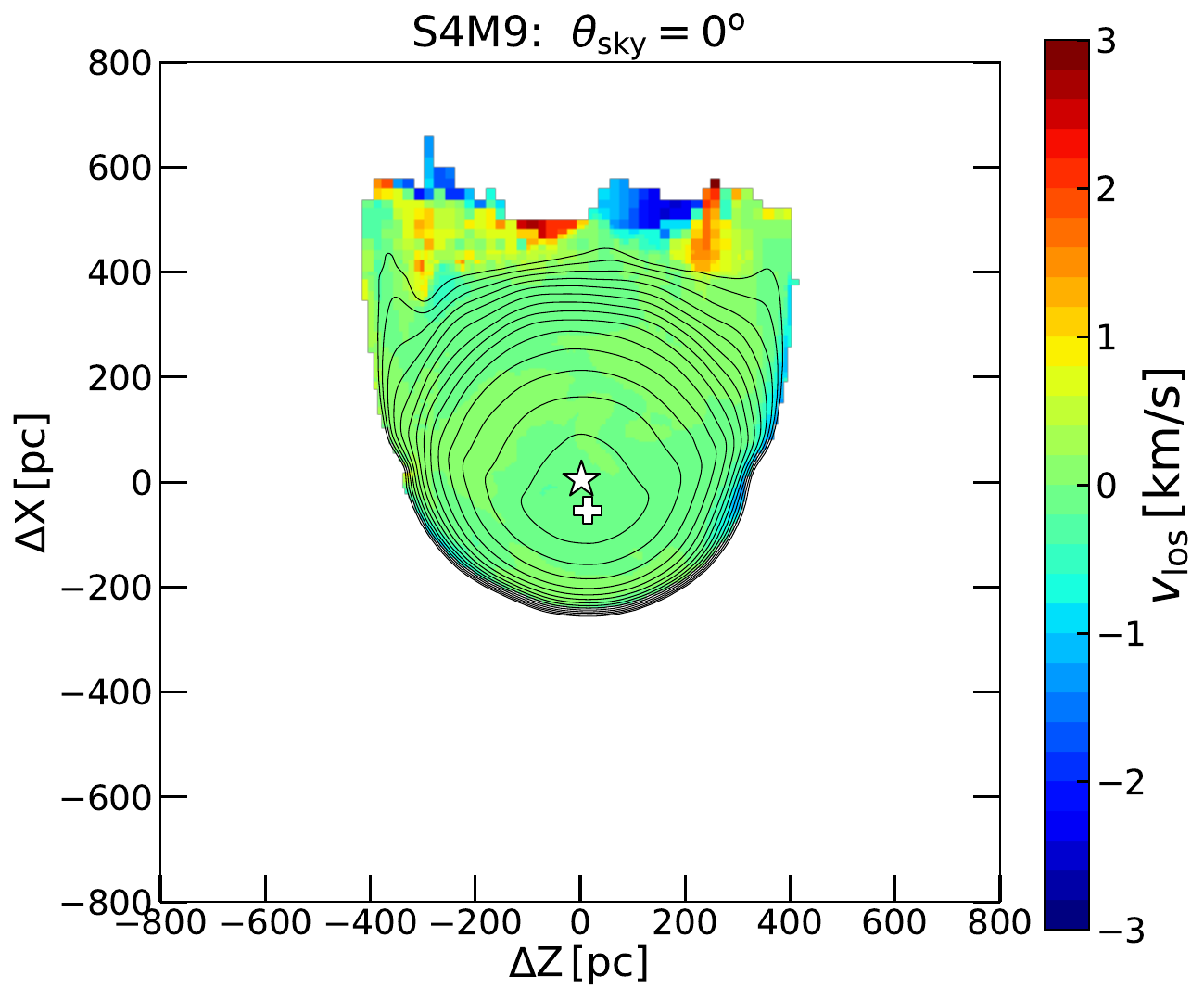}
%\vspace{-0.3cm}
\caption{\chII{Scenario I: Snapshot of the model M5 (S4M9) surface mass (left) and LOS velocity (right) projected with $\theta_{\rm sky}=18\dg$ corresponding to the BA--FI+P orbital solution with a final tangential velocity of $u_{\rm t}\!=\!100\kms$ (top rows).
This model uses the low-mass dwarf model D1.
The combined centre of mass for the stellar  and dark matter distributions is marked with the white star, and the gas peak is marked with the white cross.
Here the satellite is found passing over the perturbation added to the IGM with $\delta\rho_{\rm max}\!=\!100\times\rho^{\rm MW}_{\rm IGM}$.
We also project the model with $0\dg$ (bottom row) to better reveal the transient offset between the gas and stars; however, this causes the transient LOS velocity gradient to vanish due the projection. The $v_{\rm los}$ maps are in the rest frame of the dwarf's dark matter centre of mass. The contours on the leading side may cover the velocity map due to the strong compression, 
which can be avoided by zooming in the pdf version.
We note   that $\Sigma_{\rm gas}\!=\!1\sm\pc^{-2}$ corresponds here to $N_{\HI}\!=\!1.2\!\times\!10^{20}\icmsq$.}}
\label{fig:fig_environment_pert}
\end{center}
\end{figure*}

\chII{As the gas is stripped from the dwarf it is expected to slow down by the IGM ram pressure and mixing,
producing a gradient of the line-of-sight velocity gradient map ($v_{\rm los}$) where the material in the outer gas layers is redshifed with respect to the dwarf's centre of mass and central gas distribution.
An inspection of the kinematics in Fig.~\ref{fig:fig_environment} reveals only a slight $v_{\rm los}$ gradient in the outskirts of the dwarf models. After inspecting the models, we see that rotating them with an angle of $\theta_{\rm sky}\!=\!90\dg$ produces a stronger gradient of $v_{\rm los}$, which is stronger for fast orbits with $u_{\rm t}>300\kms$.
A consequence of the anti-correlation implied by Eq.~\ref{eq:angle} is that the faster the select satellite orbital solution is
(and larger $u_{\rm t}$), the weaker the gradient along the line-of-sight velocity will result, which confines the motion to the plane of the sky. 
Furthermore, the more dense central region within $R<200\pc$ does not show velocity gradients.
On the other hand Leo~T  shows a velocity difference of $\Delta v_{\rm los}\sim 5\kms$ between the central gas $(-2\kms)$ and the outer gas layer ($3\kms$) in Fig.~\ref{fig:LeoT_HI} (bottom right panel), which also results in a difference of $\sim2\kms$ between the systemic velocities of the CNM and WNM gas components \citepalias{Adams2018}.
Furthermore, the centre of the galaxy model shows no significant turbulent gas motions, mainly supported by pressure, whereas observations show a more turbulent internal region.}

\chII{Inspecting the region with the material trailing the dwarf models in Fig.~\ref{fig:fig_environment}, we see that the stripped colder gas of the dwarf mixes with the hot component of the IGM gas creating a warmer and turbulent region that can quickly vary its motion due to its mixing, making it difficult to identify a clear nonturbulent gradient.
Given that Leo~T shows some material that could be trailing in the northern region at the limit of detection (see Fig.~\ref{fig:LeoT_HI}), suggests that a high tangential velocity ($u_{\rm t}>200\kms$) would be required for significant stripping.}

\subsubsection{Scenario I: IGM overdensities}
\label{sec:res:env:pert}
Motivated by cosmological simulations that find gas substructures in the distribution of the CGM and IGM of galaxies, such as the splashback radius (see Sect. \ref{sec:mod:env}), we included an overdensity above the background IGM gas density
to explore how a satellite passing over these substructures would react. 
\chII{In Fig.~\ref{fig:fig_environment_pert} we show model M5 (S4M9) that used the dwarf galaxy model D1 and the orbit with a final tangential velocity of $u_{\rm t}\!=\!100\kms$, where we selected a snapshot that presented an offset between the gas and the stars. 
The offset is formed in a transient configuration, occurring only when the dwarf is in the process of leaving the IGM density perturbation and requiring a strong perturbation with an overdensity with a factor of $100$ over the background density to produce a significant offset
(i.e. $\delta \rho_{\rm IGM}\!\gtrsim\!100\,\rho_{\rm IGM}\sim 10^{4}\sm\ikpccube\,(4\!\times\!10^{-4}\icmcube))$.
The offset has a short duration ($\lesssim 50\Myr$), shifting the gas to the leading side from the centre by 50\pc when observed from the side $\theta_{\rm sky}\!=\!0\dg$, but shows a weaker offset for the correct orbital orientation of $33\dg$.
\chA{We also tested the effects of the WSRT observations beam size on the apparent morphology of the perturbed dwarf model.
For this we convolved the image of model M5 with a (2D) Gaussian function kernel with the observed beam size of 2($\sigma_{\rm z}$, $\sigma_{\rm x}$) = (15.7, 57.3)\as = (31, 113)\pc. As shown in Fig.~\ref{fig:fig_beamsize} (left column), the resulting morphology has rounder gas contours due to the asymmetric beam size. However, the gas peak still remains shifted forward from the stellar centre.
}
\newline
In the inspection of the kinematics we find that the velocity gradient with the corresponding orientation of $33\dg$ is stronger than in the case without the IGM perturbation and the same final orbital velocity $u_{\rm t}\!=\!100\kms$ shown in Fig.~\ref{fig:fig_environment_A} (middle column). The redshifted region within $R\sim400\pc$ in the model has similar qualities to the redshifted region to the north in Leo~T.
This could offer a possible solution to the velocity gradient observed in the outer region of Leo~T, and
it is an intriguing coincidence, that Leo~T is located near the splashback radius of the MW at 1.4\Rvir, where such substructures are expected. However, the central region in the model remains in hydrostatic equilibrium while in Leo~T the centre appears the as a strongly blue-shifted region with negative $\Delta v_{\rm los}$ values (see bottom panels in Fig.\ref{fig:LeoT_HI}).}

%%% =========================================================================================
%%% =========================================================================================
%%% =========================================================================================

\subsection{Scenario II: Stellar winds and internal perturbations}
\label{sec:res:int}
The observations of Leo~T of \citet{DeJong2008} reveal an old stellar population ($>7\Gyr$) and a younger population ($\sim200-1000\Gyr$) separated by 35\pc in projection,
while the \HI observations reveal a 80\pc offset with the old stellar centre \citepalias{Adams2018}, as shown in Fig.~\ref{fig:LeoT_HI}. 
In this section we explore the scenario where the younger stellar component could be oscillating around the dwarf centre while observed
AGB star candidates \citep{Weisz2012} could be associated with this component and generate stellar winds that interact with the dwarf's ISM, 
producing the \HI offset and morphology while the satellite travels trough the MW IGM.

\subsubsection{Scenario II: Gas morphology and kinematics}
\label{sec:res:int:mo}

\chII{For this scenario we used a range of dwarf galaxy models with different dark matter mass profiles, ISM composition (WNM, CNM), and AGB stellar wind strengths, which are listed in Table \ref{tab:setup} (models M8 to M14).
As we wanted to explore and include environmental effects driven by the IGM ram pressure, we used the orbital solution with a final tangential velocity of $u_{\rm t}\!=\!200\kms$.
As shown in Sect. \ref{sec:res:env} for the analysis of Scenario I, velocities slower than this results very week environmental perturbations, while larger than this it might dominate over the internal processes that we address in this scenario.}

\begin{figure*}[ht!]
\begin{center}
\includegraphics[width=8.0cm]{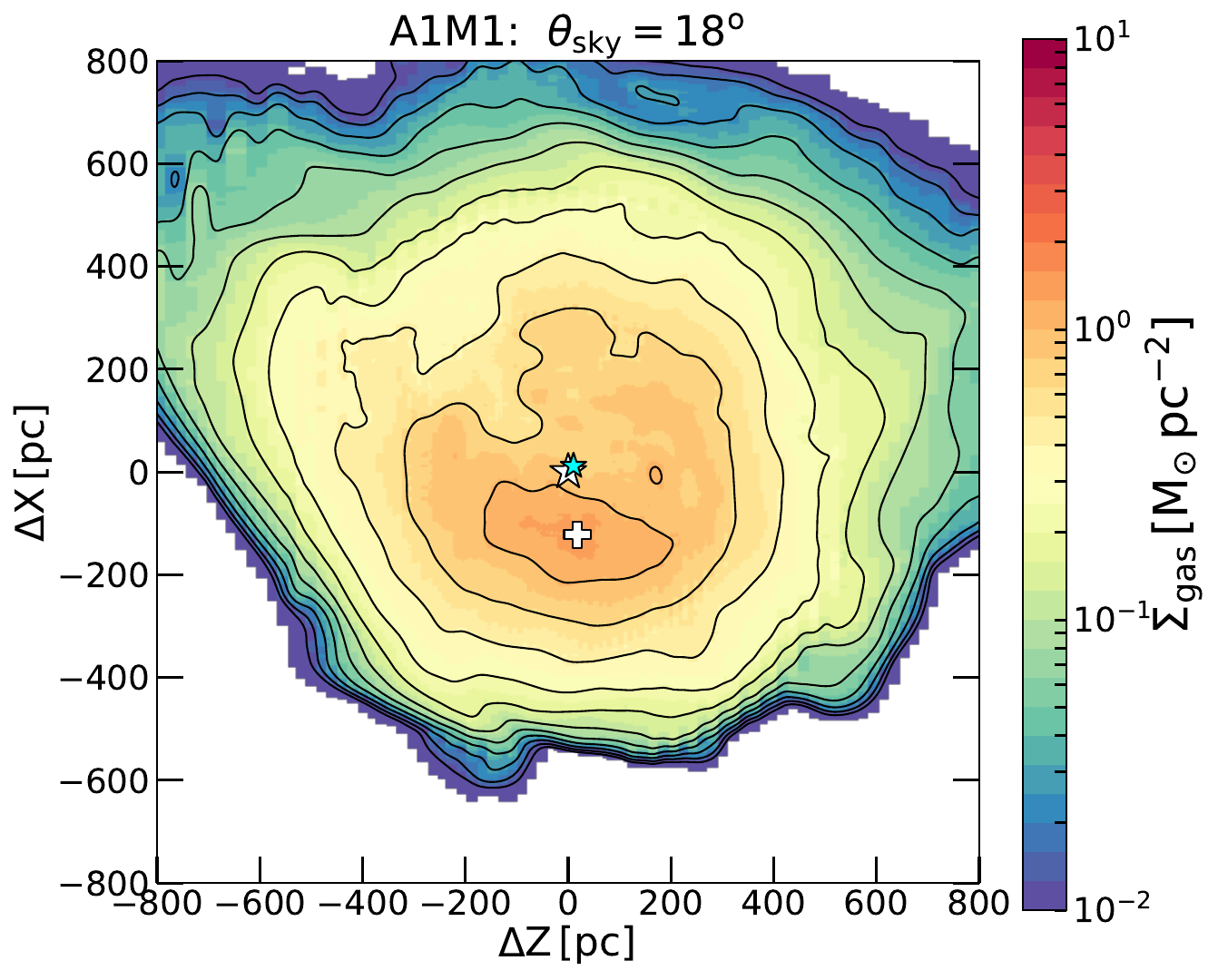}
\includegraphics[width=8.0cm]{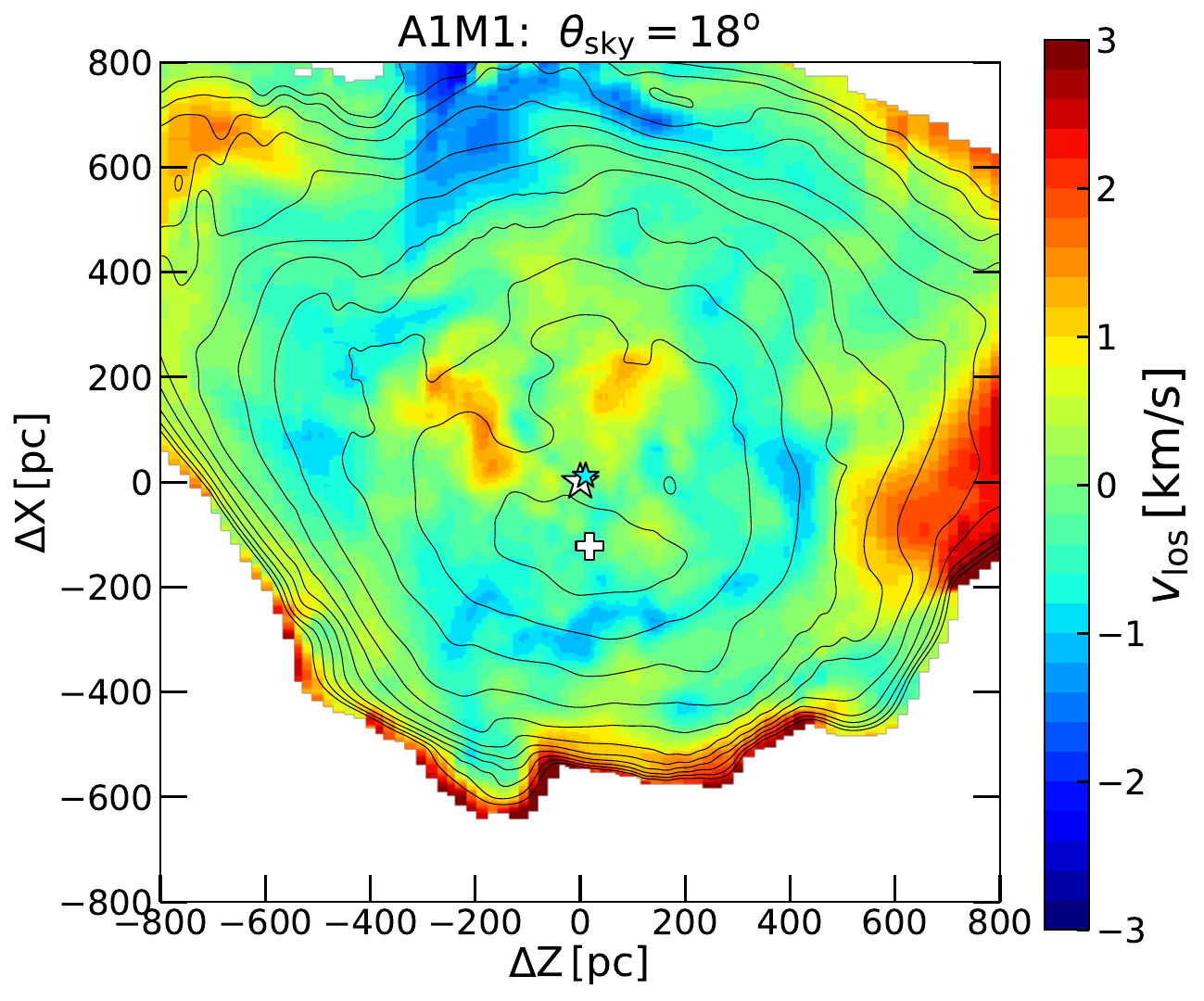}

\includegraphics[width=8.0cm]{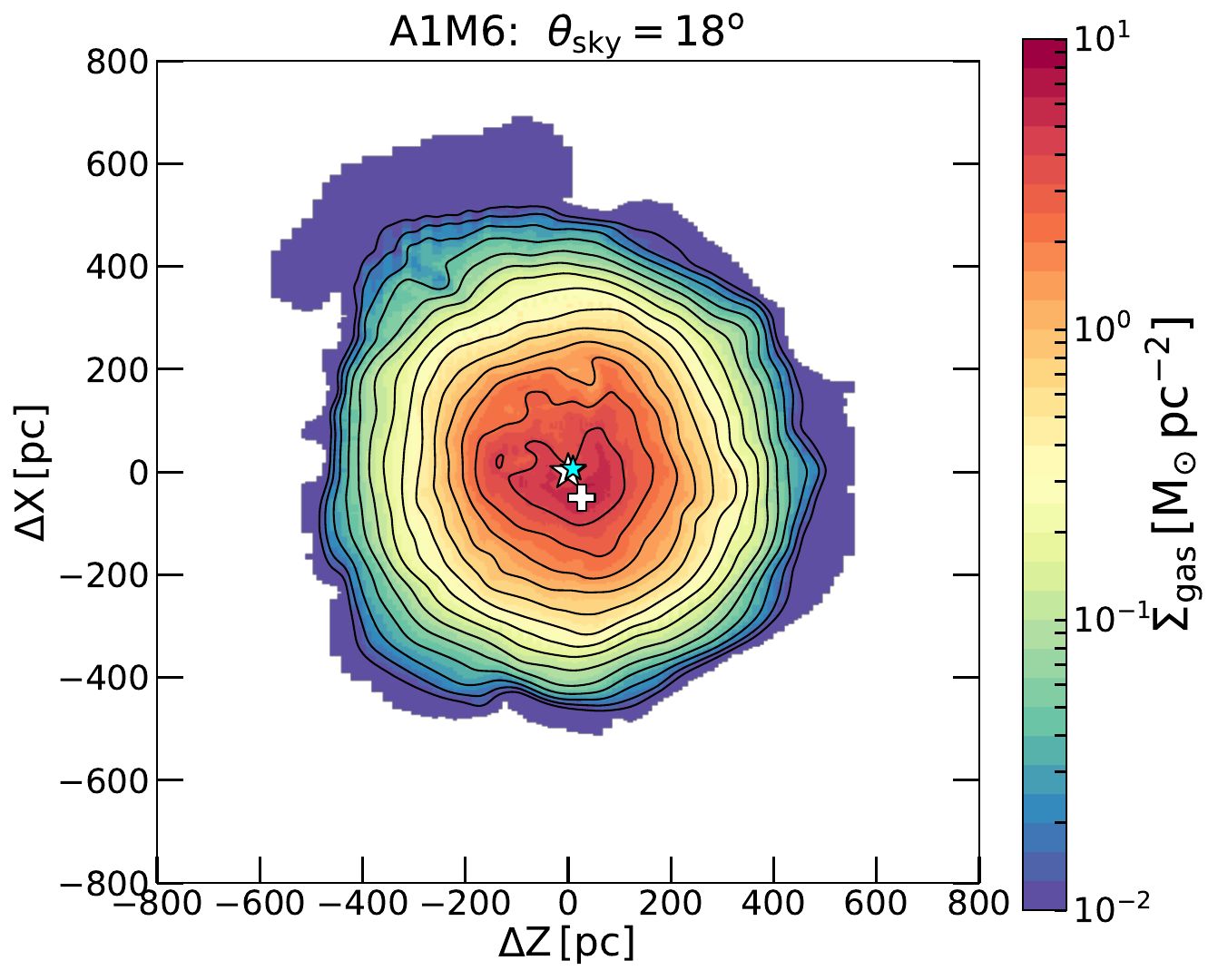}
\includegraphics[width=8.0cm]{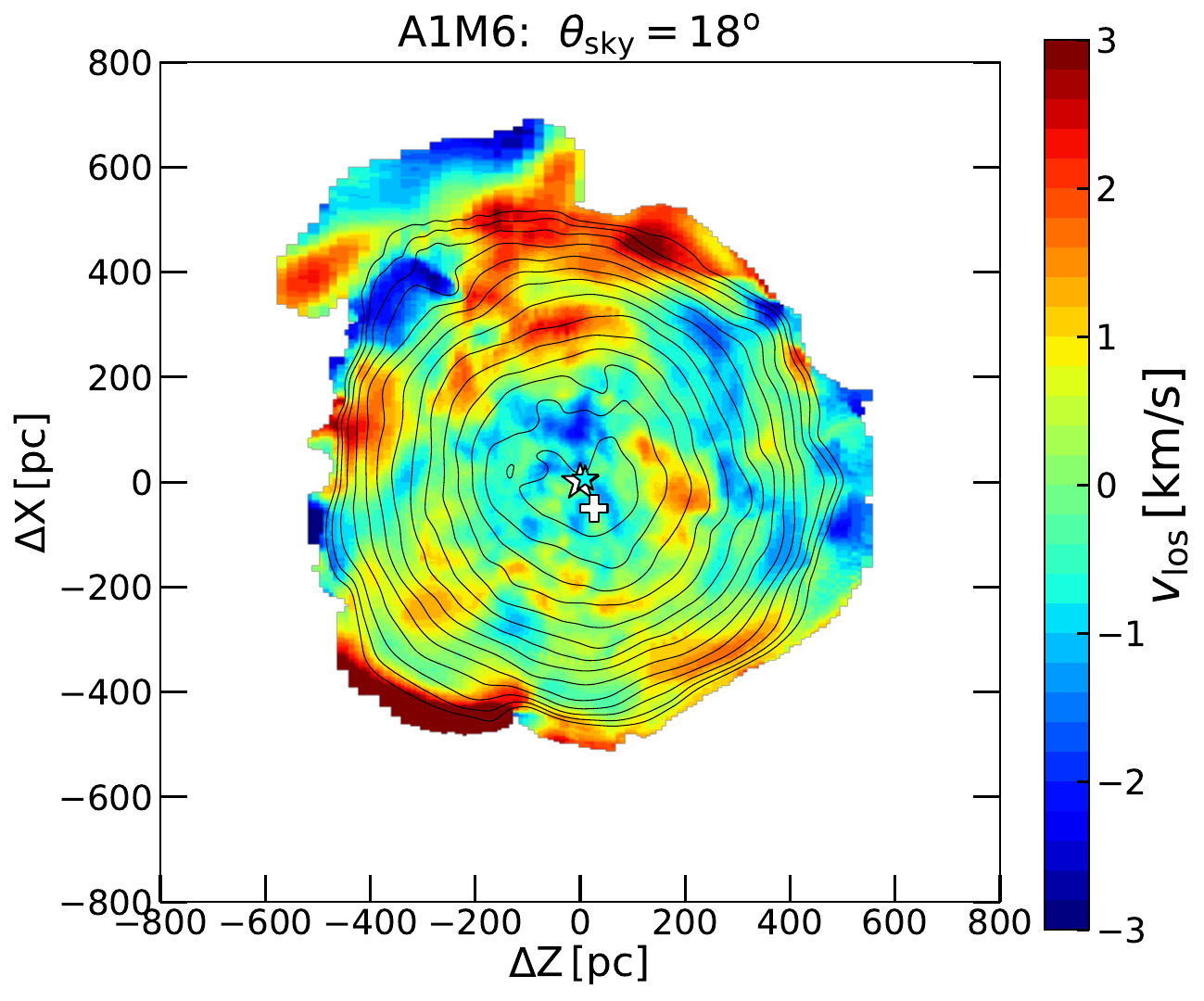}

\includegraphics[width=8.0cm]{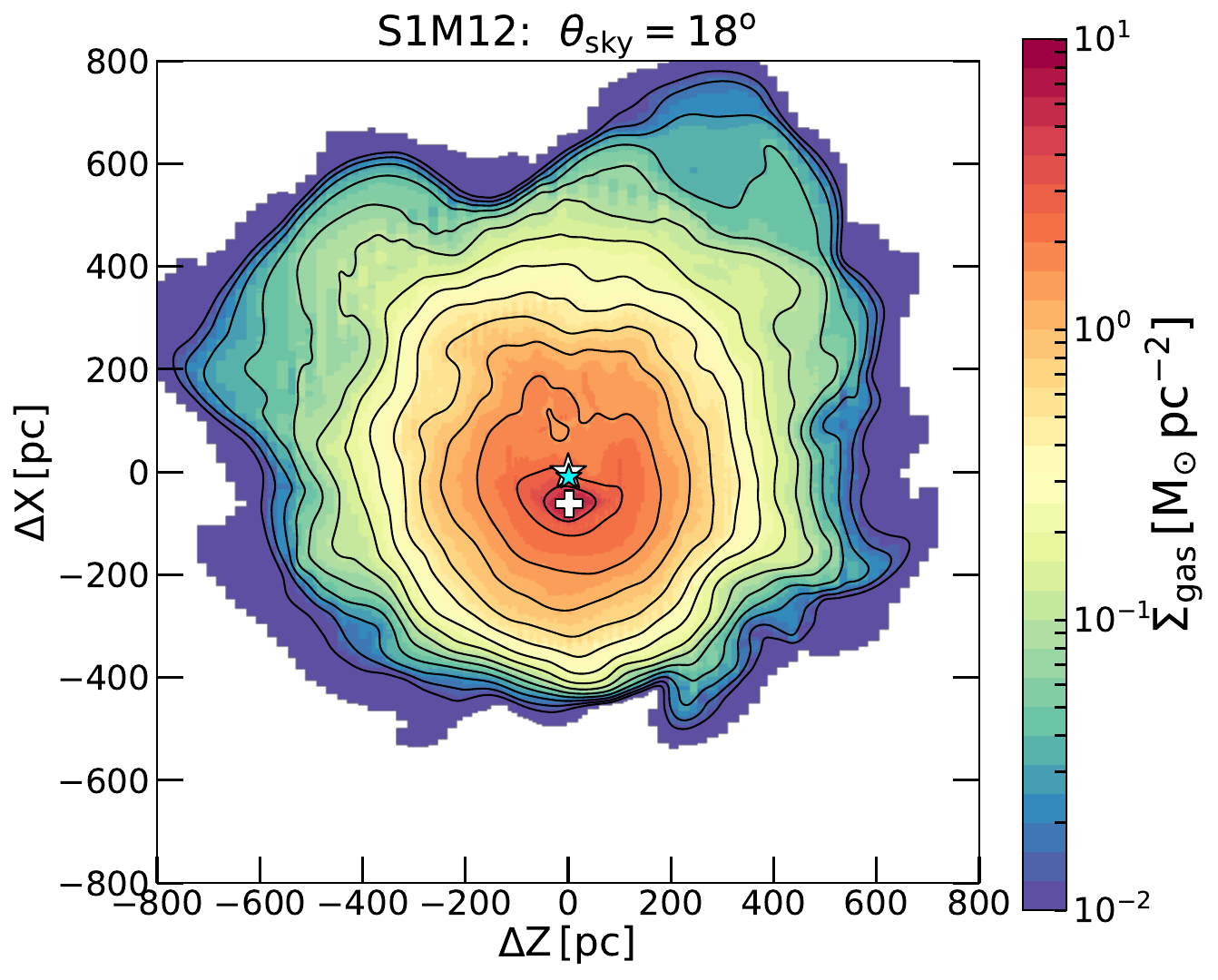}
\includegraphics[width=8.0cm]{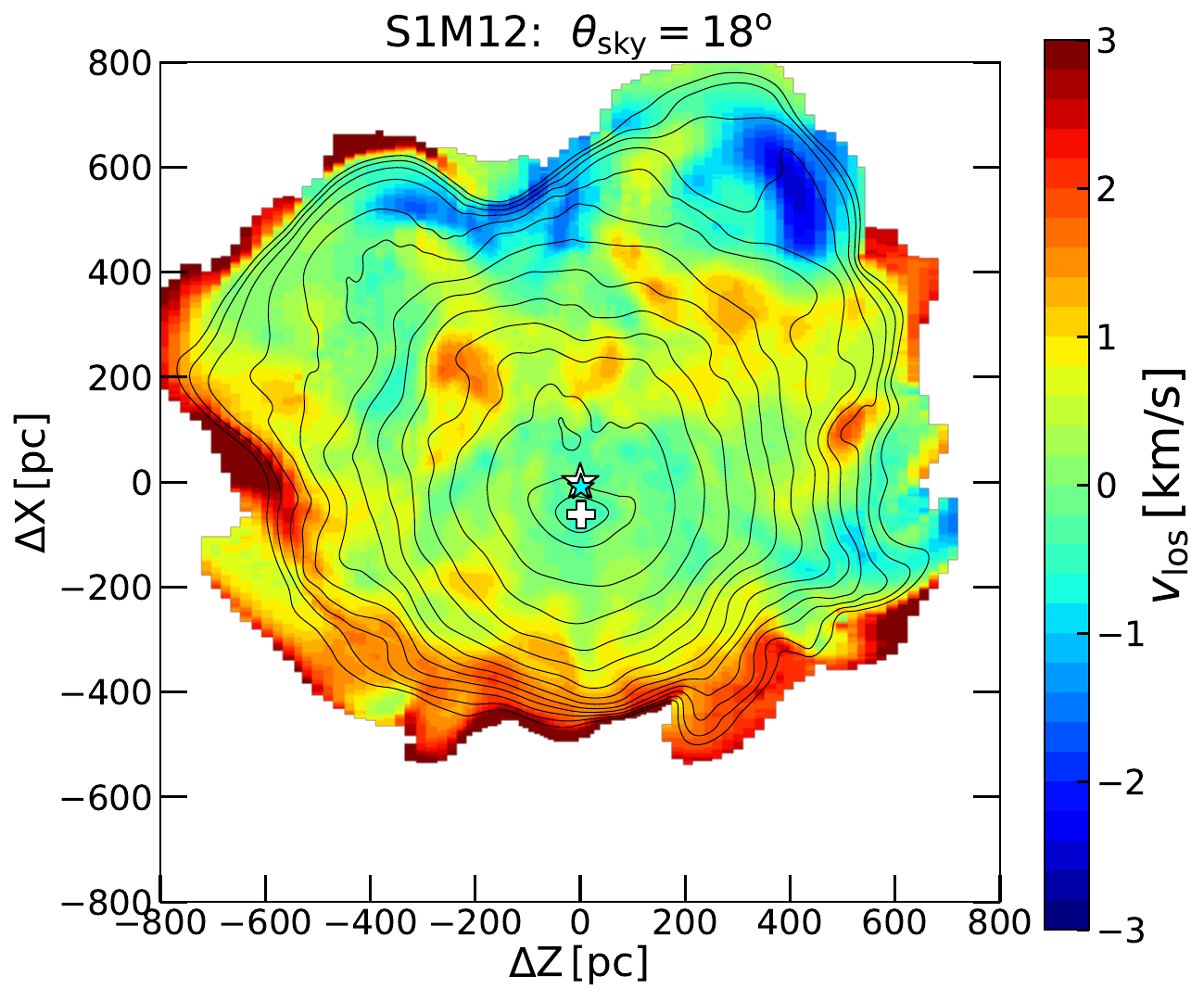}
%\vspace{-0.2cm}
\caption{\chII{Scenario II: Snapshots of model M8 (A1M1) at 957\Myr with the dwarf model D1 (top row), 
model M9 (A1M6) at 358\Myr with D2 (middle row),
and model M10 (S1M12) at 407\Myr with D1t that has \LEt{***that has? shows? } two gas components (bottom row).
Surface mass density maps (left column) and the LOS velocity maps (right column) and contours are calculated with gas colder than $T\!<\!10^4\kelvin$. 
The orbit corresponds to a first infall solution (FI) with a final tangential velocity of $u_{\rm t}\!=\!200\kms$ that corresponds to a projection angle of $\theta_{\rm sky}\!=\!18\dg$. 
The centre of mass of the old stellar and dark matter components is marked with a white star, 
while the younger stellar component centre of mass is marked with a cyan star, and the gas peak with a white cross. 
The snapshots were selected to qualitatively match the \HI morphology of Leo~T, showing the central gas isocontours displaced from the centre of mass of the dwarf, and with the younger component slightly offset as well. We note that the $v_{\rm los}$ maps are in the rest frame of the dwarf's dark matter centre of mass. 
We note  that $\Sigma_{\rm gas}\!=\!1\sm\pc^{-2}$ corresponds here to $N_{\HI}\!=\!1.2\!\times\!10^{20}\icmsq$.}}
\label{fig:fig_inner}
\end{center}
\end{figure*}
\chII{After a careful inspection of the evolution of all the simulations, we selected the models with snapshots that have gas morphologies similar to Leo~T in that they have flatter gas isocontours within 200\pc, shifted from the stellar and dark matter centres of the dwarf models.
These correspond to models M8 (A1M1), M9 (A1M6) and M10 (S1M12), which are shown in Fig.~\ref{fig:fig_inner}.}
We find that model M8, with the dwarf model D1, shows a stellar-gas offset distance similar to the one observed in Leo~T, which is produced while the younger stellar component oscillates through the dwarf's centre. In Sect.~\ref{sec:res:int:mi}, we analyse these oscillations \chII{and explain further how the offsets is generated.}
We notice that the location of the offset in the model is not perfectly aligned on the sky as in Leo~T, but this can be simply corrected by rotating the position of the initial stellar offset around the dwarf's centre.
In the figure, we include the results for model M9, which has the massive cored dwarf model D2, also showing central gas isocontours that are shifted from the centre.
This model requires slightly stronger stellar winds to be able to push the gas due to the stronger gravitational potential, but still within the AGB star wind literature estimates.
We also tested a model with no stellar winds, which also presented a periodic offset of the younger stellar component with the gas and old stellar distributions due to its oscillations, but the gas showed no offset with the older stellar component, as it is not being pushed and it remains in the dwarf centre with round gas isocontours.
The gravitational force of the younger stellar distribution alone is unable to significantly perturb the gas distribution, given that the potential is predominantly governed by the dark-matter component, as shown by the mass profiles in Fig.~\ref{fig:fig_massprof}.\\

% Kinematics
\chA{The line-of-sight velocity maps shown in Fig.~\ref{fig:fig_inner} of the simulations also reveal that the internal gas kinematics is much more complex than in the models of Scenario~I where no internal perturbations are included (see Fig.~\ref{fig:fig_environment}). 
In particular, Scenario~II models better resemble the (turbulent) variability observed in the velocity map of Leo~T (see Fig.~\ref{fig:LeoT_HI}). }
\chII{For example, model M9 (A1M6) shows in Fig.~\ref{fig:fig_inner} that the central region $R<100\pc$ has velocity differences of $\Delta v_{\rm los}\sim 4\kms$ with respect to the outer region with a blue-shifted zone near the central peak of the gas density, similar to observations.}\\

% Two gas components
\chII{In Fig.~\ref{fig:fig_inner} we also show a snapshot of model M10 (S1M12), which uses the dwarf model D1t, which has two gas components (CNM,WNM) and the same dark halo as D1 (see Table \ref{tab:setup}).
\chA{To show a case where the stellar winds have motions that oppose the perturbation produced by the IGM wind, we chose ${\rm PA}\!=\!0\dg$.}
The resulting morphology also successfully presents flatter gas isocontours in the centre and an offset between the gas peak and the younger stellar component. It takes a slightly longer time for the stellar winds to push the denser CNM component off the centre, but still faster than the age of the younger stellar component, 
which supports a scenario where Leo~T could indeed present a fraction of \HI in the form of CNM. The LOS gas velocity map shows a velocity offset between the outer layers and the central region within 100\pc.\newline

\chA{We also used our models to test the effects of the beam size on the gas morphology of the WSRT observational data and the stellar-\HI offset. 
For this we convolved the images of models M8 (A1M1) and M10 (S1M12), originally shown in Fig.~\ref{fig:fig_environment}, with a (2D) Gaussian function kernel with the beam size 2($\sigma_{\rm z}$, $\sigma_{\rm x}$) = (15.7, 57.3)\as = (31, 113)\pc.
As shown in Fig.~\ref{fig:fig_beamsize} the resulting morphology can be affected within the beam size scales, while preserving the original morphology and flattening for larger scales.
For example, Model 8 has more extended flattened contours up to $R\sim300\pc$ and aligned with the horizontal axis, while model M10 shows a central flattening aligned with the vertical axis (${\rm x}$) within the size of the longer kernel ($2\sigma_{\rm z}\gtrsim 113\pc$). 
However, contours at larger distances $R>150\pc$ recover a morphology similar to the resolved image in Fig.~\ref{fig:fig_environment}.
Moreover, the offset between the gas and the stars persists in both tests.
Given that Leo~T shows a flattened contours roughly aligned with the horizontal axis up to $R<300\pc$ would indicate that this is unaffected on beam size scales. 
The comparison with the model would indicate that the central region ($R<130\pc$) where the CNM component is located, could be in fact more elongated and aligned with the horizontal axis, following the general central distribution. 
Future higher resolution observations could potentially better reveal the CNM structure in Leo~T.}\\}

%% Offsets for Massive Burkert and NFW!!
We also explored the effects of stellar winds on the massive compact cored dwarf model D3, corresponding to model \chII{M11 (SZM4)} in Table \ref{tab:setup}.
We find that even though a gas offset is also generated, it reaches only 50\pc from the dwarf centre, even when doubling the strength of the stellar winds used in model \chII{M9 (A1M6)} with the dwarf model D2.
And finally, we explored the effects of the stellar winds in the dwarf model D4 with the cuspy (NFW) dark halos in models M12 (A1M7), M13 (S7M3) and M14 (S7M4).
Similarly to the case of the massive compact core dwarf (M11), we find that within the range of explored stellar wind parameters considered for AGB stars, the winds only weakly perturb the gas distribution in the inner region \ch{and no gas-stellar offset is observed}. 
\chII{The cuspy and compact core dwarf models have deep gravitational potentials in their centres, which confine the gas there more easily, making it difficult for the stellar winds to displace the gas from the centre.}
\ch{This could suggest that if Leo~T is more massive or if it has a cuspy dark matter distribution (e.g. NFW),\LEt{***e.g. and i.e. are never in italics. Please amend throughout } other mechanisms more powerful than (AGB) stellar winds should be responsible for the gas perturbation.}

\begin{figure*}[ht!]
\begin{center}
\includegraphics[width=5.9cm]{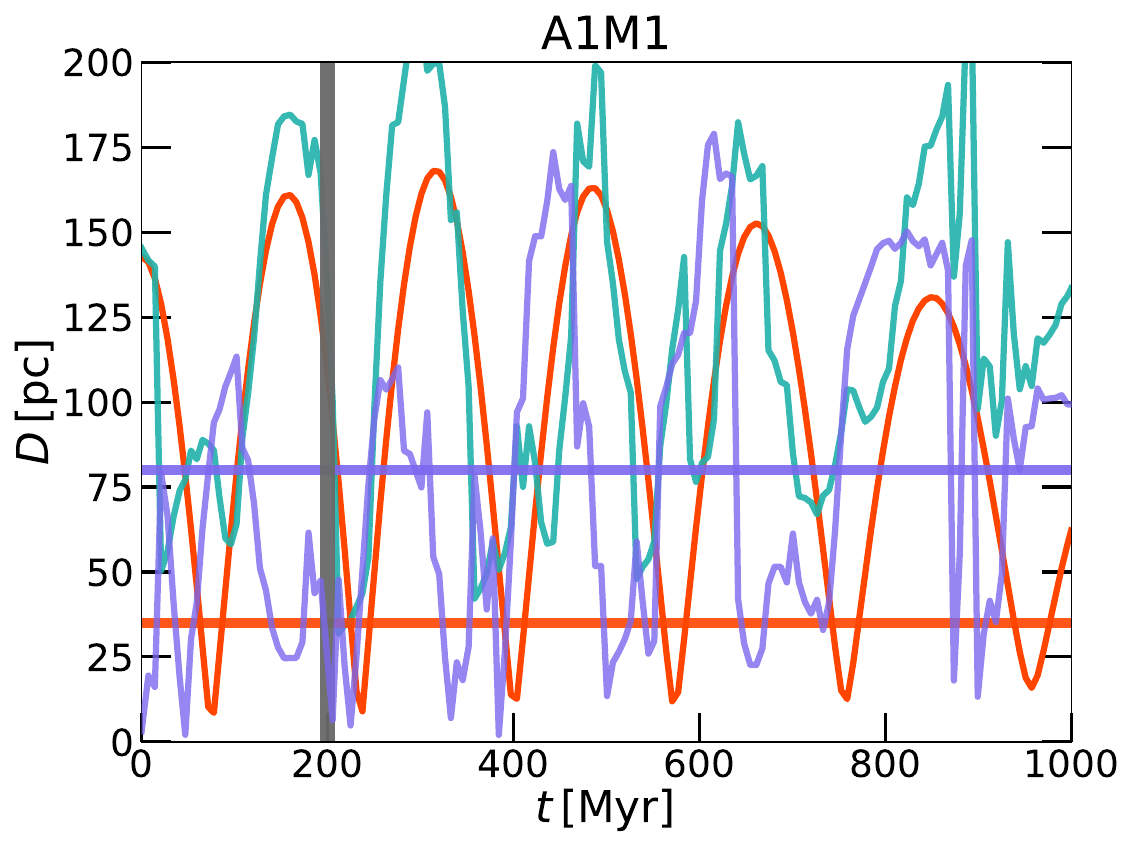}
\includegraphics[width=5.9cm]{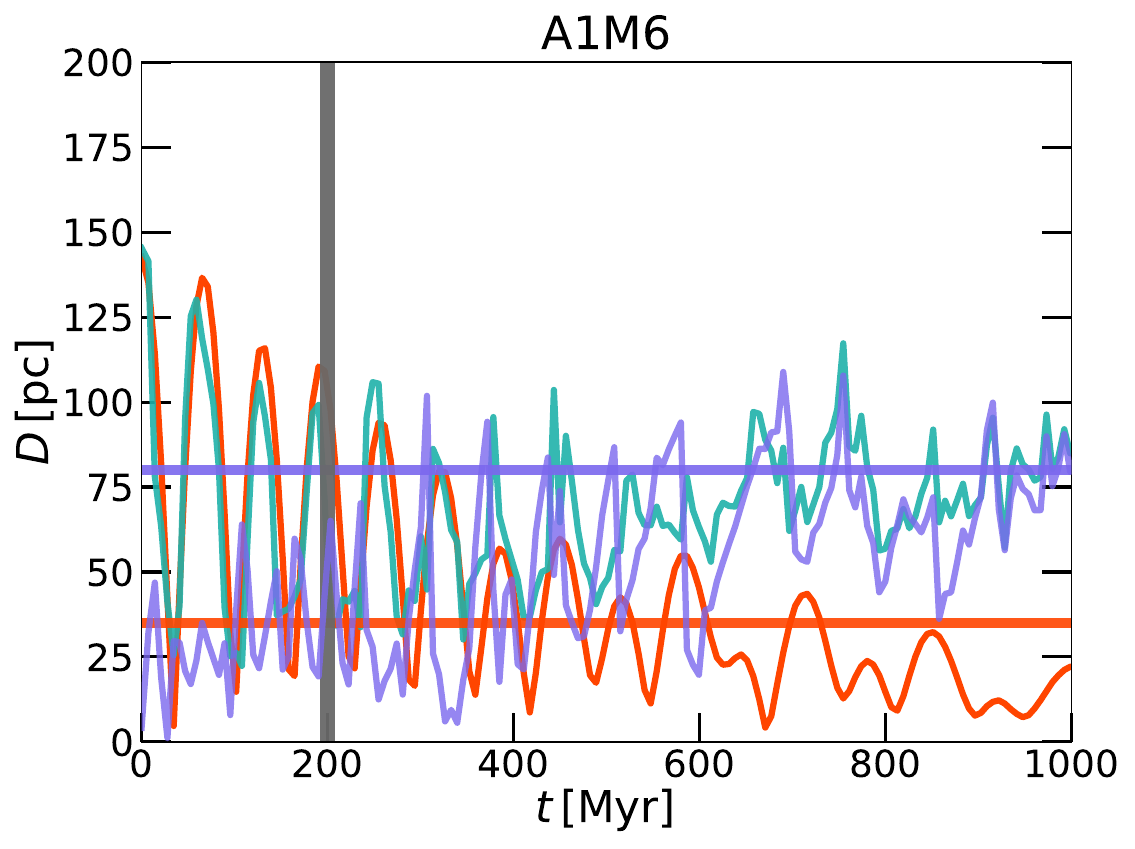}
\includegraphics[width=5.9cm]{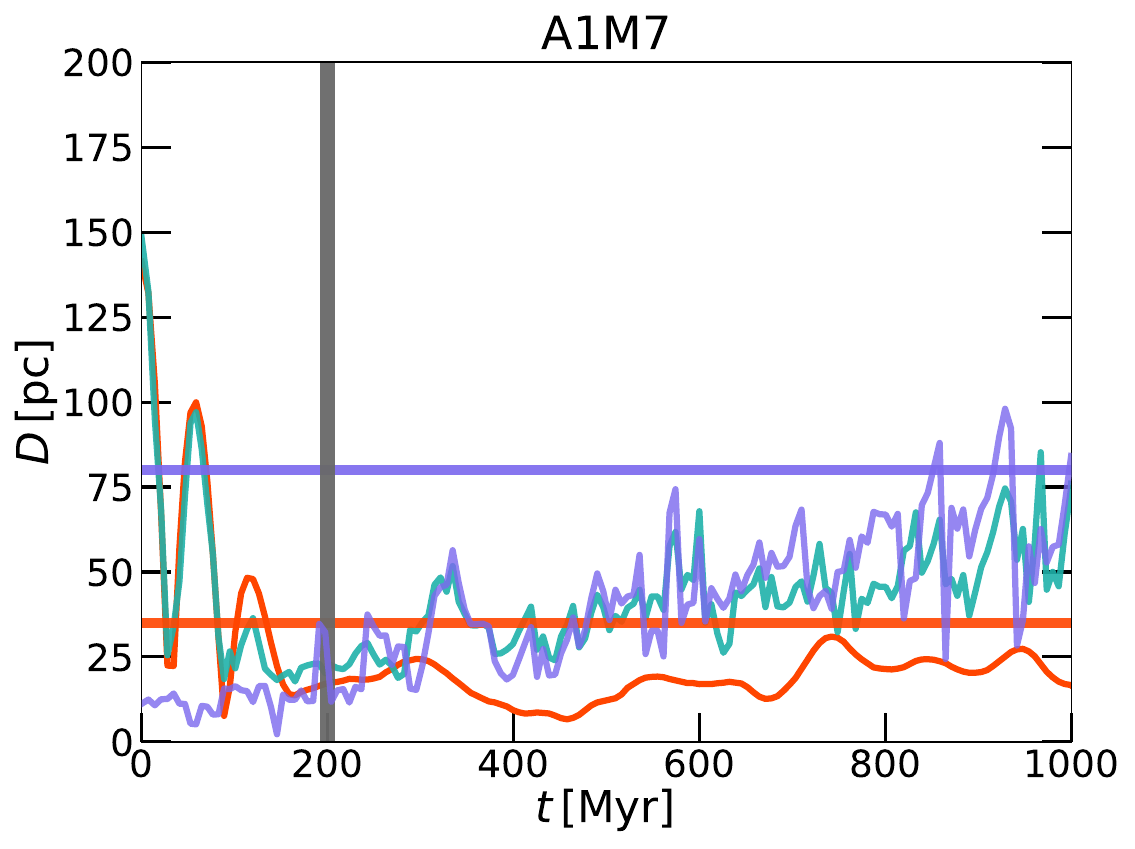}
%\vspace{-0.3cm}
\caption{\chII{Scenario~II: Temporal evolution of model M8 (A1M1) (left), which uses the dwarf model D1; model M9 (A1M6) using D2 (middle);
and model M12 (A1M7) using D4 (right).}
We show the distances with respect to the centre of mass of the dark matter halo of the younger stellar component (red curve), 
and to the peak of the gas surface mass density (blue) measured with $\theta_{\rm sky}=0\deg$ to quantify the maximum distances it could reach. 
We also show the distance between the younger stellar component centre and the gas surface density peak (turquoise). 
The observed 80\pc offset between the gas density peak and the stellar component in Leo~T is indicated with a blue horizontal line \citepalias{Adams2018}. The 35\pc offset between the younger and older stellar components is shown with the red horizontal line, while the vertical line indicates the lower age estimate of the younger stellar component \citep{DeJong2008}.}
\label{fig:fig_distance}
\end{center}
\end{figure*} 
\subsubsection{Scenario II: Mixing timescale of the younger stellar component}
\label{sec:res:int:mi}
\chII{Here we compare the orbit of the younger stellar component in the dwarf galaxy model, with the observed 35\pc offset between the older and younger stellar components in Leo~T.}
To do so, we measured the oscillations of the centre of mass of the younger stellar component with respect to the centre of mass of the dark-matter distribution as it orbits through the centre of the dwarf galaxy model while the stellar winds are being blown, showing the results of models M8 (A1M1), M9 (A1M6) and M12 (A1M7) in Fig.~\ref{fig:fig_distance}. 
\chA{We measured these distances in a frame where the younger stellar component oscillates perpendicular to the LOS to measure the maximum effect of the stellar winds, reducing  projection effects. 
We also tested the effects of the beam size on the oscillations, shown in Fig. \ref{fig:fig_distance_gauss}, which basically results in a smoother version of Fig.~\ref{fig:fig_distance}.}

In general we find that the oscillations through the dwarf centre get dampened and eventually stop, mixing the younger stellar material with the rest of the dwarf as the disruption occurs.
Given that the potential is dominated by dark matter and that no significant redistribution of dark matter occurs, it suggests that the disruption is driven by phase-space mixing and/or chaotic mixing \citep{Merritt1996,Merritt1999a}.
By comparing models, we find that model M8 with the fiducial dwarf model D1 and model M9 during a time interval, showed oscillations with maximum offsets larger than the observed 35\pc offset between the young and older stellar components. 
\chII{This, however, is not in conflict with observations, which could simply predict that the offset could reach larger distances in the future, supported by the velocity difference, or in could mean that the projection of Leo~T partially hides this oscillation along the line-of-sight axis.}\\

Furthermore, the oscillations of the younger stellar component of model M8 last longer than 1\Gyr, while model M9 with the dwarf model D2 lasts up to 600\Myr.
These long timescales are in agreement with the lifetime estimates for the younger stellar component, which is between $\sim200\Myr$ and $1000\Myr$ old.
Therefore, we consider models that last longer than $>200\Myr$ as possible solution candidates to the observed offsets between stellar components and also the gas distribution, 
while shorter timescales would have erased an offset signature by now, or the offset would strongly depend on the distance of the initial offset that we use for the younger stellar component.
Particularly on this last option, it is possible to select larger initial values, we are able to find \chII{stable solutions with the cored dwarf models D1 and D2 that naturally produce lasting oscillatory solutions that survive extremely long timescales after the initial offset is generated.
Such offsets could be generated in a previous star formation episode, as demonstrated by the observed offset between the younger and older stellar components in Leo~T.}
% Model M8 shows oscillation timescales of $\sim1\Gyr$ that decrease slowly with periodic offset distances that can match the observed offset, demonstrating the viability of this model scenario. 

\begin{figure}[ht!]
\begin{center}
\includegraphics[width=8.8cm]{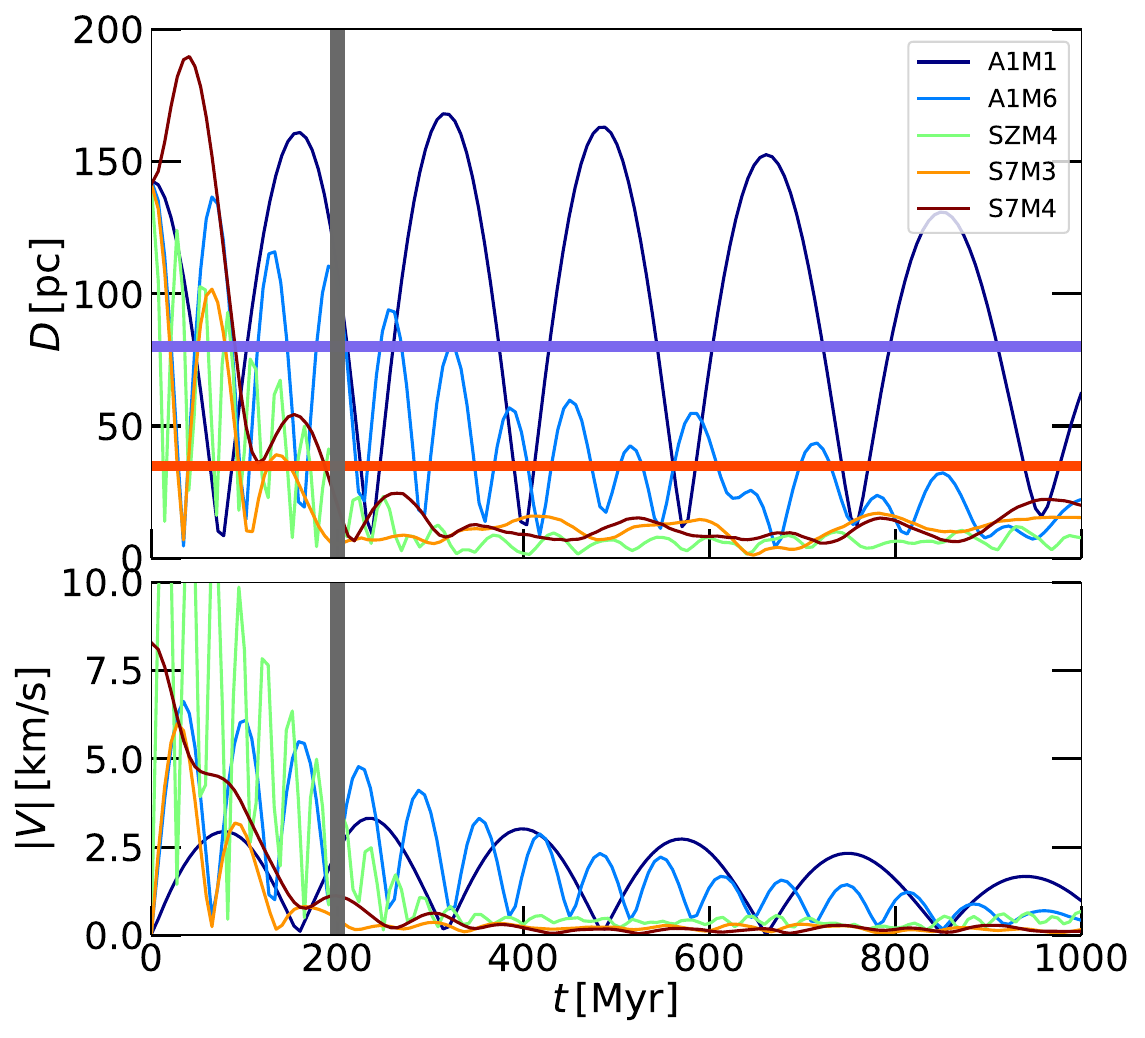}
%\vspace{-0.2cm}
\caption{Scenario II: Distance (top panel) and relative velocity (bottom panel) between the centre of mass of the dark matter component and the younger stellar component for different models as a function of time: Model M8 (A1M1) with the Burkert dwarf model D1, model M9 (A1M6), with the more massive Burkert D2, model M11 (SZM4) with the compact massive dwarf model D3, and models M13 (S7M3) and M14 (S7M4) with the NFW dark halos dwarf model D4.
The red line corresponds to the offset between the old and younger stellar components in Leo~T (35\pc), and the blue line to the 80\pc~ \HI-stellar offset. The gray  vertical line marks the lower age limit of the younger stellar component \citep{DeJong2008}.
The simulations show that the younger stellar component decays faster in the models with a compact cored \chII{M11 (SZM4) and cuspy M13 (S7M3 with radial offset), M14 (S7M4, with circular orbit offset) dark halos, than in the models with large dark matter cores M8(A1M1) and M9(A1M6).}}
\label{fig:fig_distance_velocity}
\end{center}
\end{figure}

\chII{The younger stellar component also presents a kinematic signature in its oscillation, as shown in see Fig.~\ref{fig:fig_distance_velocity} by the change of total velocity of this component with respect to the dwarf's dark-matter centre of mass. 
Depending on the dwarf model, it can reach velocities of $\Delta v\sim5\kms$ at 200\Myr and $\sim2\kms$ after 800\Myr.
Of course, the LOS velocity could be lower if the motion of the younger stellar component is in the plane of the sky, as low as the measured $\Delta v_{\rm los}\!\sim\!0.4\kms$ between both stellar components in Leo~T \citep{Vaz2023}.}

Moreover, we find that our models also generate the stellar-gas offset and it has a size comparable to the one measured in Leo~T (80\pc). As shown in Fig.~\ref{fig:fig_distance} the distance from the dwarf centre of mass to the gas surface density peak and the younger stellar component, showing a clear correlation between both of them. 
\chII{Initially as the younger component with its associated AGB wind source approaches the central gas distribution, it pushes and compresses the gas, generating a momentary shift of the overdensity away from the centre. 
Then they break through the gas distribution moving outwards to its apocenter, which allows the gas to momentarily redistribute at the centre because of the strong gravitational force of the dwarf. At this point the centre of the younger stellar component and the gas density peak can reach the largest separations (Fig.~\ref{fig:fig_distance} model M8 (A1M1) turquoise curves reaching $\sim200\pc$). After this, the younger stellar component starts to fall again inwards \chIII{due to the dwarf potential, and the oscillatory cycle of ISM gas compression and shifting by the stellar winds is repeated.}\footnote{Videos of these simulations can be found at this link \href{https://www.youtube.com/playlist?list=PLoegIUZ3yJ9FuDldEF-USJWd-o_hyLavl}{{(here)}}}}\\ 

% Fitting an exponential function, and an exponential function with an cosine therm to model S1M1, we determine a decaying timescale $\tau_{e}$ of 1000 and 768\Myr respectively, and for models S6M3 of 251 and 173\Myr. 
% We fit an exponential sinusoidal function to the temporal evolution of the distance of the younger stellar component to the dwarf centre to measure the timescale of decay $t_{\rm dec}$, finding for model M4 a value of $t_{\rm dec}=500\Myr$, while for model M5 is 200\Myr.
\begin{figure}
\begin{center}
\hspace{-0.3cm}
\includegraphics[width=4.4cm]{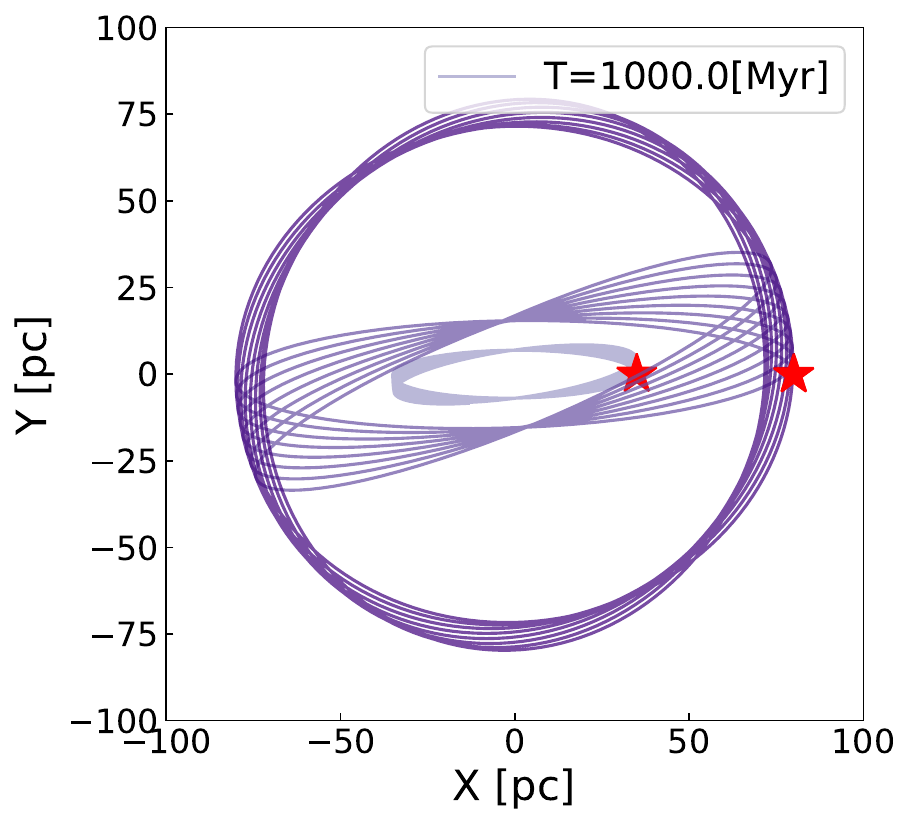}
\hspace{-0.3cm}
\includegraphics[width=4.4cm]{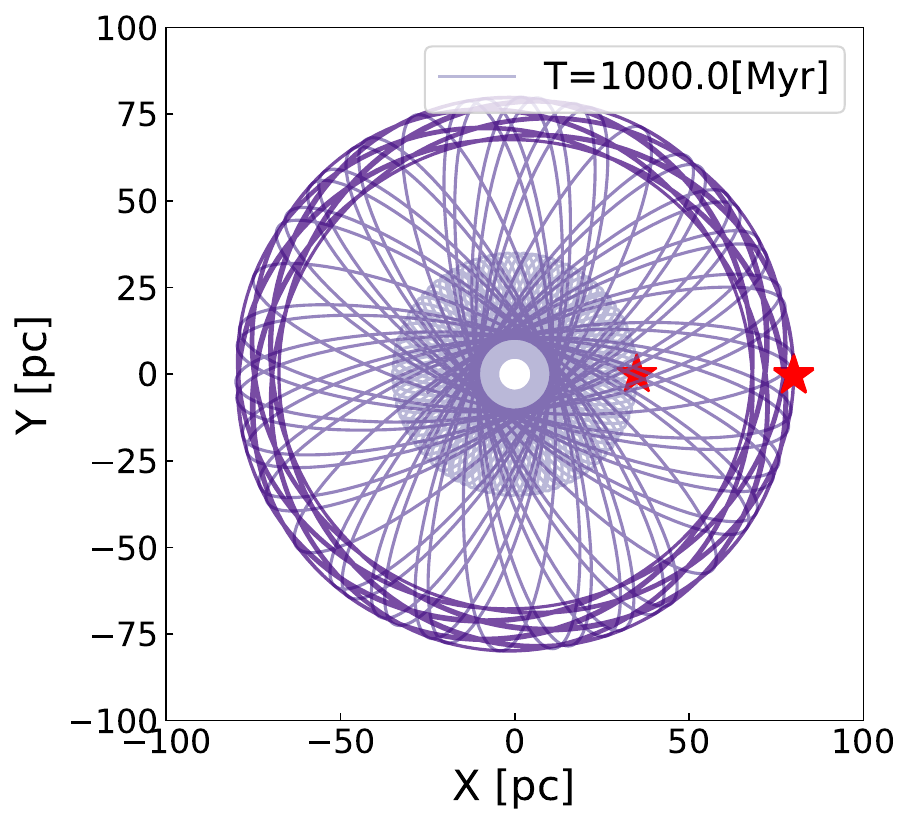}
%\vspace{-0.3cm}
\caption{Scenario II: Stellar orbits (test particles) are calculated during 1\Gyr in the potentials of dwarf models D2 (massive Burkert) in the left panel and D4 (NFW) in the right panel. Both potentials have the same mass at 300\pc, but the NFW has more mass in the inner region. The inner radial orbits in each potential are selected to have an apocentre of 35\pc, corresponding to the observed projected offset between the younger and older stellar component in Leo~T (inner red star), and the two outer orbits (one radial and another near-circular) have apocentres at 80\pc, similar to the offset between the gas density peak and the old stellar component (outer red star). The radial orbits are set with tangential velocities with 20 per cent of the local circular velocity, whereas the outer near-circular  orbit was set with 90 per cent. These orbits highlight how much faster the mixing of the material is in model D4 due to a stronger precession produced in cuspy dark-matter potentials. The orbits are calculated with \sc{delorean} \citepalias{Blana2020}.}
\label{fig:fig_orbits}
\end{center}
\end{figure}

\chII{We also detected \LEt{***to notice=to become aware of (detect, find, observe are better choices). to note=topoint out (We note that) }\coB{Ok} a dependence with the central dark matter distribution. While the cored dark matter models D1 and D2 with masses $M_{300}\!=\!1.7\times10^6\sm$ and $8\times10^6\sm$ produce lasting oscillations of the young stellar component, 
model D3 with the compact cored dark matter profile with $15.5\times10^6\sm$ and D4 with the cuspy dark matter profile with $8\times10^6\sm$, result in oscillations that are quickly dampened. 
The simulation model M11 (SZM4) with model D3 has oscillations of the younger stellar component that last roughly $50\Myr$, which is shorter than the minimum age estimate of the younger stellar population of Leo~T.}
Similarly, models M12 (A1M7), M13 (S7M3) and M14 (S7M4) that use the NFW dwarf model D4 show even shorter mixing timescales of $<100\Myr$, as shown for M12 in Fig.~\ref{fig:fig_distance}.
We can extend the oscillation timescale by 
\chII{setting the inital offset (c) of the younger stellar component in a circular orbit, or making the younger stellar component more compact, setting the half-light radius to 20\pc, extending the oscillation to $200\Myr$ after which the younger stellar component mixes with the older component changing into a more extended distribution (see 
S7M3 and S7M4 Fig.~\ref{fig:fig_distance_velocity}).}
However, as mentioned in Sect. \ref{sec:res:int:mo}, the gas offset reaches only 50\pc, because the stellar winds are marginally able to overcome the central gravitational force that confines the gas.
\ch{Stronger winds would better reproduce the gas shift; however, the wind parameters would exceed the estimates for AGB stars.}

We also notice in Fig.~\ref{fig:fig_distance} the large gas offset at $900\Myr$ for models M9 and M12. 
This is produced when the younger stellar component settles in the centre of the dwarf, and
the stellar winds push the central gas, lowering the central density generating a cavity and a sporadic semi-circle shaped gaseous distribution in projection that is shifted from the dwarf centre. Moreover, usually the semi-circle's outer edge is more prominent at the leading side of the dwarf due to the ram pressure compression with the IGM. 
After inspecting its gas distribution, we discard these solutions as its morphology differs from the flattened \HI isocontours observed in Leo~T.

\chII{Based on the comparisons between different dwarf models and their dark matter profiles, we suggest two main factors that could contribute to the perdurance of the stellar offset observed in Leo~T:}
\chII{i) for a given mass at a fixed radius (e.g. $M_{300}$), the region within $<100\pc$ will be less massive in cored models than in cuspy (NFW) models (see Fig.~\ref{fig:fig_massprof}), which naturally extends the orbital timescales,
and ii) the orbits confined within the cored region} in cored potentials have near closed orbits given that the potential resembles a harmonic potential (constant density), which can extend the mixing timescale of the new material.
We illustrate this in Fig.~\ref{fig:fig_orbits} with the test particle orbits in the potentials of the dwarf model D2, which have a Burkert model, and the dwarf model D4 with the NFW model, which are calculated semi-analytically with {\sc delorean} \citepalias{Blana2020}. 
The potentials of the gas and the old stellar distributions are also included.
We placed test particles at the projected position of the younger stellar component and the gaseous peak to visualise the precession of their orbits. 
The Rosetta orbits in the NFW models are clearly revealed with rapid orbital precessions. 
On the other hand, the Burkert models show slower precessions, which naturally increase when we place the test particles at $300\pc$, since the Burkert profile behaves as an NFW for distances close to or larger than the core radius, but still with a precession that is much slower than with the NFW profile. 
\chII{We also calculated the orbits for the dwarf model D1. Given that this potential is just a re-scaling of model D2, it results in the same orbital family as D2, but with much longer orbital periods.}

%%% ====================================================================
%%% =====================================================================
\begin{figure*}[ht!]
\begin{center}
\includegraphics[width=8.2cm]{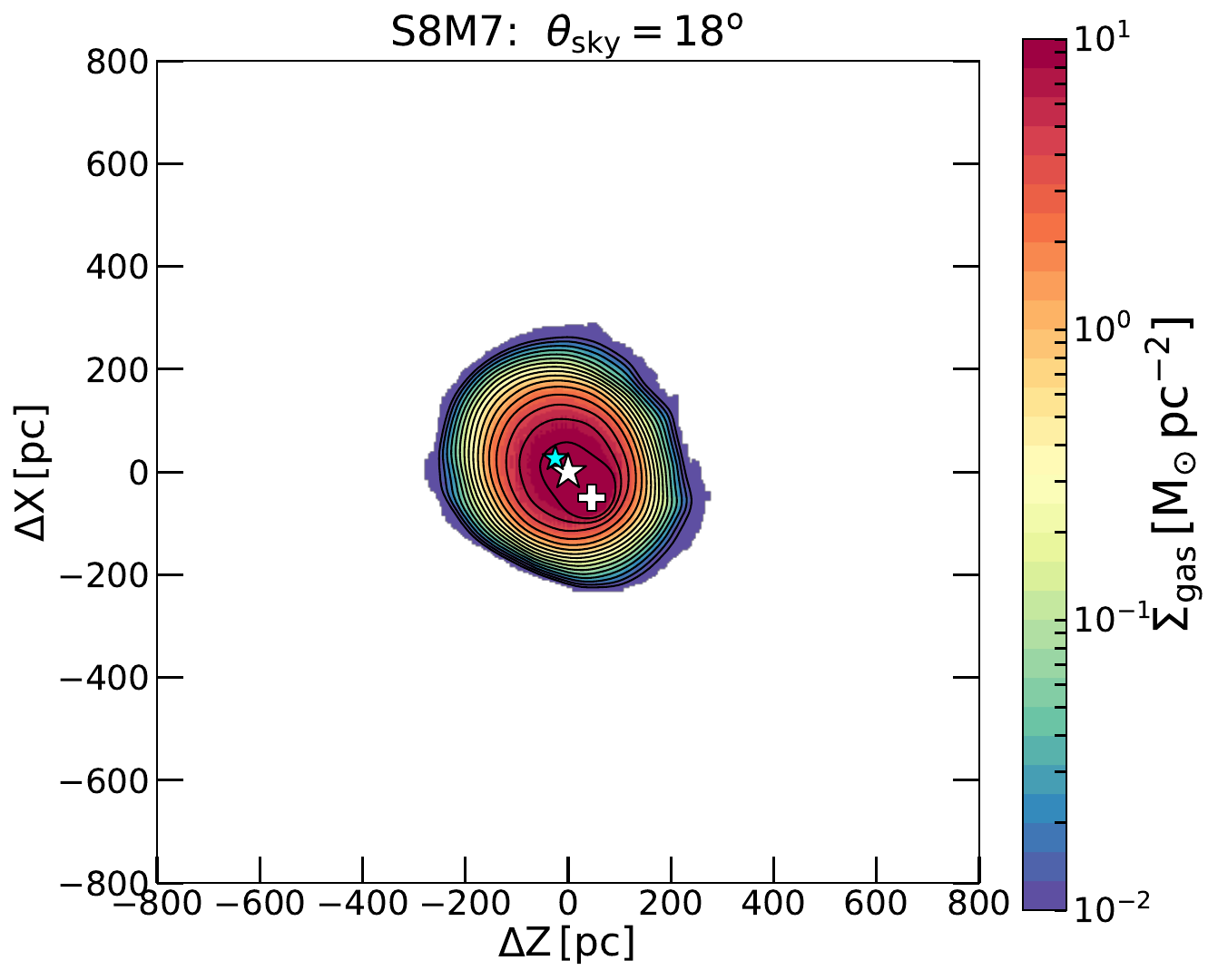}
\includegraphics[width=8.2cm]{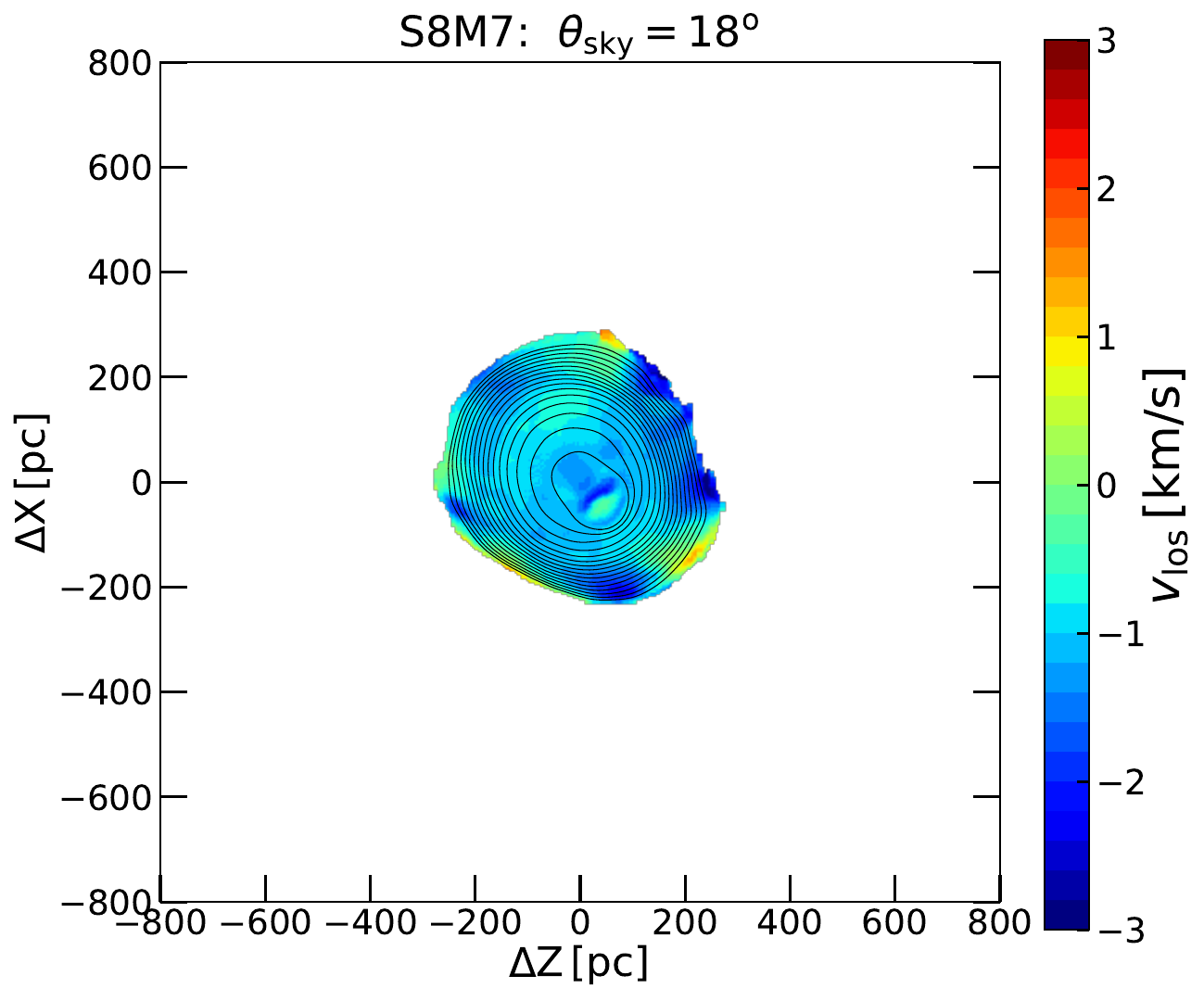}\\
% \noindent\rule{\textwidth}{0.5pt}
% %\vspace{-1.5cm}
\includegraphics[width=8.2cm]{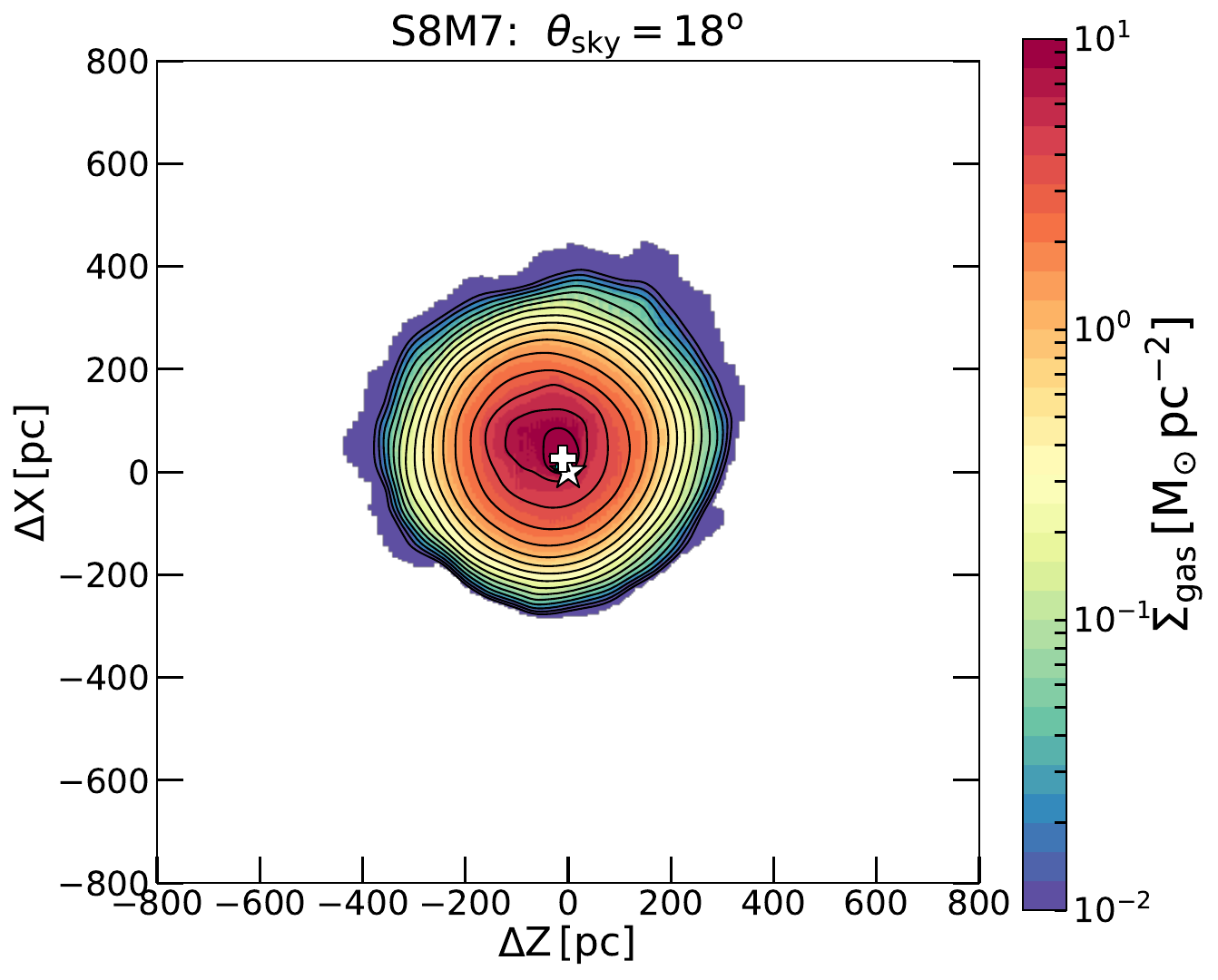}
\includegraphics[width=8.2cm]{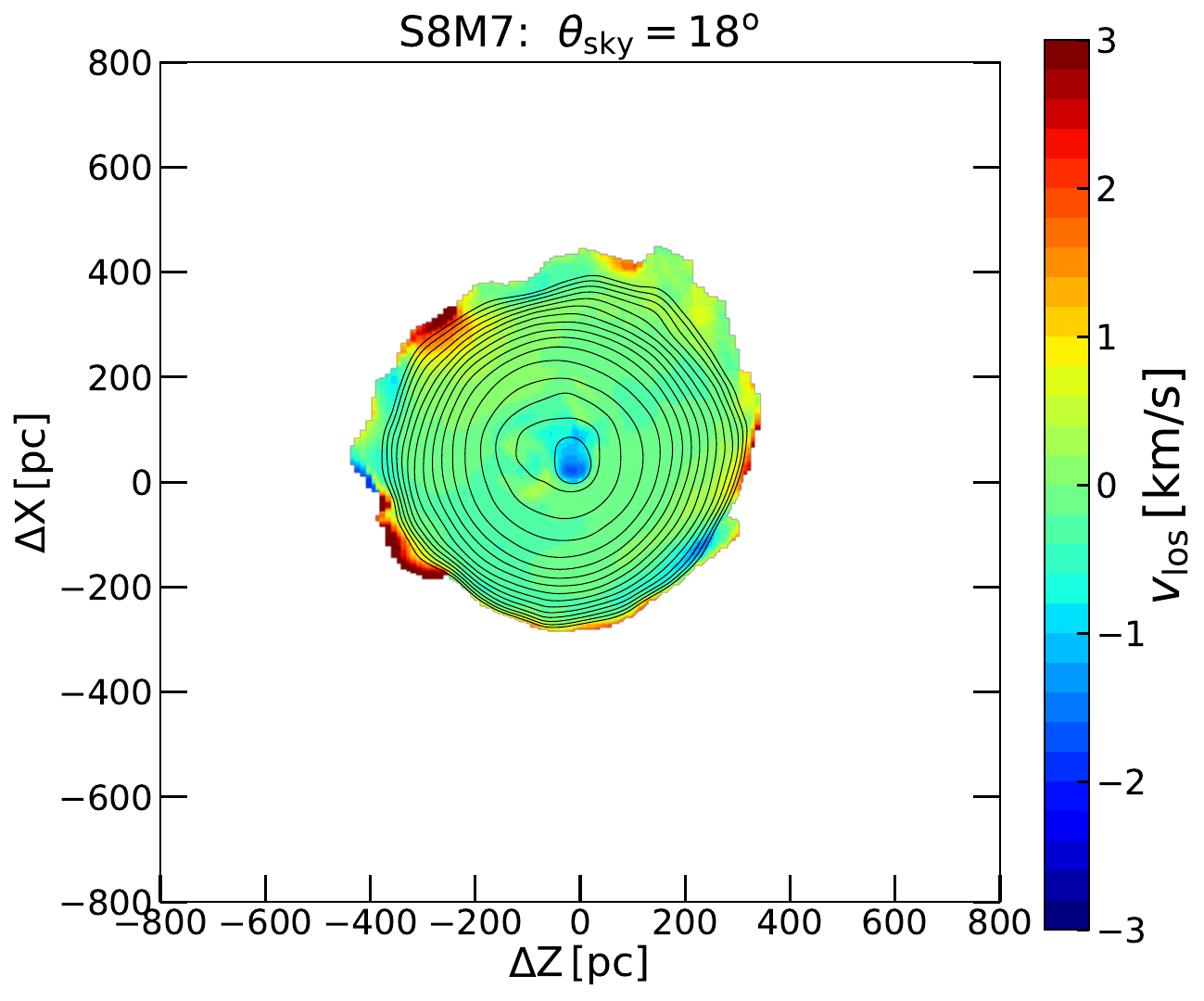}\\
% \noindent\rule{\textwidth}{0.5pt}
\includegraphics[width=8.2cm]{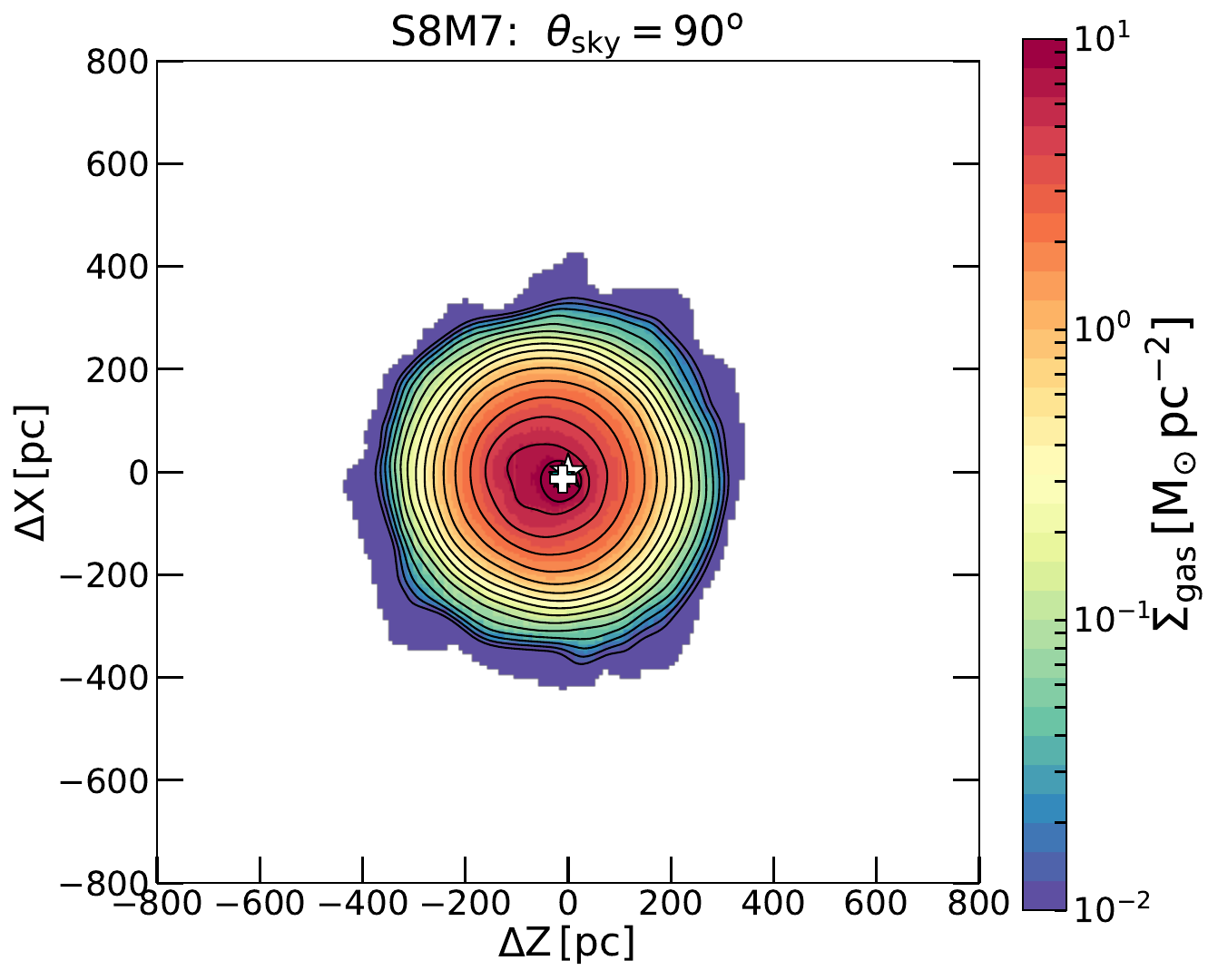}
\includegraphics[width=8.2cm]{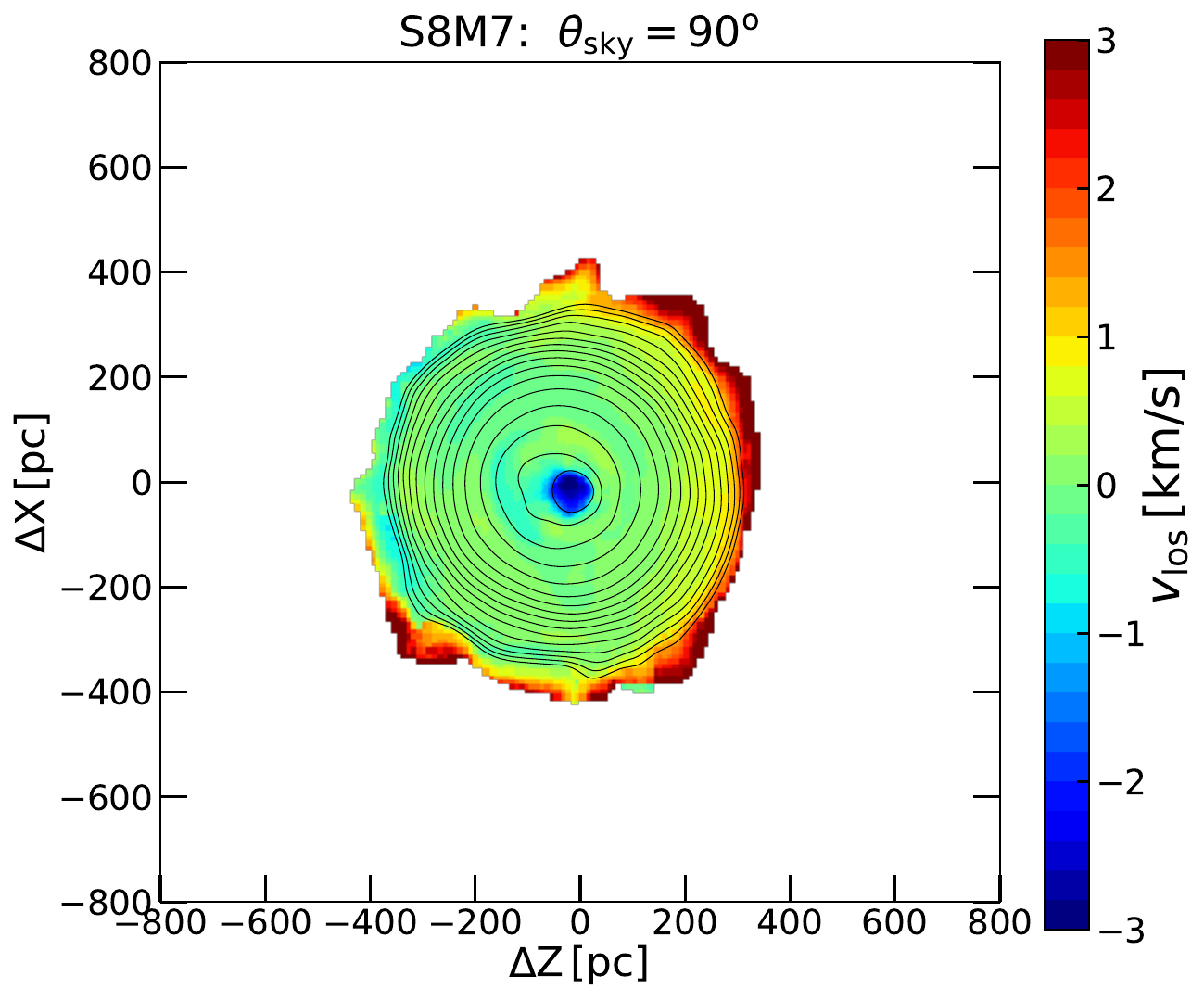}
%\vspace{-0.3cm}
\caption{\chII{
Scenario III: Snapshots of the gas surface mass density map of the model M19 (S8M7) taken at two different times; at 173\Myr (top row) and at 277\Myr (middle and bottom row) after the initial gas offset. 
This model uses the dwarf model D2t, similar to model D2, but with two gas components.
This uses a first infall orbital solution with a final tangential velocity of $u_{\rm t}\!=\!200\kms$,
corresponding to a projection angle of $\theta_{\rm sky}=18\dg$ (top and middle rows). 
We also include a projection of 90\dg to analyse the kinematic features of this solution (bottom row).
The combined centre of mass of the stellar and dark matter components is marked with a white star, 
while the gas surface density peak is marked with a white cross. The $v_{\rm los}$ maps are in the rest frame of the dwarf's dark matter centre of mass. We note  that $\Sigma_{\rm gas}\!=\!1\sm\pc^{-2}$ corresponds here to $N_{\HI}\!=\!1.2\!\times\!10^{20}\icmsq$.\coB{COMMENT: Is it fine to keep this figure with this size?}}}
\label{fig:fig_maptwogas}
\end{center}
\end{figure*}

\subsection{Scenario III: Initial gas sloshing}
\label{sec:res:int:gas}
% NOTE: add figure gas 1 component map: no morphological match, too round isocontorus!!
\chII{We explored a third scenario to see whether the whole gas distribution could be sloshing in the overall potential, caused by a displacement formed by stellar feedback from a previous star formation episode as, for example, it has been suggested for the more massive dwarf Phoenix~I to explain a $500\pc$ offset between the \HI and stellar components \citep[see][]{Young2007}.
We need to know how plausible this is, at least in energetic terms. 
For example, a supernova (SNe) (Type Ia) and core-collapse supernova (Type II) can output radiation and kinetic energies in the range of $E_{\rm SNe}\!=\!10^{49 - 53}\,{\rm erg}\!=\!5\!\times\!10^{5-9}\sm \km^2\s^{-2}$ \citep[][and references therein]{Woosley1986,Smartt2009,Janka2012}. 
Given the measured gas mass of Leo~T ($5.5\!\times\!10^{5}\sm$) and considering an initial bulk motion velocity with a value of the $7.5\kms$ dispersion, results in a motion kinetic energy of $E_{\HI}=1.5\!\times\!10^{7}\sm \km^2\s^{-2}$, to move the whole distribution of \HI to that velocity. This value falls within the energy output of a SNe, making it at least feasible. However, the complexity of supernova explosions and its effects on the ISM make accurate estimations difficult.}\\

\chII{Nonetheless, what we can study here is: given an initial displacement for the whole gas distribution, 
how would be the dynamics of the infalling gas, the timescales of the instabilities and relaxation of gas while it settles in the centre of the dwarf's potential.}
Therefore, in the set-ups of \chII{models M15 to M21}, shown in Table \ref{tab:setup},
we shift at the beginning of the simulations the gas distribution of the dwarf models from the dark-matter centre of mass by $140\pc$ at ${\rm PA}=45\dg$ and measure the resulting oscillations and time decay into the dwarf's centre. 
\chII{In these set-ups we do not include stellar winds to explore only the hydrodynamical response.}
Furthermore, given that \citetalias{Adams2018} infers the possible presence of a central CNM component in Leo~T, we also set-up models with both gas components (\eg D1t, D2t, D3t, and D4t). 
In Fig.~\ref{fig:fig_maptwogas} we present the resulting gas morphology of model M19 (S8M7), \ch{with the dwarf model D2t, showing two snapshots; at 173 and 277\Myr, and at different orientations.} We choose to show this model considering that: i) D2t is more massive than D1t and reacts more strongly to the gas sloshing, and ii) the CNM gas component generates a transient elongated structure within the less dense WNM component as they slosh at different periods within the dwarf potential. However, models with only one gas component always maintain a   simpler gas morphology with rounder iso-contours, unlike Leo~T.\\

We also show the oscillations of the gaseous and stellar components for models M15 to M20 in Fig.~\ref{fig:fig_distance_shiftgas}.
 We find that it can generate an offset for a long period while the gas oscillates around the centre with timescale as long as $\sim800\Myr$ for simulations that use the dwarf model D1, \chII{while simulations that use the dwarf models D4 and \chII{D4t} with NFW profiles (M17,M20,M21) the oscillations stopped before 200\Myr.}
An interesting effect is that, given the relatively high gas mass content of $5.5\times10^5\sm$, a factor of 2.8 larger than the stellar mass ($1.4\times10^5\sm$), the stellar distribution reacts to the gas distribution as it is gravitationally attracted by the gas, starting to oscillate while both orbit in the dark-matter-dominated potential. 
For the cored models, the effect is stronger, where model M15 with the dwarf model D1 shows its stellar components oscillating and reaching a maximum distance of 90\pc at 450\Myr and decreasing afterwards.
An important implication here is that if such a shift for the whole gas distribution had occurred in the past, then both stellar components would be equally perturbed and out of equilibrium. Moreover, we find that the massive dark-matter models are only slightly affected.\\

In general, we find that dwarf models with cored dark-matter profiles can sustain oscillations for long periods and reproduce the offset between the gas and the stellar distribution, while these are quickly suppressed in the NFW models.
However, as the gas oscillates through the centre of the dwarf, the morphology of the gas maintains near-circular isocontours, contrary to the flatter shape observed in the centre of Leo~T ($<200\pc$). 
Only the cored models with two gas components (D1t and D2t) present a flatter isocontorus, as the dense cold gas sloshes in the centre. 
\ch{While at first the WNM and the CNM oscillate together, the slow mixing of the cored models and the different periodicity that develops for the oscillation of each gas component result in the CNM component having a \chII{strong kinematic motion  that can be decoupled from the WNM component even by $\Delta v_{\rm los}\sim 3\kms$} \chA{for perturbations along the LOS ($\theta_{\rm sky}\!=\!90\dg$)}, as shown in Fig.~\ref{fig:fig_maptwogas}, and as it is also observed in Leo~T. 
\chII{However, the predominantly rounder morphology suggests that additional mechanisms, such as the stellar winds explored in Scenario II in Sect.~\ref{sec:res:int}, could better reproduce to flatter morphology and more complex hydrodynamics in the observations.}}

%%% ==============================================================================================
%%% ================================================================================================
%%% ===============================================================================================
%%% ================================================================================================
% \newpage
\section{Summary and conclusions} 
\label{sec:sum}
\chII{We developed models of the distant ($409\kpc$) Milky Way (MW) gas-rich dwarf galaxy satellite Leo~T 
to study its observed properties and current dynamical state, such as the three offsets between its old and younger stellar components \citep{DeJong2008,Vaz2023}, and the \HI gas component \citep[][\citetalias{Adams2018}]{Adams2018}. 
For this, we set up hydrodynamic N-body simulations\footnote{Videos of these simulations can be found at this link \href{https://www.youtube.com/playlist?list=PLoegIUZ3yJ9FuDldEF-USJWd-o_hyLavl}{{(here)}}} 
for Leo~T conducted in a wind tunnel configuration, 
modelling the last 2\Gyr and 1\Gyr\LEt{***yes?}\coB{Ok} of orbital history until the present in a MW IGM environment.
We set up three main scenarios where we explored dwarf galaxy models with a range of properties based on estimates from the literature, 
such as different dark matter masses and profiles (cusped, cored), gas components (WNM without CNM), first infall (FI), and backsplash (BA) orbital solutions with different GSR tangential velocities ($u_{\rm t}$).
These models are listed in Table \ref{tab:setup}.
The summary and conclusions are as follows:}

\begin{enumerate}
\item \chII{In Scenario~I we used dwarf galaxy models to explore the environmental ram pressure perturbations produced by the MW intergalactic medium (IGM). 
As expected, dwarf ISM gas is compressed on the leading side, thus producing round gas isophotes
for slow BA orbits ($u_{\rm t}\!\!\!<\!\!\!100\kms$), while fast FI orbits ($>\!200\kms$) generate a bullet-shaped gas morphology,
with an extended trailing region.
The well-constrained LOS velocity in Leo~T implies that fast orbits must have high tangential (sky) velocities,\LEt{***Generally, for rates, ratios, and masses use high(er) and low(er); for sizes and
magnitudes use large(r) and small(er). Please check throughout }\coB{Ok.} which as a consequence produce a weak gas LOS velocity gradient.
We also conclude that if the faint material observed to the north of Leo~T is trailing stripped material, then fast FI orbits are required to generate significant amounts of trailing cold gas, finding $u_{\rm t}\!>\!200\kms$ for the dwarf model D1 ($M_{300}\!=\!1.7\!\times\!10^6\sm$),
and $>\!300\kms$ for D2 ($8\!\times\!10^6\sm$).
Moreover, slower (BA--FI)\LEt{***ratio here? Slashes ( / ) are used in equations, and to denote ratios, instrument pairings, and
wavelength ranges (e.g., optical/UV). All other appearances should be removed
and the sentence rephrased. Please check throughout. You can substitute "and",
"or", "and/or", or a double hyphen (which can be used to indicate a range or dual
nature: Hertzsprung--Russell diagram). For more details, see Sect. 2.9 of the
language guide  }\coB{Ok.} orbits $\sim100\kms$ could be considered if the MW IGM has large overdensities 
of $\delta \rho_{\rm IGM}\!\gtrsim\!100\,\rho_{\rm IGM}\sim 10^{4}\sm\ikpccube\,(4\!\times\!10^{-4}\icmcube)$, 
which can momentarily increase the gas stripping and velocity gradients and generate a central offset between the stellar and gas distributions. 
However, this offset is smaller than the 80\pc offset observed in Leo~T \citepalias{Adams2018} and quickly settles in equilibrium as soon as the satellite leaves the perturbation ($\lesssim 30\Myr$).} 
\newline
\item 
\chII{In Scenario~II we explored states out of equilibrium for the younger stellar component motivated by the observed offset of 35\pc between the old and younger stellar components in Leo~T \citep{DeJong2008}. 
We find that models M8, M9, and M10 with extended dark matter cores (dwarf models D1, D1t, and D2) can better reproduce Leo~T's observed gaseous and stellar offsets simultaneously, where the younger stellar component oscillates around the dwarf centre for periods as long as the age estimates of the younger stellar population in Leo~T $\lesssim\!200\!-\!1000\Myr$ \citep{DeJong2008,Weisz2012,Vaz2023}. 
This is compatible with a scenario where the younger lower mass stellar component ($\sim\! 10^4\sm$) could have been formed in a gas cloud displaced from the centre of the dwarf, accreting and mixing the star cluster with the old massive stellar component later \citep[$\sim1.3\!\times\!10^5\sm$,][]{Irwin2007,DeJong2008,Weisz2012}.
Dwarf galaxy evolution studies indicate that cored halos can sustain primordial complex stellar substructures for longer timescales than cuspy halos \citep{Kleyna2003,Penarrubia2009,AlarconJara2018,Aravena2019,Lora2019}, like the MW dwarf Eridanus II with its star cluster separated by 23--45\pc in projection \citep{Crnojevic2016,Simon2021}.
\chA{Moreover, we find, as expected, that oscillations aligned with the sky plane rather than the LOS axis increase the visible offsets between the gas and younger stars, while decreasing the $v_{\rm los}$ signature, and vice versa. This suggests that Leo~T central perturbations are likely along an intermediate axis (this also applies for Scenario III models).}}
\newline
\item \chII{The best matching models (M8, M9, M10) also reproduce the flat morphology of the \HI isocontours observed in Leo~T ($R\!\!<\!\!200\pc$).
We included a mild stellar wind source ($\sim\!10\!-\!20\kms$) attributed to an observed AGB stellar population candidate \citep{Weisz2012}. As this wind source oscillates with the younger stellar component, it can push the gas off centre, compressing it and generating the flatter distribution periodically. Moreover, models with one or two gas components (CNM, WNM) manage to reproduce this flat morphology. 
\chA{Our beam size tests also revealed the possibility that the CNM core component in Leo~T could be in fact more elongated and reach higher surface densities, which could be probed in future observations.}}
\ch{We also find that the periodic stirring of the gas delays its settling until the younger stellar component fully phase mixes, sinking the wind source to centre where it cannot overcome the gravitational force and high densities. 
This provides a viable mechanism that could modulate the bursty (variable) star formation rate history estimated for some low-mass dwarf galaxies \citep{Grebel2000}.}
\chII{Furthermore, we find that stellar winds induce a turbulent interior similar to Leo~T 
\chA{(unlike models in Scenario III)}, 
generate mild outflows that perturb the external gas layers
facilitating gas stripping, 
and drive gas ejections misaligned with the orbital trajectory.}
\newline
\item \chII{We explored dwarf galaxy models with cored (Burkert) and cuspy (NFW) dark matter halos with a range of parameters based on observational estimates from the literature. We find that only cored halos with core sizes comparable to the stellar and gas distribution ($\sim\!200\!-\!400\pc$) allow oscillations of the younger stellar component to persist for $200\Myr$ or longer, which is consistent with the age of this component in Leo~T.}
On the other hand, cuspy and compact cored halos favour a quick mixing of the gaseous and stellar material in the centre due to faster timescales and the rapid orbital precession, while cored halos have near closed orbits that contribute to delaying the mixing of the material.
\newline
\item 
In Scenario~III,
motivated by \chII{larger stellar-gas offset configurations observed in other dwarf galaxies, such as Phoenix~I \citep{Young2007}, and by the last star formation event of Leo~T,
we explored the gas sloshing generated by a preceding global larger offset of the total gas with respect to the dark matter and stellar mass distributions.
This scenario also generated a periodic offset between the gas and the stellar components, which is enhanced if we include a denser cold gas component. 
In particular, models with two gas components (e.g. model M19, with dwarf model D2t) produced transient slightly flattened iso-contours of the CNM dense component sloshing and kinetically decoupled within the WNM for long timescales ($>200\Myr$) (models M18, M19), as the \HI observations in Leo~T suggest \citepalias{Adams2018}. 
Moreover, we also find that a global gas offset is able to perturb both the older and the younger stellar populations out of equilibrium as the system relaxes.
However, we find that the gas iso-contours of the models of this scenario are not flat enough to match the \HI morphology in the central $200\pc$ of Leo~T,
\chA{nor its turbulence,} and therefore, the models of Scenario~II still more closely  match the overall main properties of Leo~T.}
\end{enumerate}
\chII{Here we model the current dynamical state of Leo~T,
finding a rich dynamical coupling between the stellar, gaseous, dark matter substructures and the IGM environment. 
The models reveal that stellar populations recently formed in low-mass dwarfs require long mixing timescales to reach equilibrium, proving how helpful spectral and chemo-dynamical information is to disentangling tracers of different stellar populations used to measure dynamical properties \citep{Vaz2023}.
Moreover, these findings strongly motivate future observations to better understand the evolution of gas-rich and recently quenched dwarf galaxies, starting with the better-resolved dwarfs in the Local Group \citep{Spekkens2014,Rubio2015,Putman2021}.
While deeper observations of \HI, ionised gas (e.g. OIII) and X-ray data would improve the estimates of dwarf gas stripping and mixing with the IGM environment,
observations of the dwarf interstellar medium (ISM) from molecular gas, dust properties (IR, NIR) and gas chemical abundances will 
improve star formation and cooling models in low-mass dwarf galaxy simulations ($M_{\star}\!=\!10^2\!-\!10^6\sm$); this  will be addressed in a follow-up publication  (Bla\~na\,et\,al.\,in\,prep.).
}

\begin{acknowledgements}
The authors deeply thank Elizabeth Adams and Tom Oosterloo for making the observational data available, 
and they also thank Elizabeth for the additional details and comments regarding the observations. 
The authors sincerely thank the referee for the insightful comments and constructive suggestions, which have enhanced the clarity and quality of the manuscript.
MB also thanks Thomas Puzia and Daniel Vaz for their insightful comments and the delightful discussions.
This research was supported by the ORIGINS Excellence Cluster, which is funded by the Deutsche Forschungsgemeinschaft (DFG, German Research Foundation) under Germany's Excellence Strategy - EXC-2094-390783311.
The authors thank the Max Planck Computing and Data Facility (MPCDF) for providing excellent computing services.
MB thanks ANID for funding through the awarded postdoctoral fellowship ``Beca de Postdoctorado en el Extranjero Convocatoria 2018 folio 74190011 resoluci\'on exenta N$\dg$ 8772/2018", and
the Fellowship FONDECYT POSTDOCTORADO 2021 Nr 3210592, and deeply thank the people in Chile for their tax contributions to science that supported this project during the difficult times of the covid crisis.
MF acknowledges funding through FONDECYT regular N1180291 and Basal ACE 10002 FB210003.
%The research of DC has been supported by Horizon 2020 ERC Starting Grant `Cat-In-hAT' (grant agreement no. 803158). 
The research of DC has been supported by the Deutsche Forschungsgemeinschaft (DFG, German Research Foundation) under Germany’s Excellence Strategy - EXC 2121 - 'Quantum Universe' - 390833306
\end{acknowledgements}
% \clearpage
% ~
% \newline\newline
% ~
% \appendix
\bibliographystyle{aa}
%\bibliography{paper_references_acron_v2}
\bibliography{Blana_LeoT_v3.bib}
% \clearpage
% \clearpage
~
% \newline\newline
% ~
% \appendix
% \newpage
\onecolumn
\begin{appendix}

\section{Additional figures}
\label{sec:app}
% \begin{minipage}{\textwidth}
% \newpage
% \newpage
% \begin{appendix}
% \appendix

% \vspace{-20cm}
% \begin{minipage}{\textwidth}
\begin{figure*}[h]
\begin{center}
\includegraphics[width=0.32\textwidth]{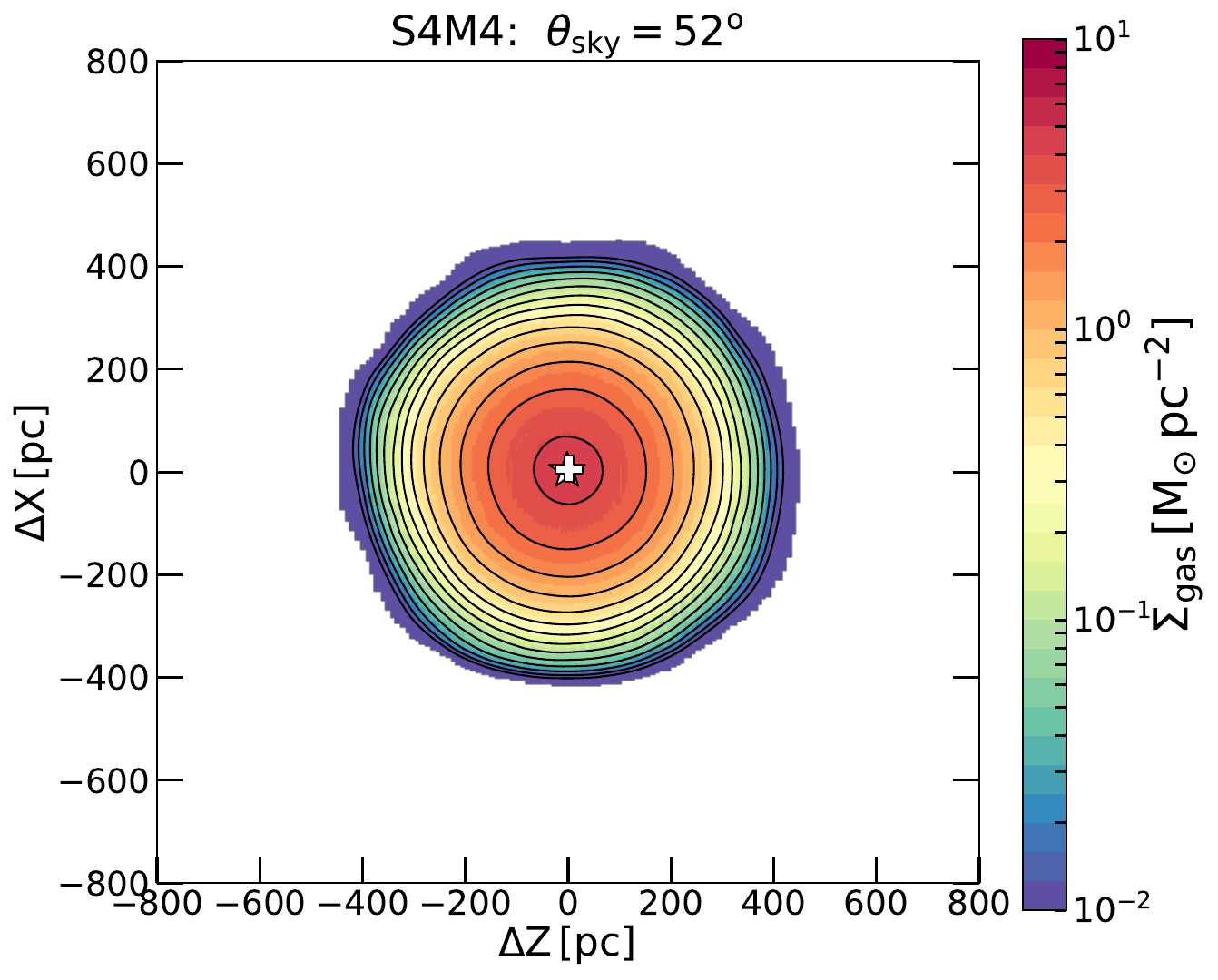}
\includegraphics[width=0.32\textwidth]{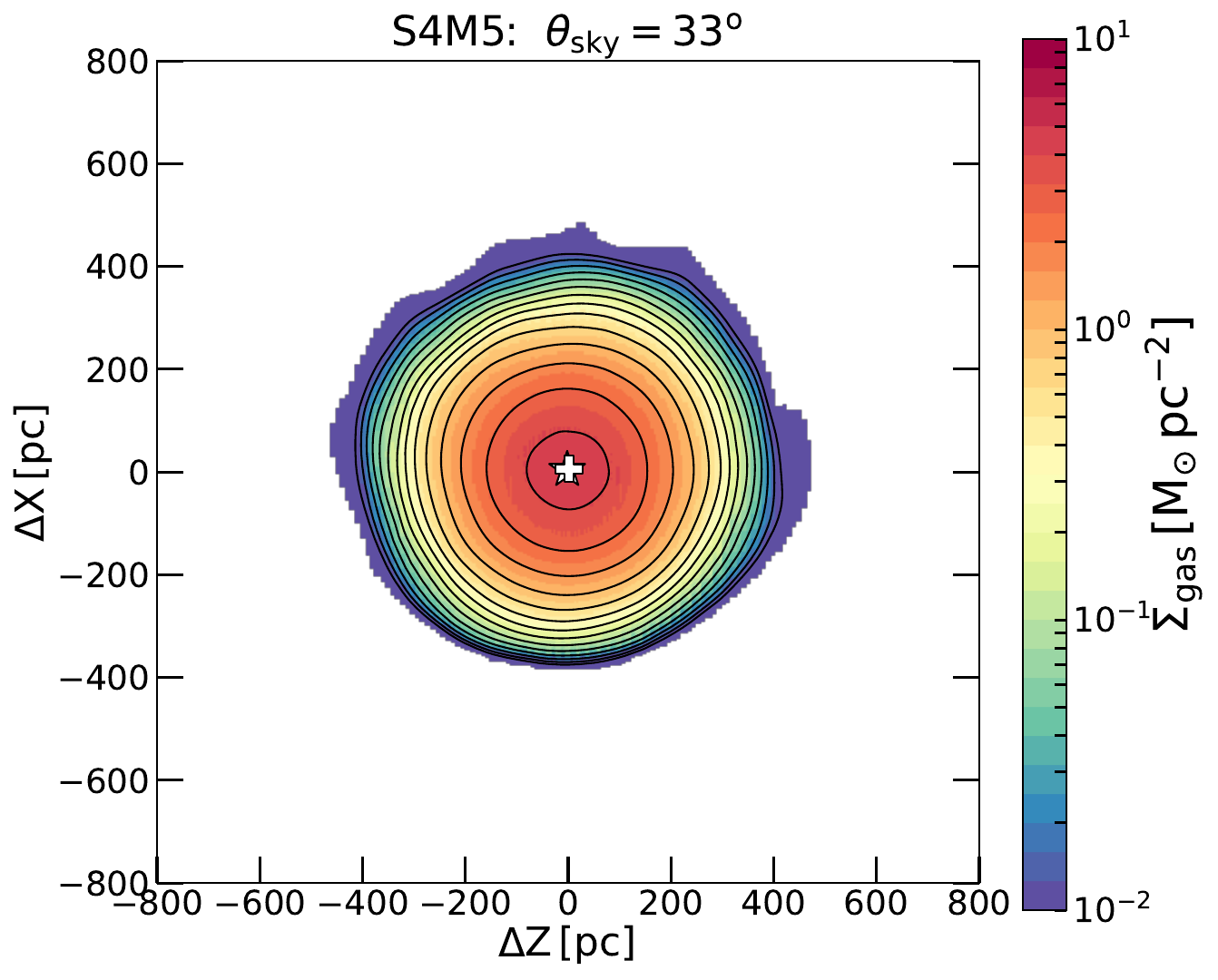}
\includegraphics[width=0.32\textwidth]{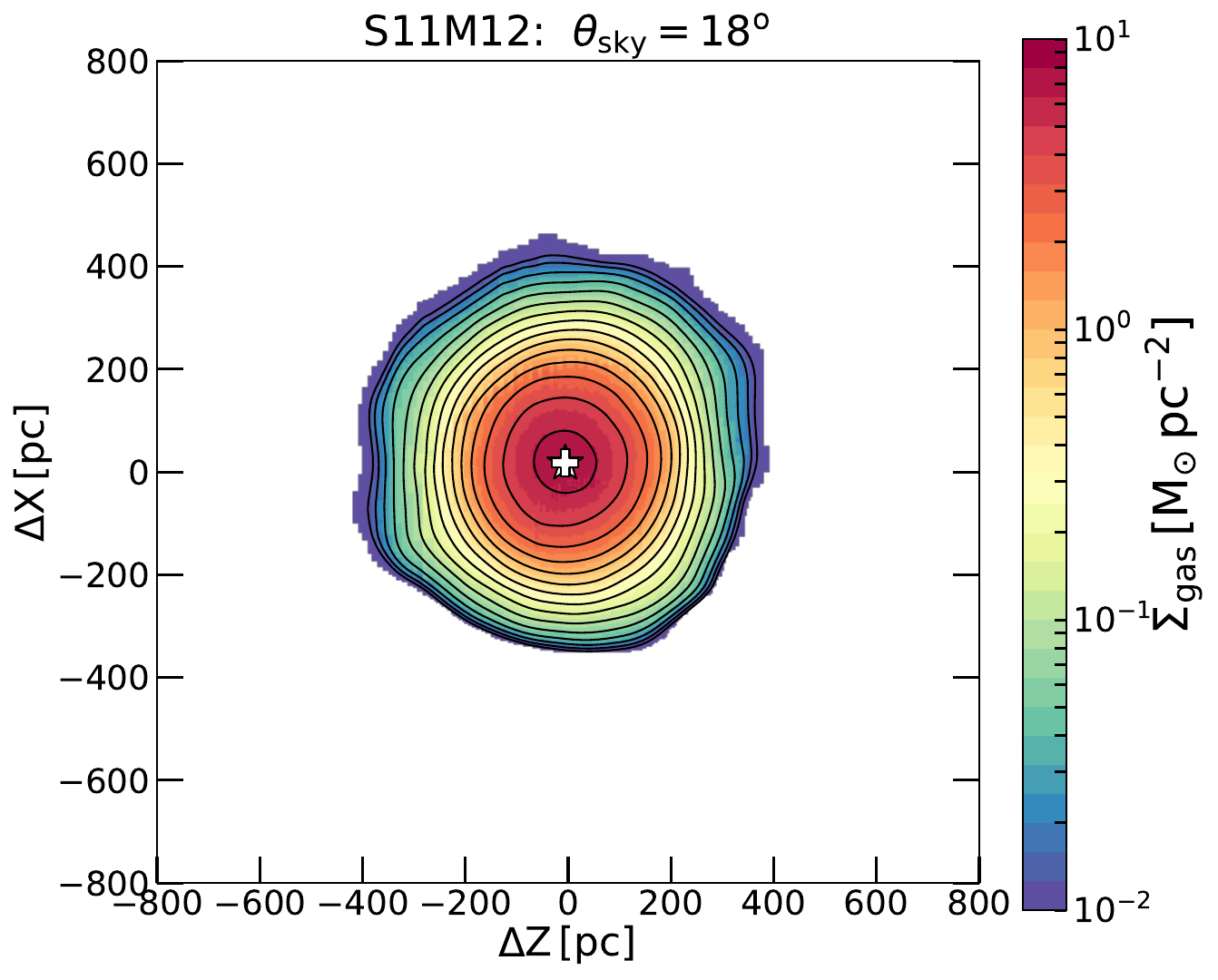}\\
\includegraphics[width=0.32\textwidth]{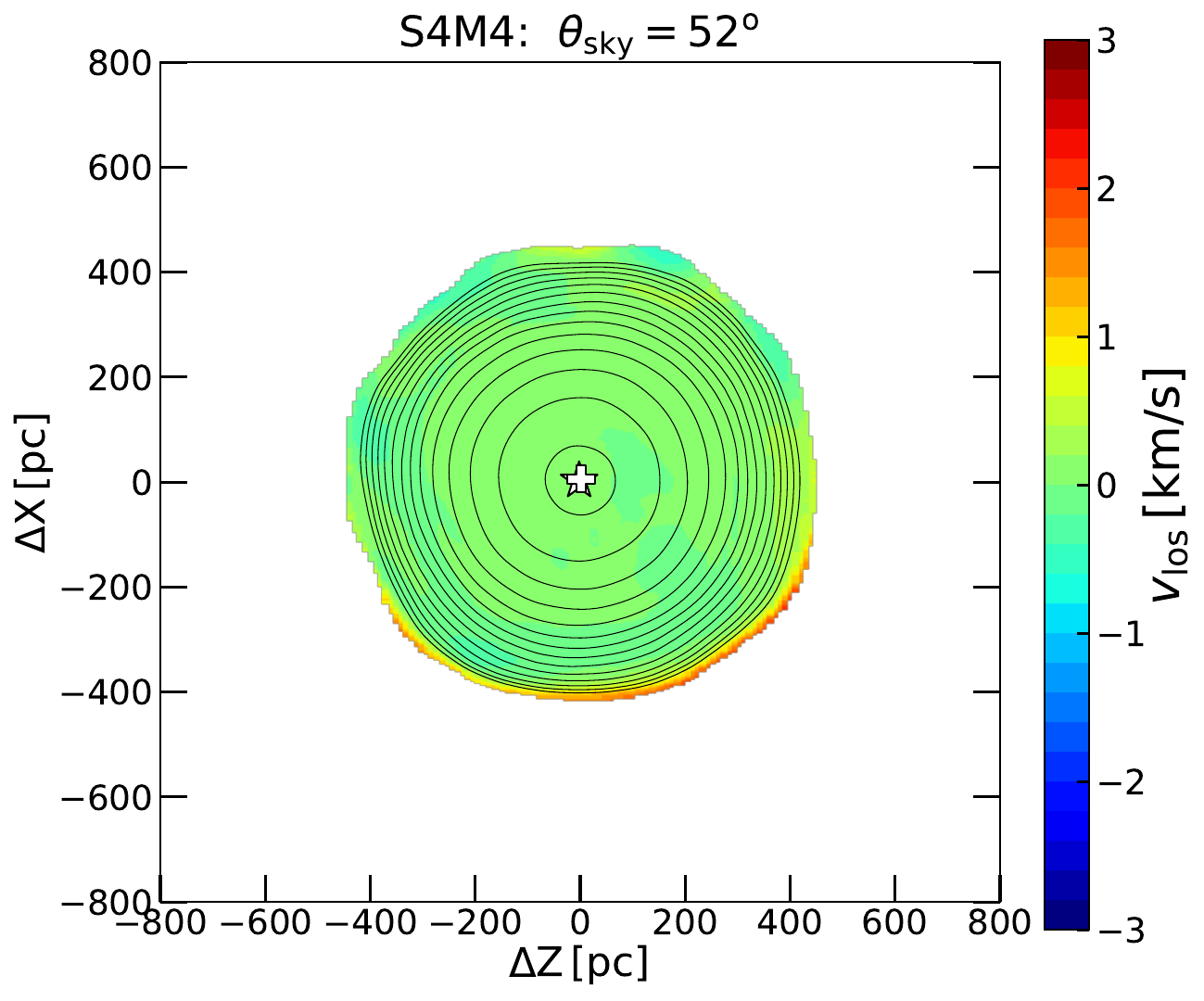}
\includegraphics[width=0.32\textwidth]{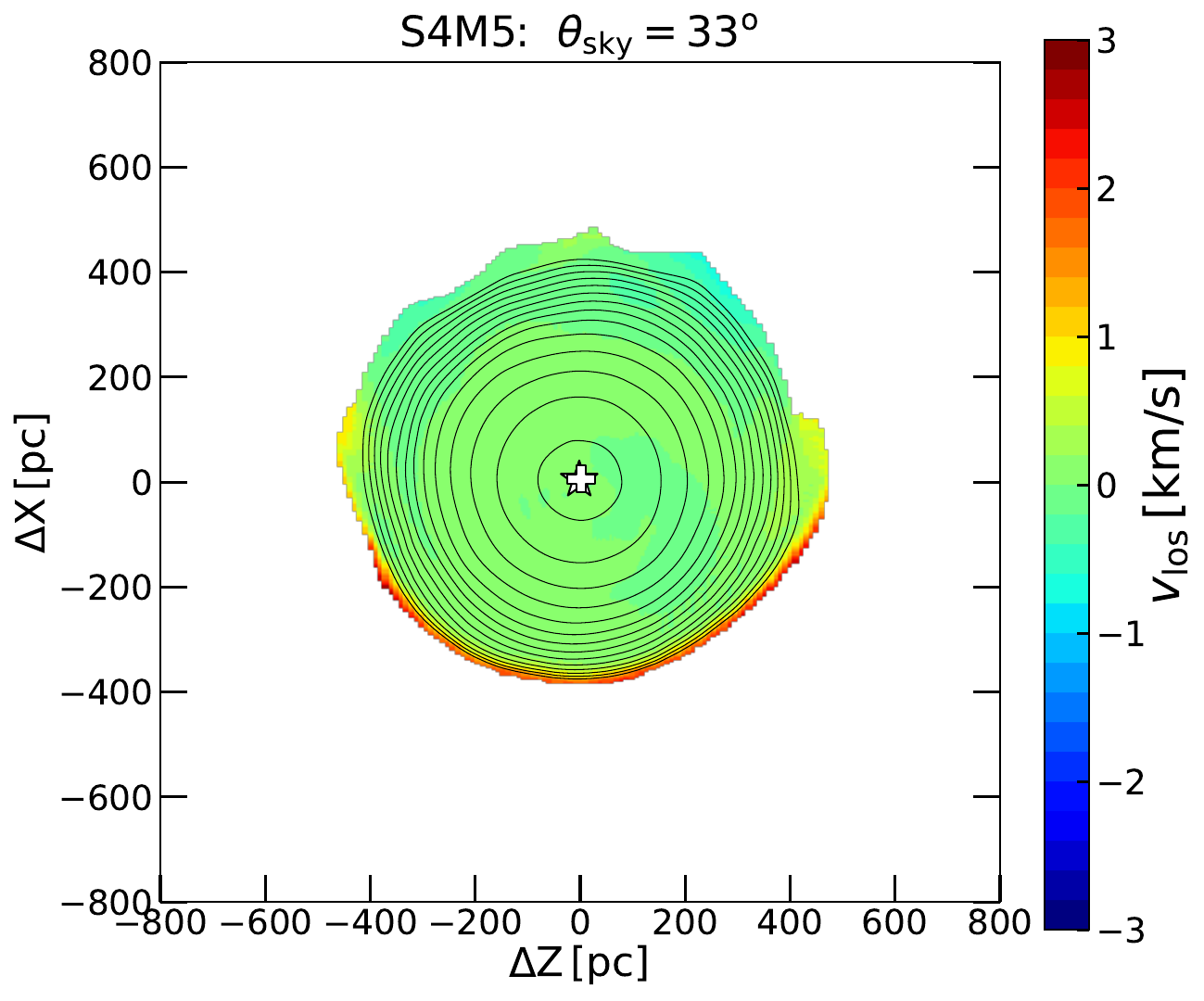}
\includegraphics[width=0.32\textwidth]{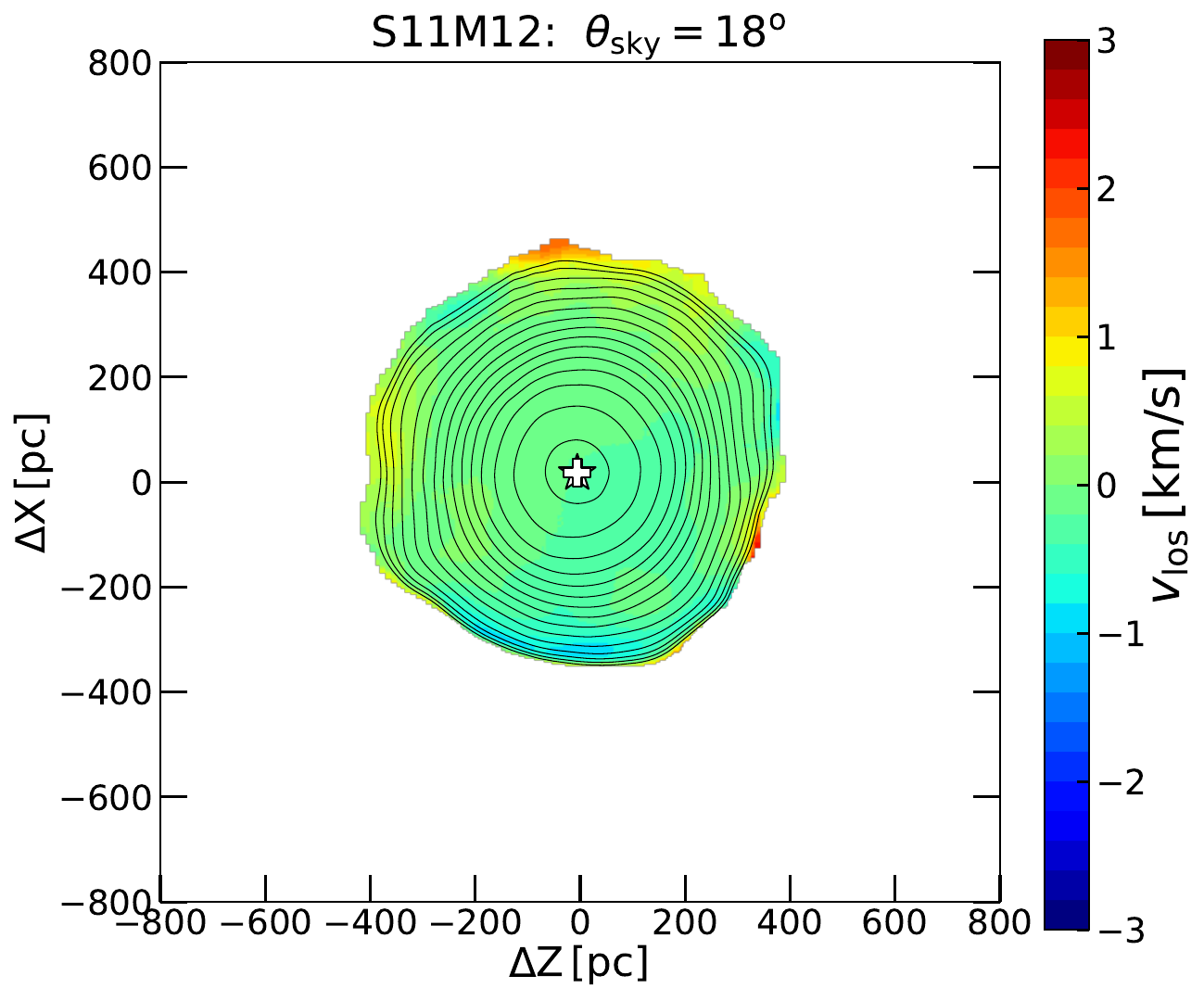}
\caption{Scenario I: Surface densities (top) and LOS velocity maps (bottom) of snapshots of models that use the dwarf model D1 with two different orbits, showing model M1 (S4M4) with a tangential velocity of $u_{\rm t}\!=\!50\kms$ that projects with $\theta\!=\!54\dg$ (left) and model M2 (S4M5) with $100\kms$ and $33\dg$ (middle). We also show model M6 (S11M12), which uses the more massive dwarf model D2 with a final velocity of $200\kms$ and an angle of $18\dg$. The $v_{\rm los}$ maps are in the rest frame of the dwarf's dark matter centre of mass.
We note  that $\Sigma_{\rm gas}\!=\!1\sm\pc^{-2}$ corresponds here to $N_{\HI}\!=\!1.2\!\times\!10^{20}\icmsq$.}
\label{fig:fig_environment_A}
\end{center}
\end{figure*}
% \end{minipage}
~~~~~~~~~~~~~~~~~~~
~~~~~~~~~~~~~~~~~
~~~~~~~~~~~~~~~~~~~
~~~~~~~~~~~~~~~~~~~
~~~~~~~~~~~~~~~~~~~
~~~~~~~~~~~~~~~~~~~

\begin{figure*}[htb]
\begin{center}
\includegraphics[width=5.8cm]{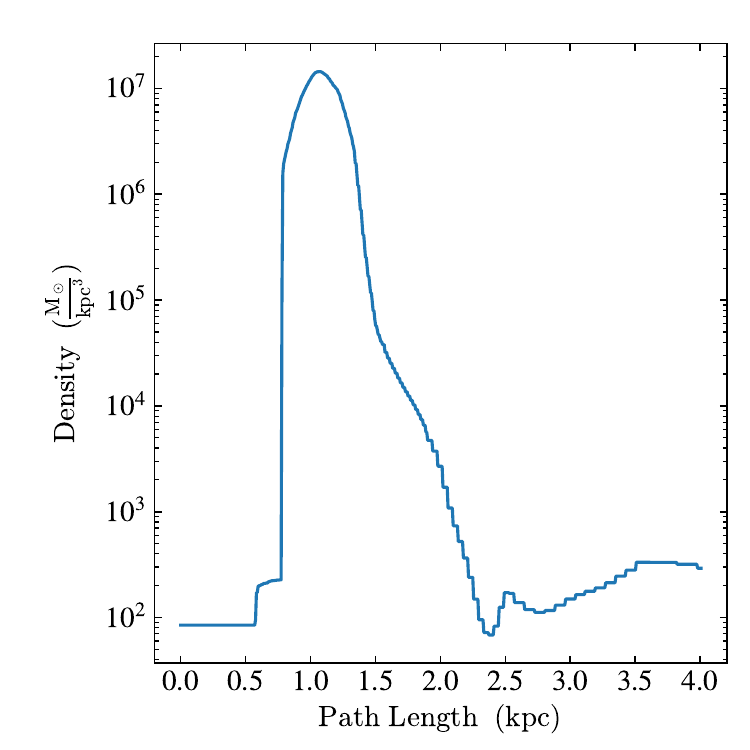}
\includegraphics[width=5.8cm]{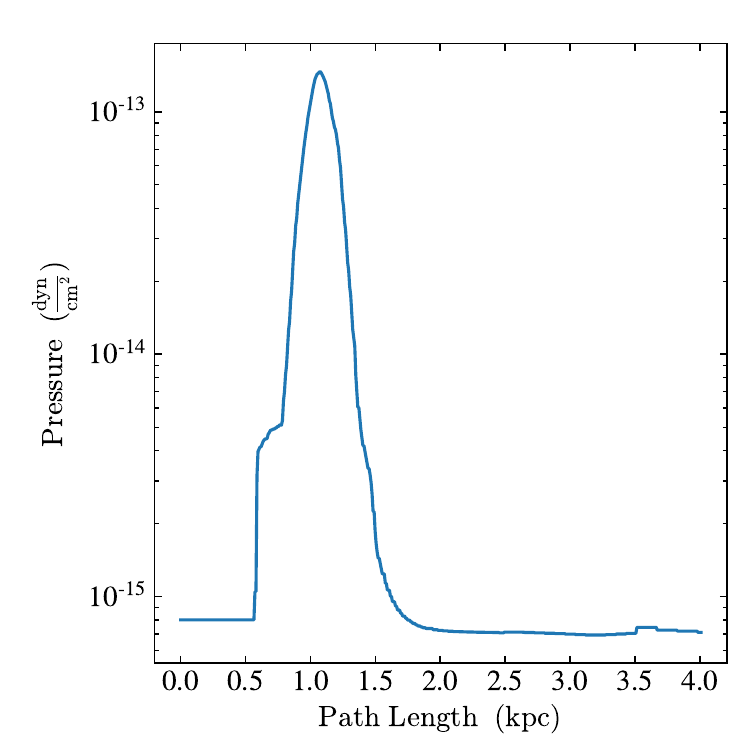}
\includegraphics[width=5.8cm]{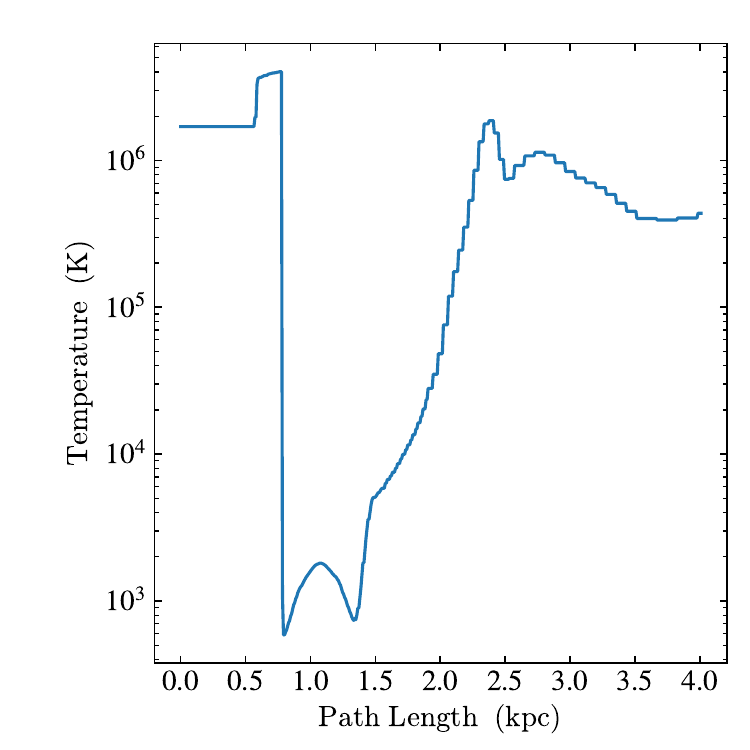}
%\vspace{-0.3cm}
\caption{\chII{Scenario I: Snapshot of model M4 (S4M7) evolved from -2\Gyr until the current position of Leo~T, where we show slit cuts along the Declination axis through the dwarf's centre for the gas density (left panel), pressure (middle) and temperature (right). The orbit corresponds to a first infall solution, reaching 300\kms tangential GSR velocity. The dwarf centre is located at 1.1\kpc, and the IGM wind is injected from the left side of the panels, following a beta density profile that depends on the distance between Leo~T and the Milky Way. 
We note that a gas density of $\rho_{\rm gas}\!=\!10^7\sm\kpc^{-3}$ here corresponds to $n_{\HI}\!=\!0.4\icmcube$.}}
\label{fig:fig_profcut}
\end{center}
\end{figure*}
~~~~~~~~~~~~~~~~~~~
~~~~~~~~~~~~~~~~~
~~~~~~~~~~~~~~~~~~~
~~~~~~~~~~~~~~~~~~~
~~~~~~~~~~~~~~~~~~~
~~~~~~~~~~~~~~~~~~~

\begin{figure*}[htb]
\begin{center}
\includegraphics[width={0.32\textwidth}]{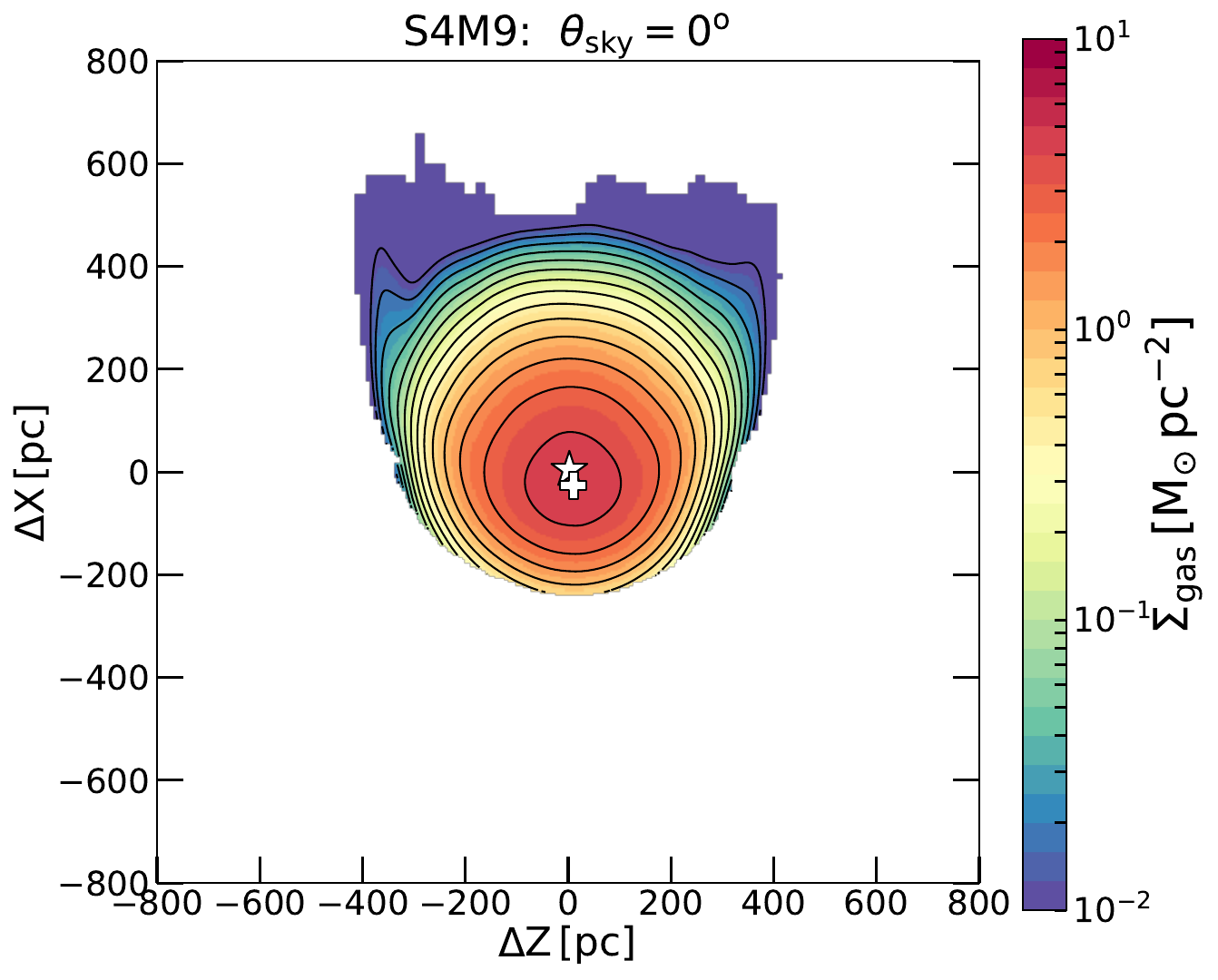}
\includegraphics[width={0.32\textwidth}]{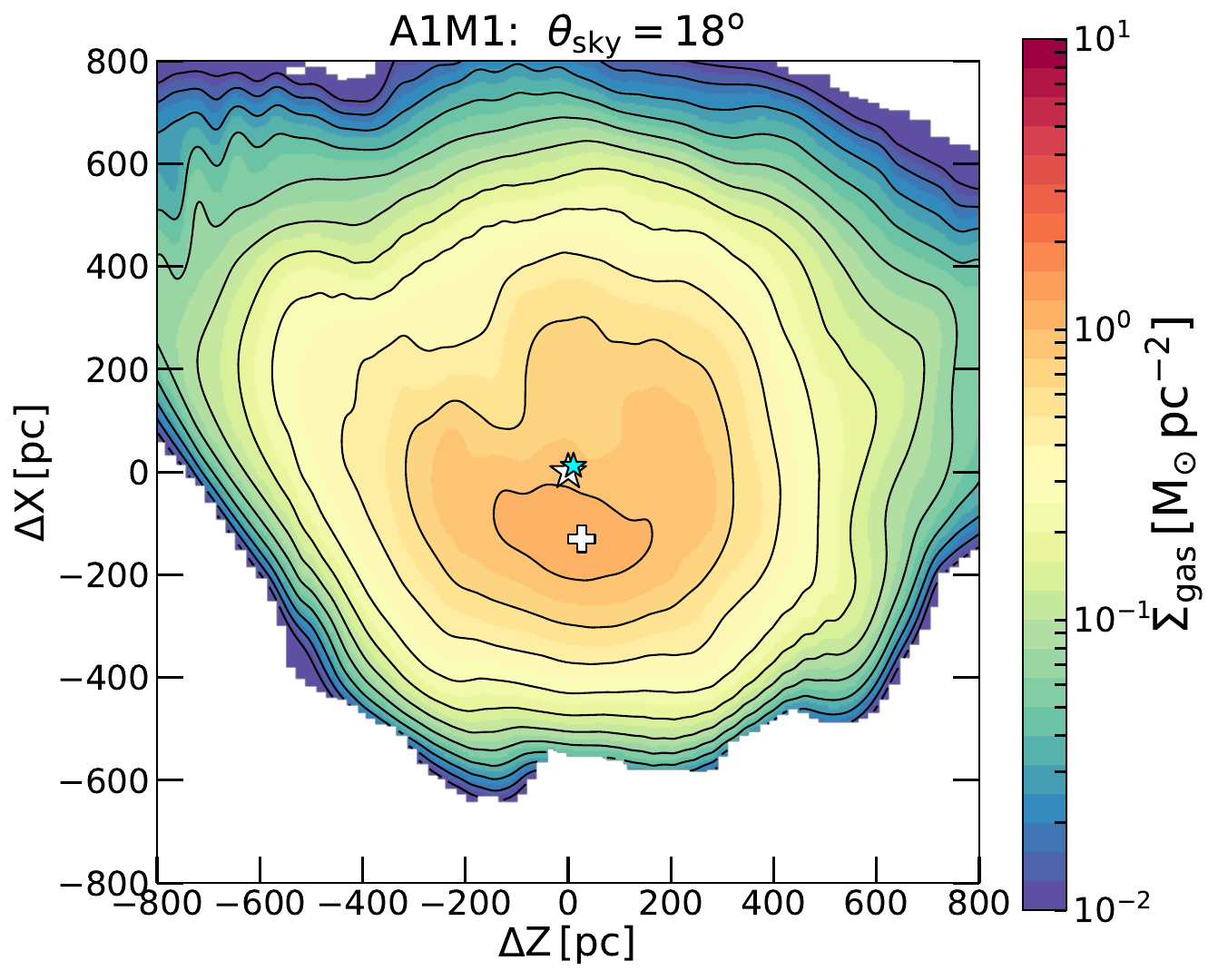}
\includegraphics[width={0.32\textwidth}]{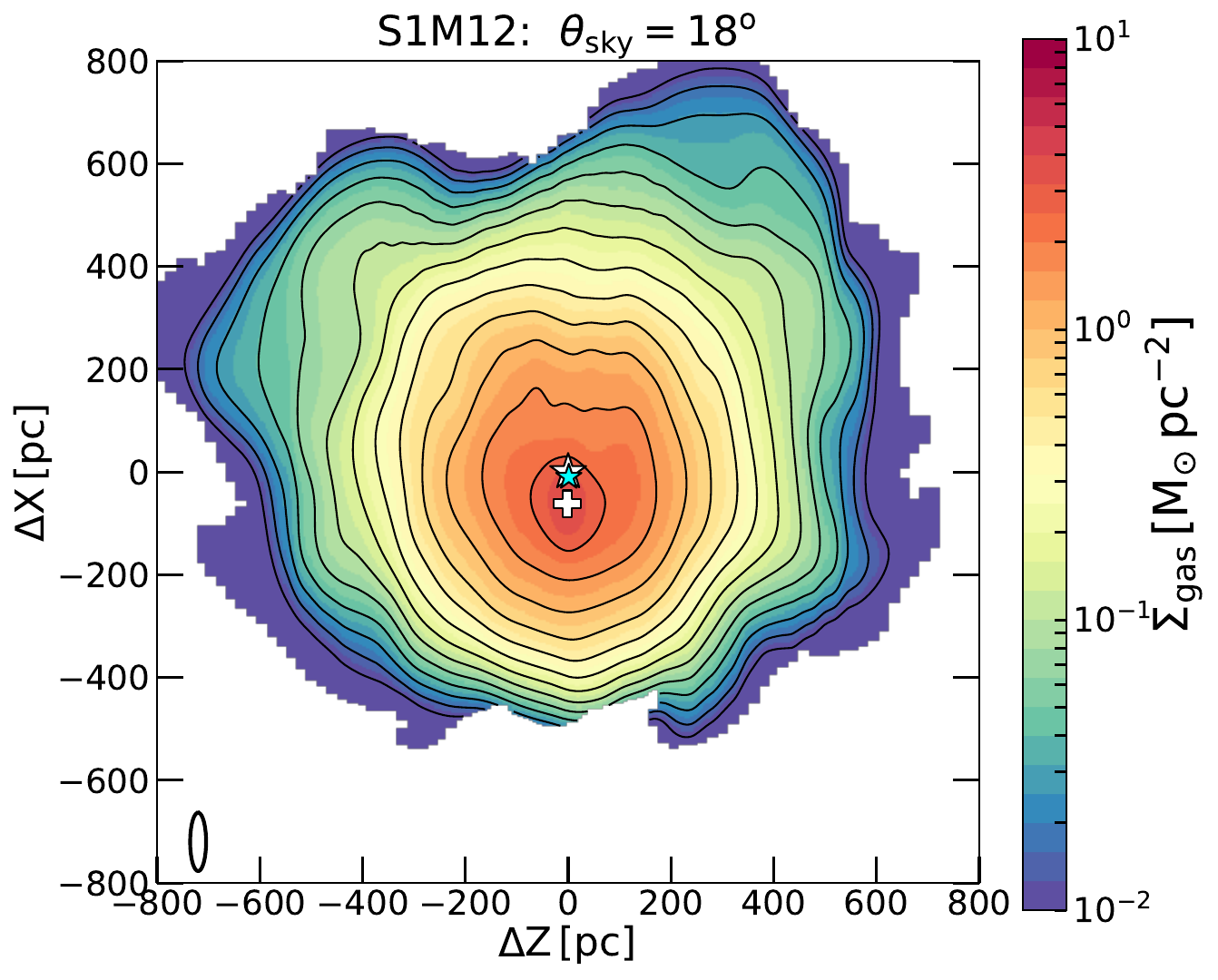}
\includegraphics[width={0.32\textwidth}]{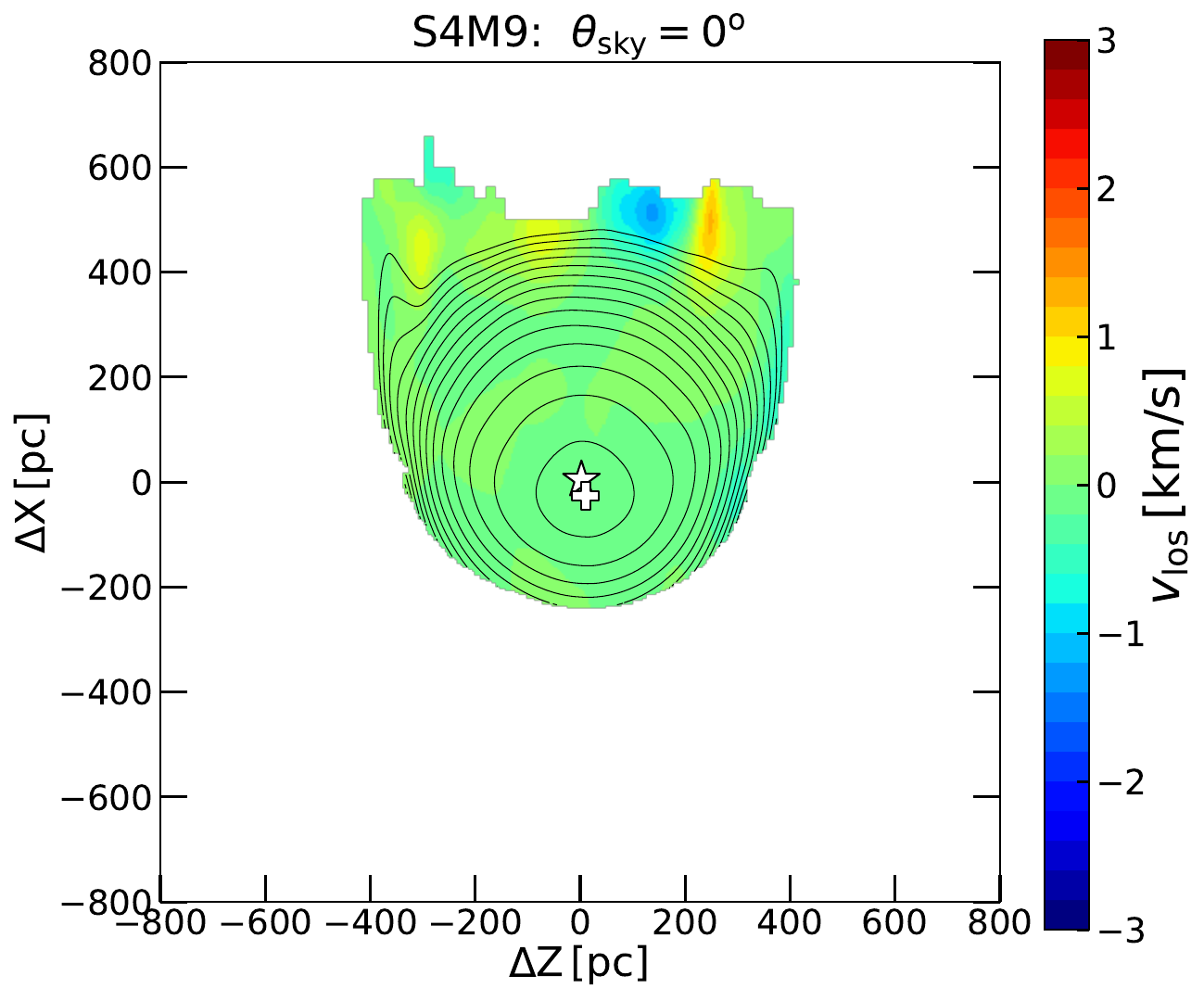}
\includegraphics[width={0.32\textwidth}]{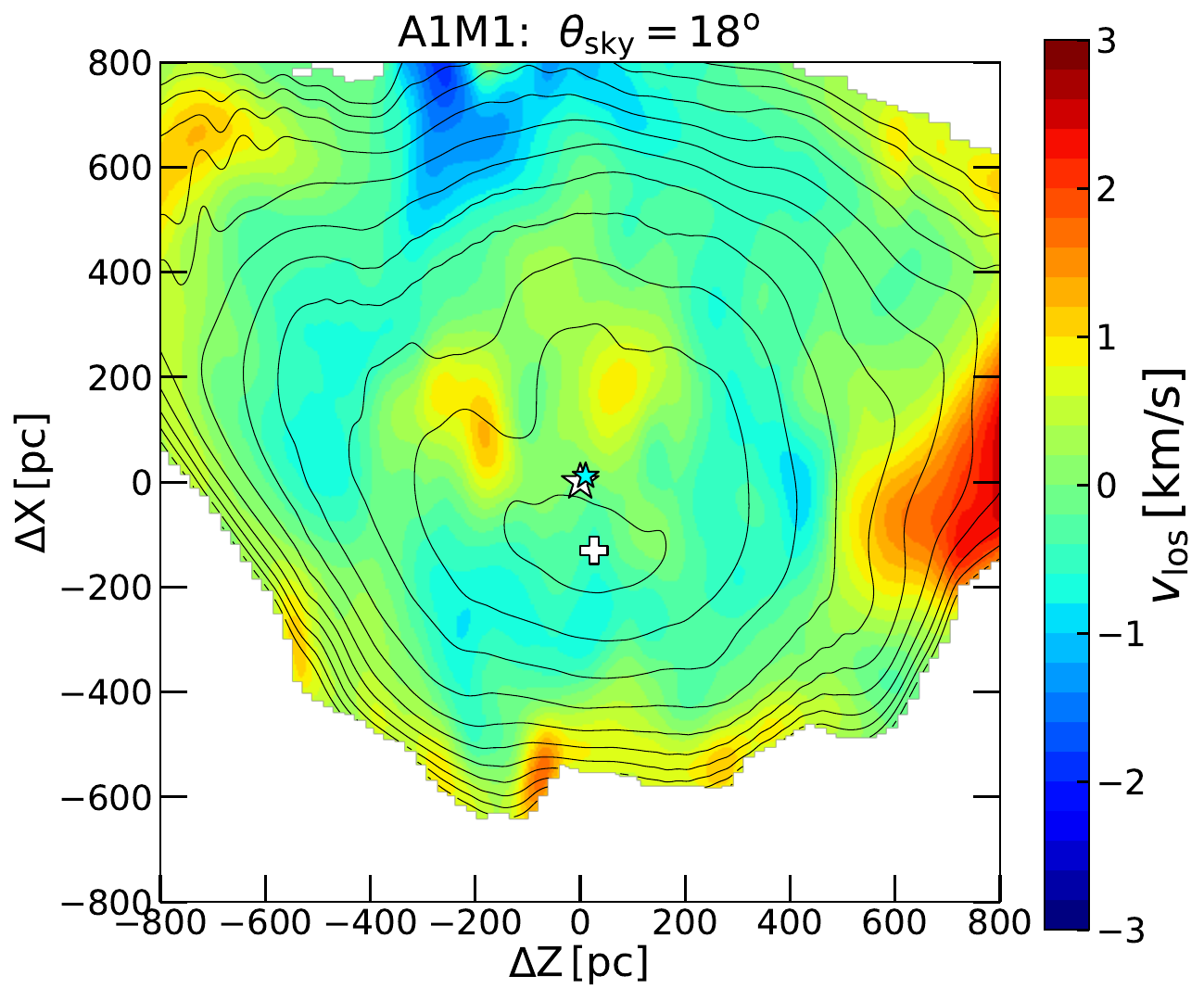}
\includegraphics[width={0.32\textwidth}]{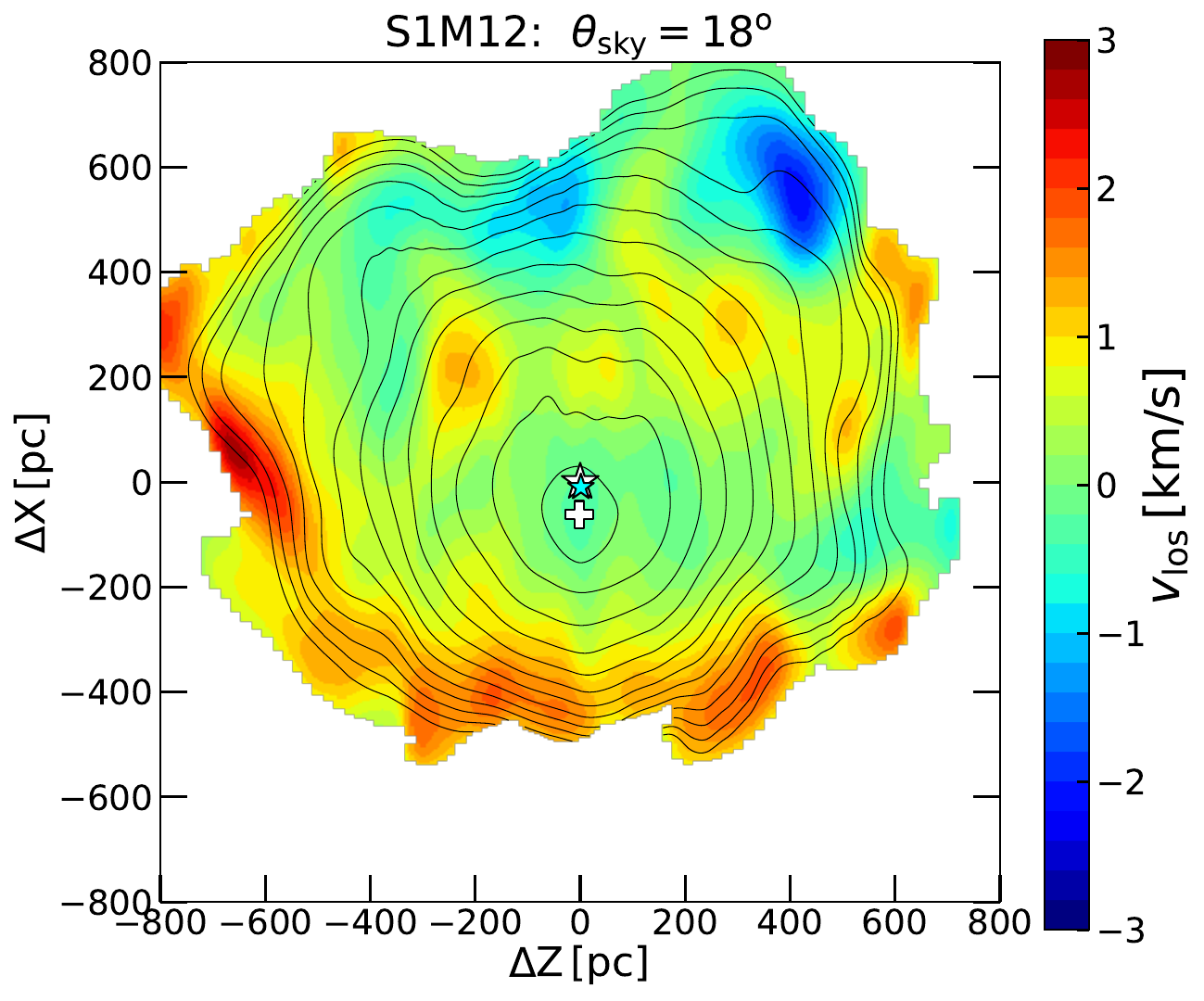}
%\vspace{-0.5cm}
\caption{\chA{Testing the effects of the WSRT data beam size in models from Scenario I (left column) with the IGM perturbed model M5 (S4M9) and from Scenario II with models M8 (A1M1) and M10 (S1M12) in the middle and right columns, respectively. 
The maps of the gas surface mass density (top row) and the velocity field (bottom row) are convolved with a Gaussian (2D) kernel with the observed beam size \citepalias{Adams2018} of 2($\sigma_{\rm z}$,$\sigma_{\rm x}$) =(15.7, 57.3)\as = (31, 113)\pc, showed in the top right panel (black ellipse).
The centre of mass of the dark-matter/old stellar population (white star),
the younger stellar population (cyan star), and the gas surface density peak (cross) are marked.
We note  that 1) the Gaussian smoothing with the masked hot gas results in some masked cold gas cells as well, and 2) $\Sigma_{\rm gas}\!=\!1\sm\pc^{-2}$ corresponds here to $N_{\HI}\!=\!1.2\!\times\!10^{20}\icmsq$.}}
\label{fig:fig_beamsize}
\end{center}
\end{figure*}
~~~~~~~~~~~~~~~~~~~
~~~~~~~~~~~~~~~~~
~~~~~~~~~~~~~~~~~~~
~~~~~~~~~~~~~~~~~~~
~~~~~~~~~~~~~~~~~~~
~~~~~~~~~~~~~~~~~~~
\begin{figure*}[htb]
\begin{center}
\includegraphics[width=9cm]{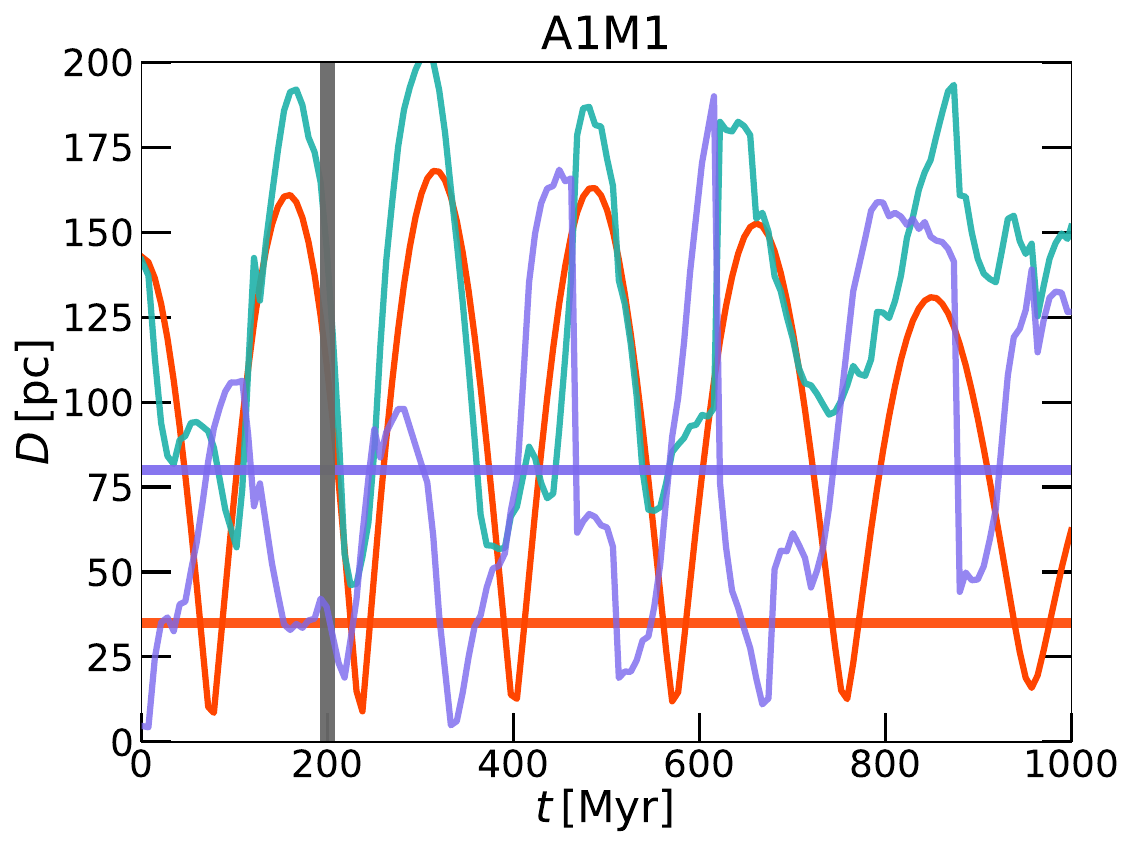}
\includegraphics[width=9cm]{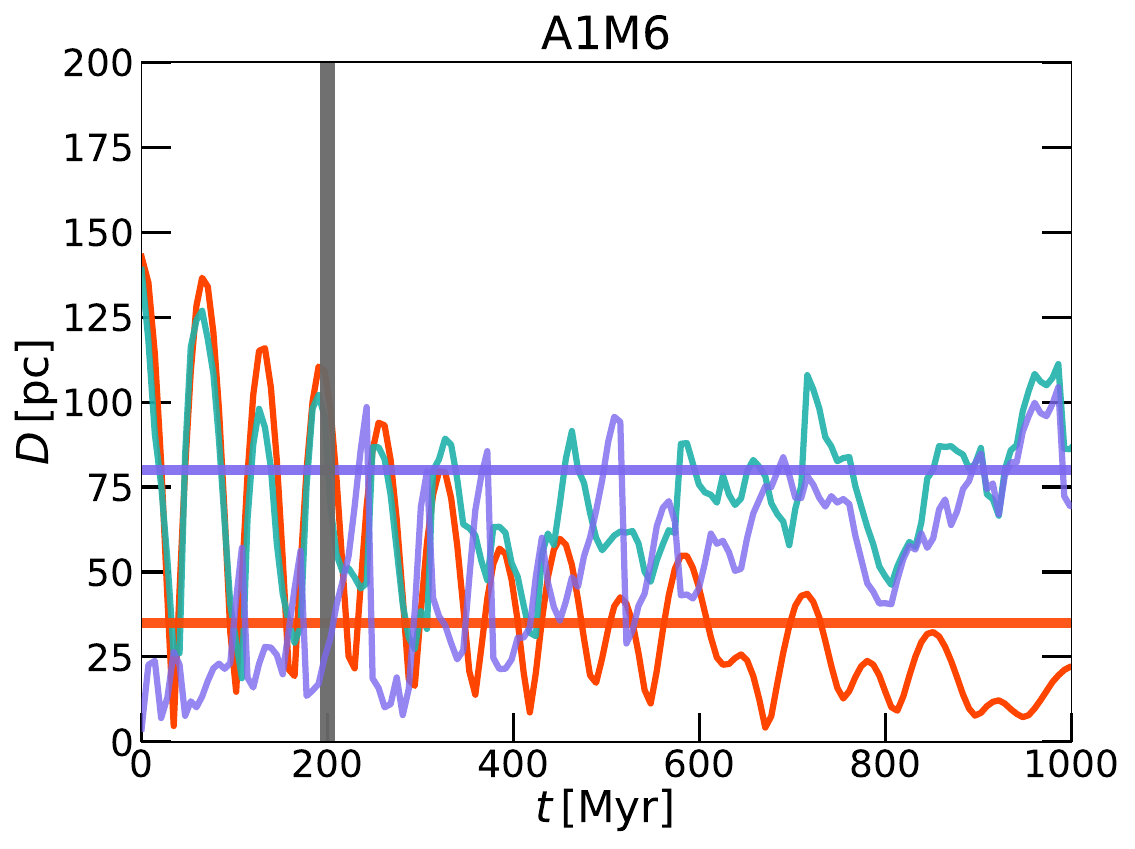}
%\vspace{-0.3cm}
\caption{\chA{Scenario~II: Temporal evolution of model M8 (A1M1) (left panel), which uses the dwarf model D1, model M9 (A1M6) using D2 (right panel) measured after convoluting the images with the WSRT observations beam size of 2($\sigma_{\rm z}$,$\sigma_{\rm x}$) =(15.7, 57.3)\as = (31, 113)\pc \citepalias{Adams2018}. The measurements without the beam size effect are shown in Fig.~\ref{fig:fig_distance}.
We show the distances with respect to the centre of mass of the dark-matter halo of the younger stellar component (red curve)
and of the peak of the gas surface mass density (blue) measured with $\theta_{\rm sky}=0\deg$ to quantify the maximum distances it could reach. 
We also show the distance between the younger stellar component centre and the gas surface density peak (turquoise). 
The observed 80\pc offset between the gas density peak and the stellar component in Leo~T is indicated with a blue horizontal line \citepalias{Adams2018}. The 35\pc offset between the younger and older stellar components is shown with the red horizontal line, while the vertical line indicates the lower age estimate of the younger stellar component \citep{DeJong2008}.}}
\label{fig:fig_distance_gauss}
\end{center}
\end{figure*} 

~~~~~~~~~~~~~~~~~~~
~~~~~~~~~~~~~~~~~
~~~~~~~~~~~~~~~~~~~
~~~~~~~~~~~~~~~~~~~
~~~~~~~~~~~~~~~~~~~
~~~~~~~~~~~~~~~~~~~
\begin{figure*}[htb]
\begin{center}
\includegraphics[width=5.9cm]{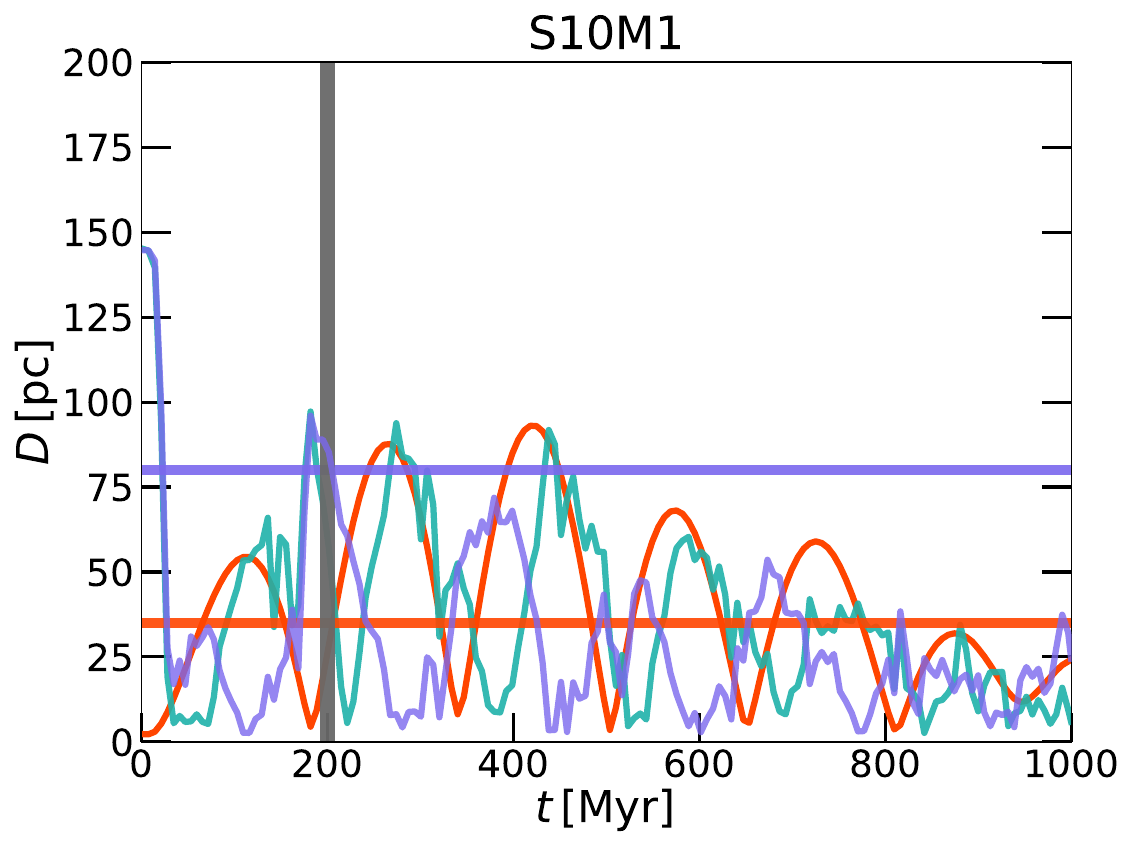} % 1 comp
\includegraphics[width=5.9cm]{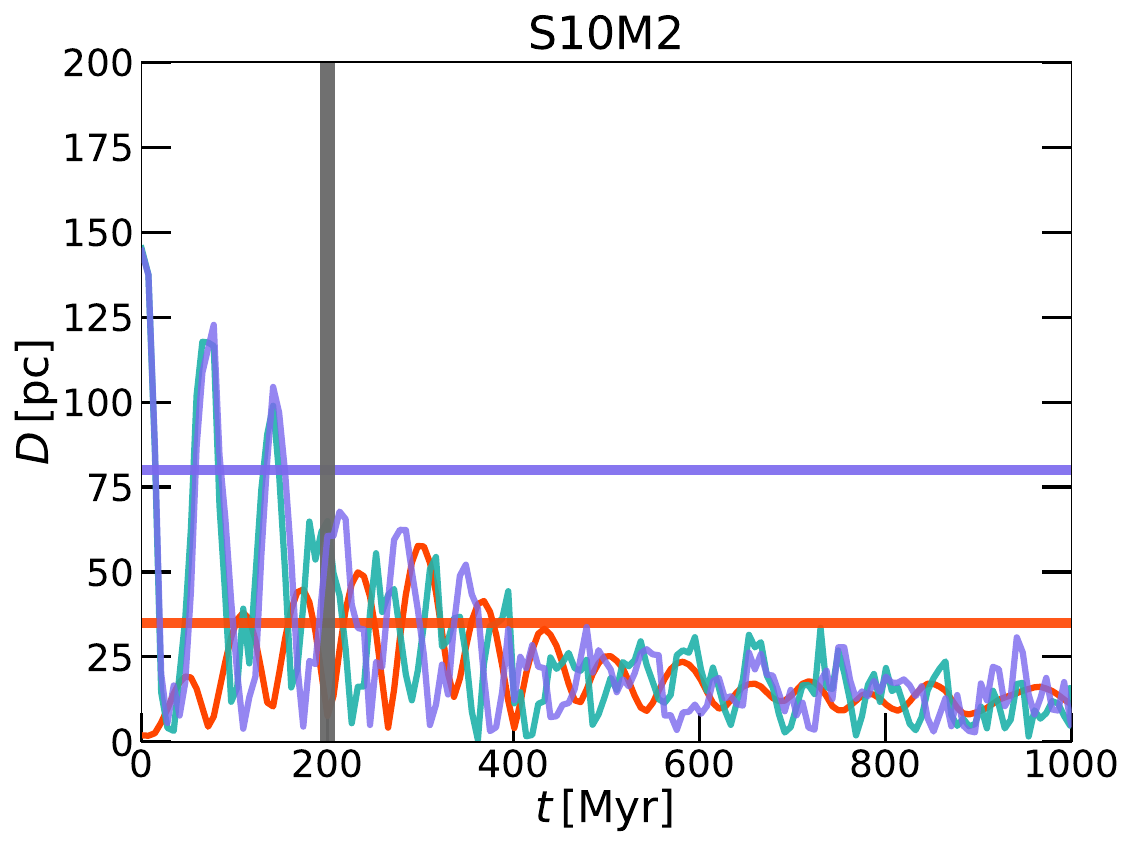} % 1 comp
\includegraphics[width=5.9cm]{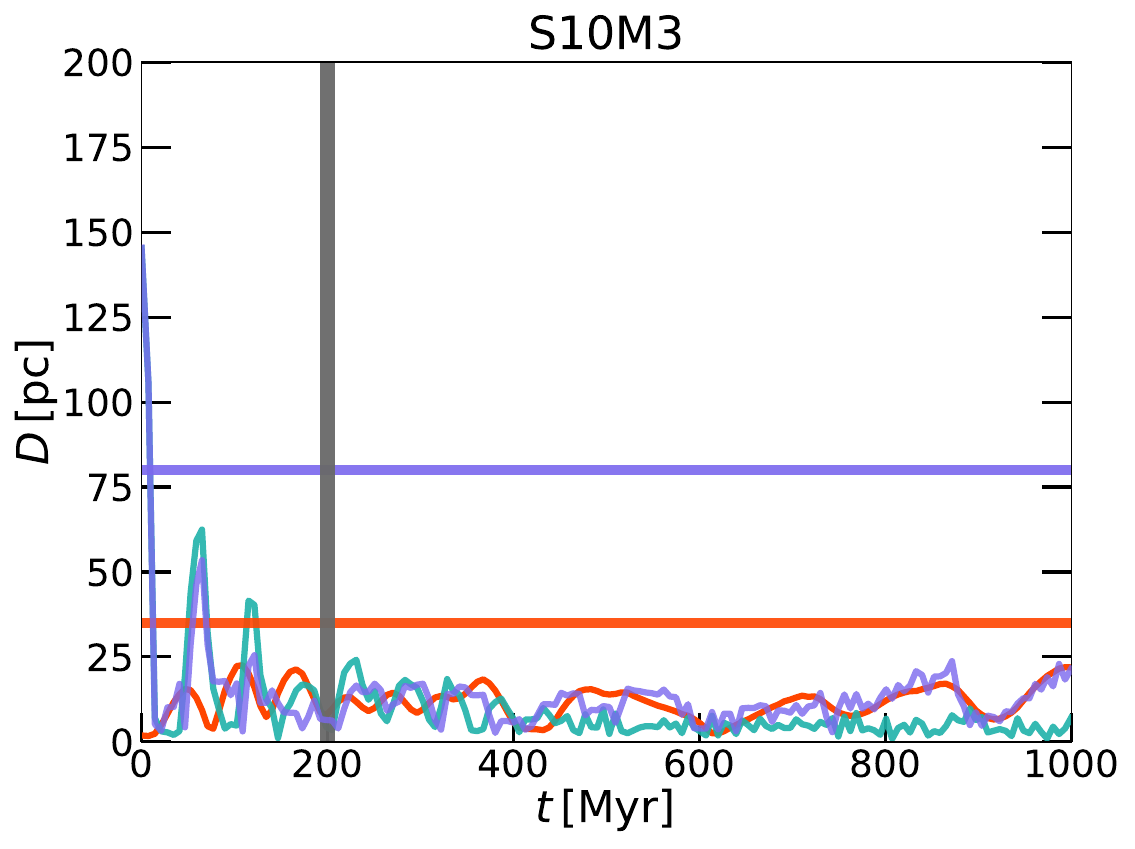} % 1 comp
\includegraphics[width=5.9cm]{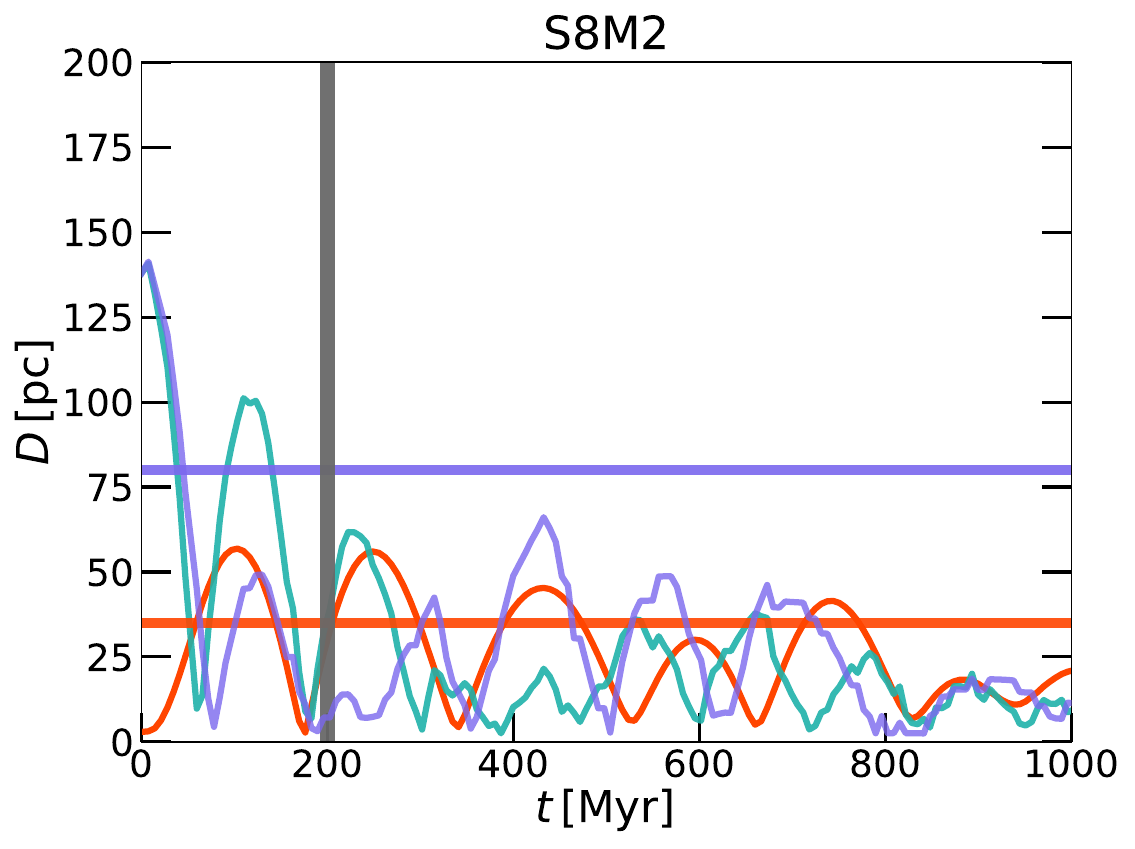} % 2 comp
\includegraphics[width=5.9cm]{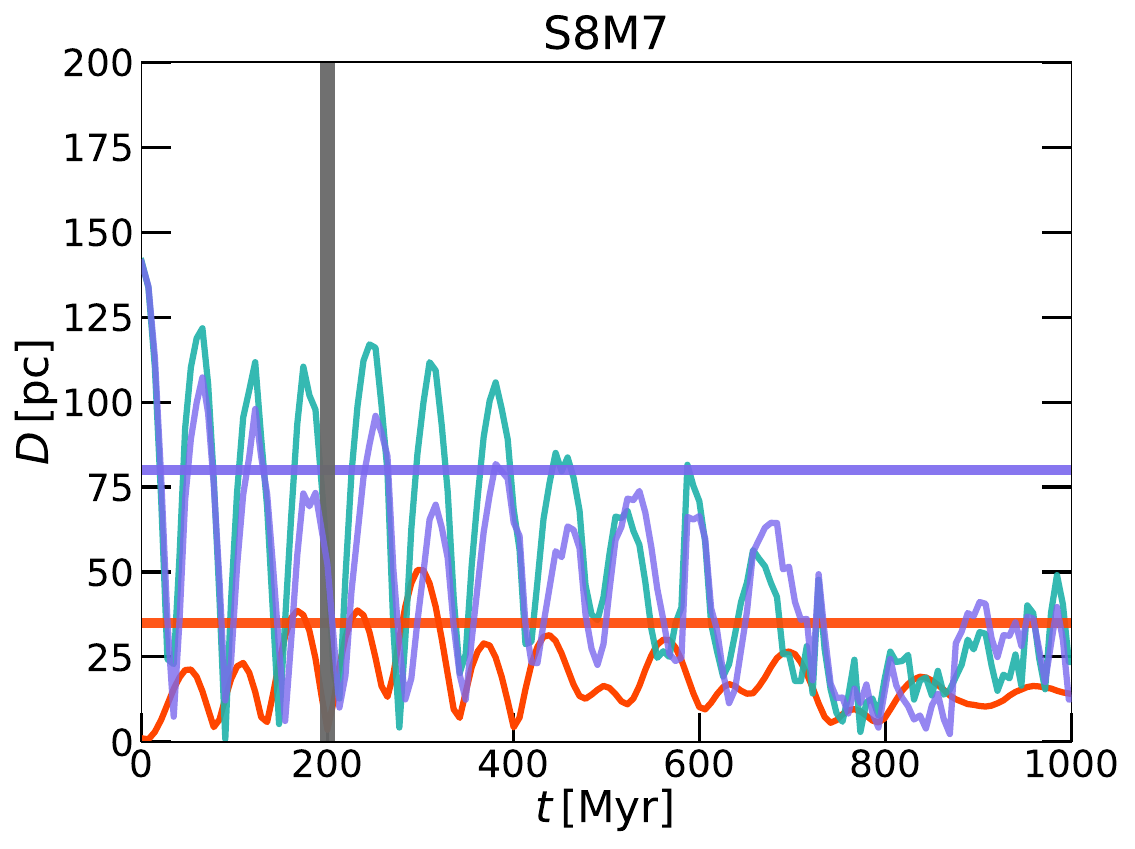} % 2 comp
\includegraphics[width=5.9cm]{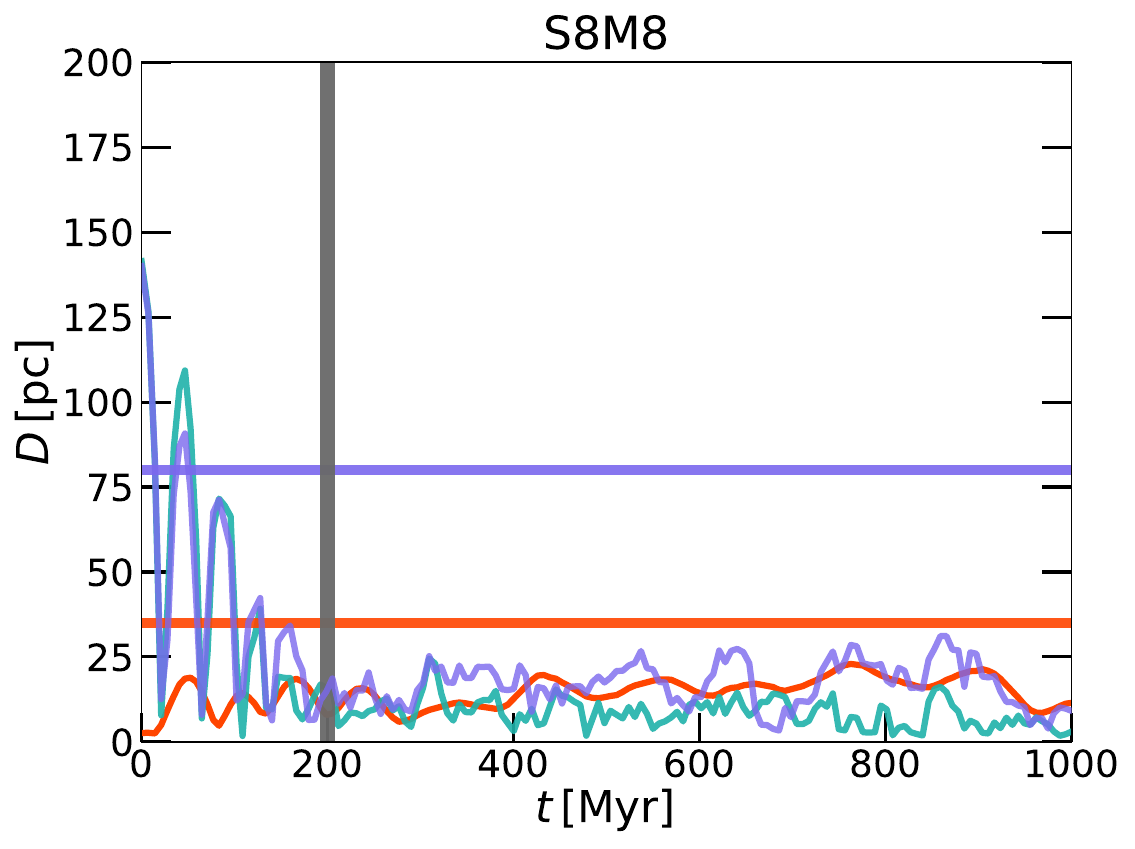} % 2 comp
%\vspace{-0.2cm}
\caption{Scenario III: Gas offset models, showing the temporal evolution of the centre of mass of the stellar components (red), the peak of the surface gas density (blue) to the dark matter centre of mass. 
The top panels show models with one gas component, whereas the bottom panels show models with two gas components.
We also show the distance between the stellar centre of mass and the gas surface density peak (turquoise). The respective simulations and used dwarf models are listed in the Table \ref{tab:setup} correspond to model M15 with the dwarf model D1 (top left, S10M1), M16 with D2 (top middle, S10M2), M17 with D4 (top right, S10M3), M18 with D1t (bottom left, S8M2), M19 with D2t (bottom middle, S8M7) and M20 with D4t (bottom right, S8M8).}
\label{fig:fig_distance_shiftgas}
\end{center}
\end{figure*}
% \end{minipage}

~~~~~~~~~~~~~~~~~~~
~~~~~~~~~~~~~~~~~
~~~~~~~~~~~~~~~~~~~
~~~~~~~~~~~~~~~~~~~
~~~~~~~~~~~~~~~~~~~
~~~~~~~~~~~~~~~~~~~
\end{appendix}

\end{document}